\documentclass[smalltextended, 12pt, final]{svjour3}[2020]
\usepackage{fullpage}
\usepackage{graphicx}
\usepackage{lineno}
\usepackage{rotating}
\usepackage{multirow}
\usepackage{xcolor}
\usepackage{hyperref}
\usepackage{tikz}
\usepackage{tikz-3dplot}
\usepackage{circuitikz}
\usepackage{siunitx}
\usetikzlibrary{decorations.pathmorphing}

\newcommand{\zero}{\phantom{0}}
\newcommand{\dg}{\phantom{$^\dagger$}}
\newcommand{\ddg}{\phantom{$^\ddagger$}}

\usepackage{etoolbox}

\newcommand*{\affmark}[1][*]{\textsuperscript{#1}}
\usepackage{latexsym}
\usepackage[gen]{eurosym}

\journalname{Experimental Astronomy}

\begin{document}

\title{A chemically etched corrugated feedhorn array for D-band CMB observations}

\author{S. Mandelli\affmark[1, 2, *] \and
        E. Manzan \affmark[1] \and
        A. Mennella \affmark[1, 2] \and
        F. Cavaliere \affmark[1, 2] \and
        D. Vigan\`o \affmark[1, 2]\and
        C. Franceschet \affmark[1, 2] \and
        P. de Bernardis \affmark[3, 4] \and
        M. Bersanelli \affmark[1, 2] \and
        M. G. Castellano \affmark[5] \and
        A. Coppolecchia \affmark[3, 4] \and
        A. Cruciani \affmark[4] \and
        M. Gervasi \affmark[6, 7] \and
        L. Lamagna \affmark[3, 4] \and
        A. Limonta \affmark[6, 7] \and
        S. Masi \affmark[3, 4] \and
        A. Paiella \affmark[3, 4]\and
        A. Passerini \affmark[6, 7] \and
        G. Pettinari \affmark[5] \and
        F. Piacentini \affmark[3, 4] \and
        E. Tommasi \affmark[8] \and
        A. Volpe \affmark[8] \and
        M. Zannoni \affmark[6, 7]
}
\authorrunning{S. Mandelli \and
        E. Manzan \and
        A. Mennella \and
        F. Cavaliere \and
        D. Vigan\`o \and
        C. Franceschet \and
        P. de Bernardis \and
        M. Bersanelli \and
        M. G. Castellano \and
        A. Coppolecchia \and
        A. Cruciani \and
        M. Gervasi \and
        L. Lamagna \and
        A. Limonta \and
        S. Masi \and
        A. Paiella \and
        A. Passerini \and
        G. Pettinari \and
        F. Piacentini \and
        E. Tommasi \and
        A. Volpe \and
        M. Zannoni
        } 

\institute{1. UNIMI - Department of Physics - Via Celoria 16 - Milan\and \\
           2. INFN - Milano - Via Celoria 16 - Milan \and \\
           3. Universit\`a di Roma La Sapienza - Department of Physics - Piazzale Aldo Moro, 5, - Roma \and \\
           4. INFN - Roma 1 - Piazzale Aldo Moro, 5, - Roma \and \\
           5. CNR/IFN - Via Cineto Romano, 42 - Roma \and \\
           6. UNIMIB - Universit\`a Milano Bicocca - Department of Physics - Piazza della Scienza, 3 - Milan \and \\
           7. INFN - Milano Bicocca - Piazza della Scienza, 3 - Milan \and \\
           8. ASI - Agenzia Spaziale Italiana - Via del Politecnico snc - Roma \and \\
           *. stefano.mandelli@unimi.it
}

\date{Received: date / Accepted: date}

\maketitle

\begin{abstract}

    We present the design, manufacturing, and testing of a 37-element array of corrugated feedhorns for Cosmic Microwave Background (CMB) measurements between 140 and 170\,GHz. The array was designed to be coupled to Kinetic Inductance Detector arrays, either directly (for total power measurements) or through an orthomode transducer (for polarization measurements). We manufactured the array in platelets by chemically etching aluminum plates of $0.3$\,mm and $0.4$\,mm thickness. The process is fast, low-cost, scalable, and yields high-performance antennas compared to other techniques in the same frequency range. Room temperature electromagnetic measurements show excellent repeatability with an average cross polarization level about $-20$\,dB, return loss about $-25$\,dB, first sidelobes below $-25$\,dB and far sidelobes below $-35$\,dB. Our results qualify this process as a valid candidate for state-of-the-art CMB experiments, where large detector arrays with high sensitivity and polarization purity are of paramount importance in the quest for the discovery of CMB polarization $B$-modes.


\end{abstract}
\section{Introduction}
\label{sec:intro}
    The Cosmic Microwave Background (CMB) is one of the most powerful probes that allows us to study the early Universe and constrain cosmological parameters to sub-per-cent precision\cite{aghanim2018planck}. In the context of the Big Bang cosmology, the CMB is a relic radiation from the early stage of our Universe. This radiation coupled with the primordial hot baryonic plasma for the first $\sim 380\,000$ years of the Universe evolution. When the plasma temperature fell below $\sim 3\,000$\,K, matter became neutral and radiation decoupled, propagating freely in the expanding universe. Today, we detect this radiation as a black-body emission at $\sim 2.73$\,K, with a brightness peak at $\sim 160$ GHz. The CMB intensity anisotropies, $\Delta T / T\sim 10^{-5}$, trace the primordial density fluctuations, while polarization anisotropies, $\Delta P / P\sim 10^{-6}$, were generated at the last scattering surface by variations both in the matter density (scalar fluctuations) and in the gravitational field (tensor fluctuations).

    We can decompose the polarization pattern in the sky in two different components, the so-called $E$-modes and $B$-modes \cite{Kamionkowski1997}. $B$-modes, in particular, arise from tensor fluctuations, like those generated on small scales by gravitational lensing and on large scales by primordial gravitational waves, as is predicted by inflationary models.

    The BICEP2/Keck, POLARBEAR and, more recently, the South Pole Telescope (SPT) teams have measured the lensing $B$-modes power spectrum  \cite{ade2016bicep2,polarbear2014,spt2019}, allowing them to determine the dark matter lensing potential with unprecedented precision. The large-scale $B$-modes from primordial gravitational waves, instead, have not been detected yet.

    The detection of cosmological $B$-modes is the primary goal of several ongoing and planned observations from the ground, stratosphere, and space  (see \cite{Staggs_2018} for a review). In all cases, the challenge is to develop large polarization-sensitive detector arrays with ultra-low noise and negligible systematic effects.

    In this context, there is great interest in developing technologies to produce large antenna arrays to be coupled to low-noise detectors like Transition Edge Sensors (TES, \cite{irwin1995application,Bastia_2019}) and Kinetic Inductance Detectors (KIDs, \cite{day2003broadband,masi2019kinetic,Paiella_2019,Steinbach2018}). Critical assets are scalability, low-cost, and high optical performance in the frequency range 95--220\,GHz, which is currently the most exploited CMB observational window.

    In this paper we address this challenge and propose chemical-etching combined with the so-called platelet technique to build high-performance arrays of corrugated feedhorns in the D-band (110-170\,GHz) with very low-cost and processing time. The platelet technique (described, for example, in \cite{del2011w}) consists of piercing  thin metallic (or metal-coated) plates that are subsequently overlapped and mechanically assembled to produce the desired horn profile.

    Traditional fabrication techniques for corrugated feedhorns include electroforming and direct milling. Electroforming ensures excellent mechanical accuracy but is expensive and not scalable to large arrays. Milling also yields excellent accuracy, with a precision of about 0.03\,mm, and it is scalable to large numbers, and although it is limited by long processing times.

    Chemical-etching can solve scalability, cost and time issues, at the price of tight requirements in the control of the etching process (acid composition and erosion time) to guarantee the necessary mechanical accuracy. In our paper we show that these issues can be successfully dealt with allowing one to produce antennas with state-of-the-art performance.

    Our prototype consists of a 37-elements array of corrugated feedhorns in hexagonal configuration, designed to be coupled to kinetic inductance detectors (KIDs) both directly (with the total intensity illuminating the detector) and through an orthomode transducer, to discriminate polarization. In section~\ref{proto} we present the electromagnetic and mechanical design, detail the manufacturing process, and describe the metrological measurements assessing the achieved mechanical tolerance. In section~\ref{El:Ch} we discuss the radiation patterns measured in the anechoic chamber of our laboratory in Milan. In section~\ref{discuss} we summarize the main performance and compare our prototype with others in the literature in the same working frequency range. We also discuss possible ways to improve the achieved performance.

\section{The D-band array prototype}
\label{proto}
The prototype was not developed to be installed in the focal plane of a particular telescope, so we set the requirements listed in Table~\ref{feed:req} considering the optics of typical space CMB experiments like Planck \cite{Tauber2010} and COrE \cite{bouchet2015core}.

\begin{figure}[htbp]
  \centering
  \includegraphics[width=0.8\textwidth]{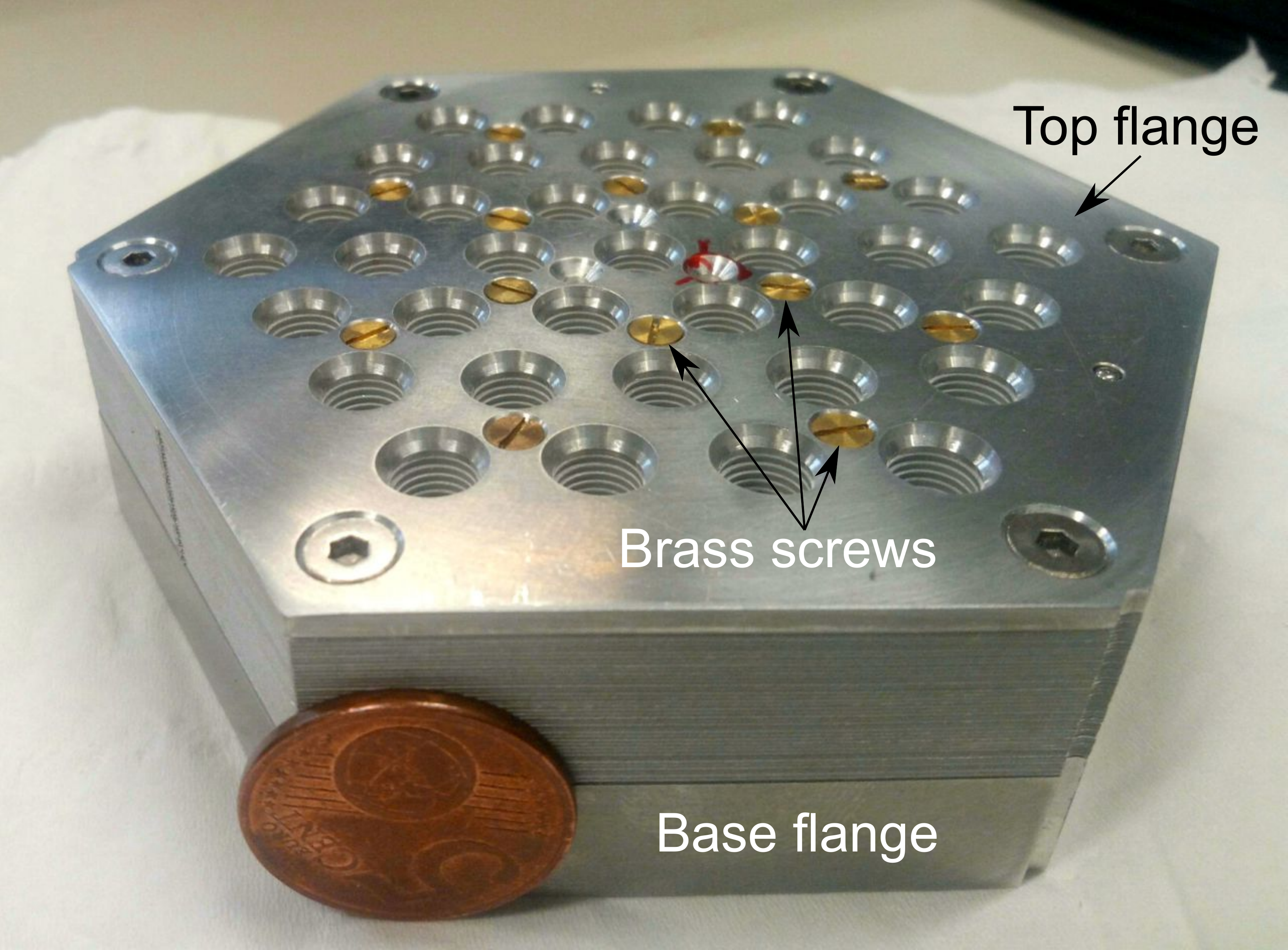}
  \caption{\label{KID:ARR}Picture of the integrated 37-elements corrugated feedhorn array. Fourteen brass screws tighten the middle plates between the top and bottom flanges.}
\end{figure}

    \subsection{Electromagnetic Design}
    \label{sec_electromagnetic_design}
        In our prototype we implemented a dual-profiled, corrugated design that allowed us to obtain compact horns with highly symmetric and Gaussian main beams, low sidelobes, and low cross polarization \cite{Clarricoats}. The first four corrugations mix the two fundamental modes of propagation TE11 and TM11 to create a hybrid mode HE11, characterized by a low level of cross polarization.

        The feedhorn design is inspired by the W-band prototype proposed by Del Torto \& al.\cite{del2011w}, which was adequately rescaled for the D-band. The optimization work was focused on designing a new top-flange suitable to ensure an adequate stiffness of the prototype and avoid degradation of the electromagnetic performance. The thickness of the plates is constrained by the aluminum manufactured companies which produce a minimum standard thickness of $0.3\,$mm and $0.4\,$mm.

        The electromagnetic simulations and optimization were perfomed with the software CST Microwave Studio using the \textit{Time Domain} solver based on \textit{Finite Integration Technique}. 

            \begin{table}[htbp]
                \renewcommand{\arraystretch}{1.5}
                \centering
                \caption{\label{feed:req}Electromagnetic requirements of the feedhorn array.}
                \begin{tabular}{p{5cm} r}
                    \hline
                    Parameter & Requirement  \\
                    \hline
                    \hline

                    FWHM\dotfill & $20$ deg \\
                    Max. cross polarization\dotfill& $< - 35$ dB    \\
                    First Sidelobes level\dotfill & $< - 30$ dB    \\
                    Far Sidelobes level\dotfill & $< - 50$ dB \\
                    Return Loss\dotfill & $< - 30$ dB    \\
                    Bandwidth\dotfill & $1:1.31$     \\
                    \noalign{\smallskip}\hline
                \end{tabular}
            \end{table}

        \subsubsection{Corrugations profile}
        \label{sec_corrugations_profile}

            The horn has a linear aperture of $6\,deg$ and is profiled with a squared sinusoid section followed by an exponential flare.

\begin{figure}[htbp]
                \centering
                \includegraphics[width=0.95\textwidth]{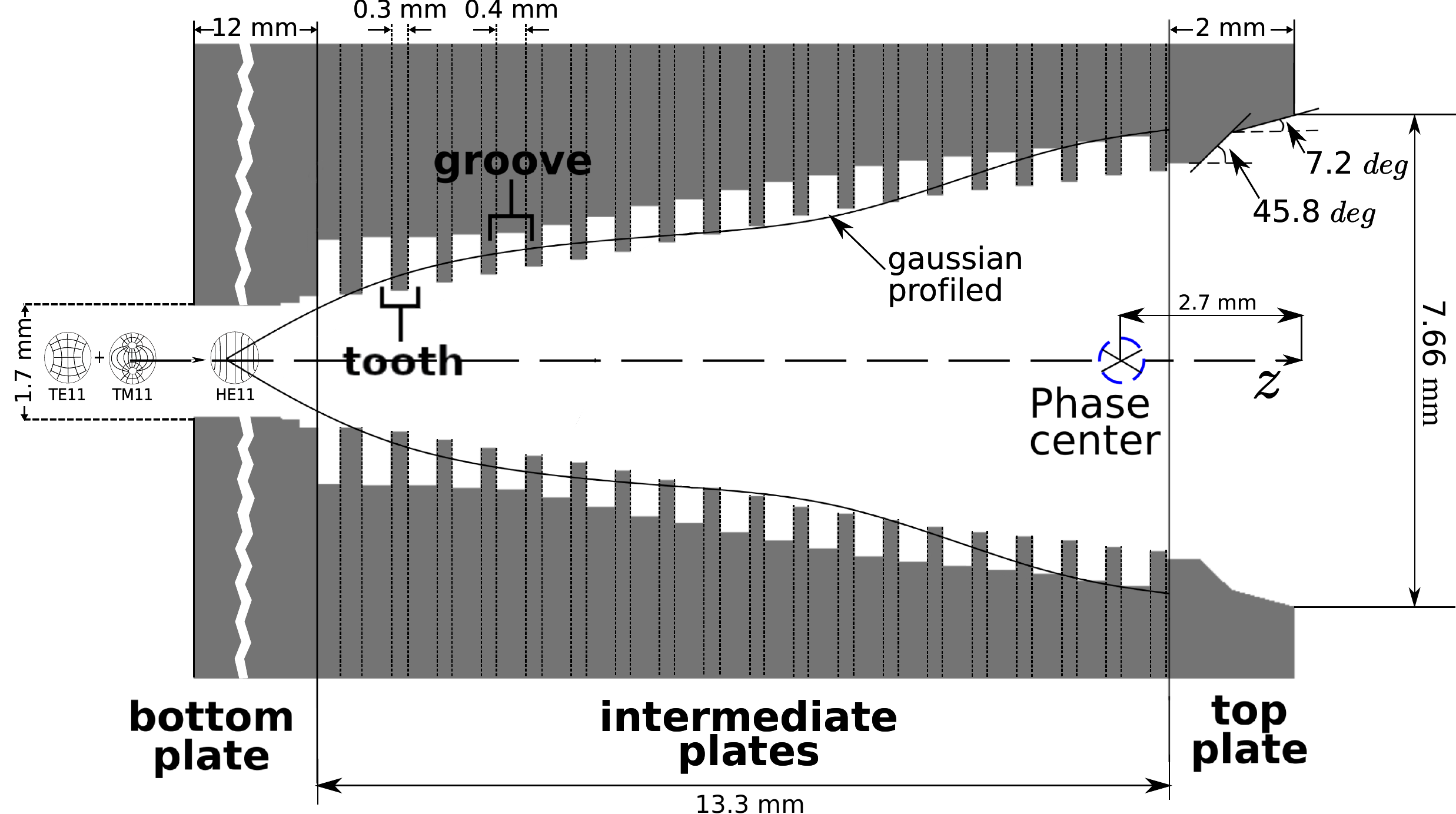}
                \caption{\label{model:cad}Details of the feedhorn profile. The inner section is made of 0.3\,mm and 0.4\,mm stacked plates clamped between a pair of flanges containing the horn aperture and the output waveguide. In the figure, the bottom flange is not drawn to scale.}
            \end{figure}

            Figure~\ref{model:cad} shows the geometrical details of the feedhorn profile. The inner structure is made of $0.3$\,mm and $0.4$\,mm stacked platelets that reconstruct, respectively, the teeth and grooves sequence. The stack is clamped mechanically with brass screws between a pair of thick plates (the 2\,mm top plate, containing the horn aperture, and the 12\,mm bottom plate, containing a two-steps impedance matching transition to minimize reflections and terminating with a 1.7\,mm diameter circular waveguide). This process does not need any glues or soldering medium.

        \subsubsection{Thermo-mechanical assessment}
        \label{sec_thermal_assessment}
\begin{figure}[htbp]
                \centering
                \begin{tabular}{c  c}
                    \includegraphics[width=0.45\textwidth]{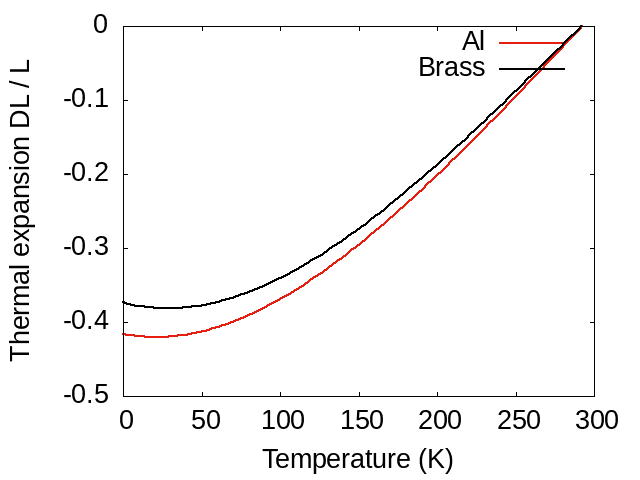}&
                    \includegraphics[width=0.45\textwidth]{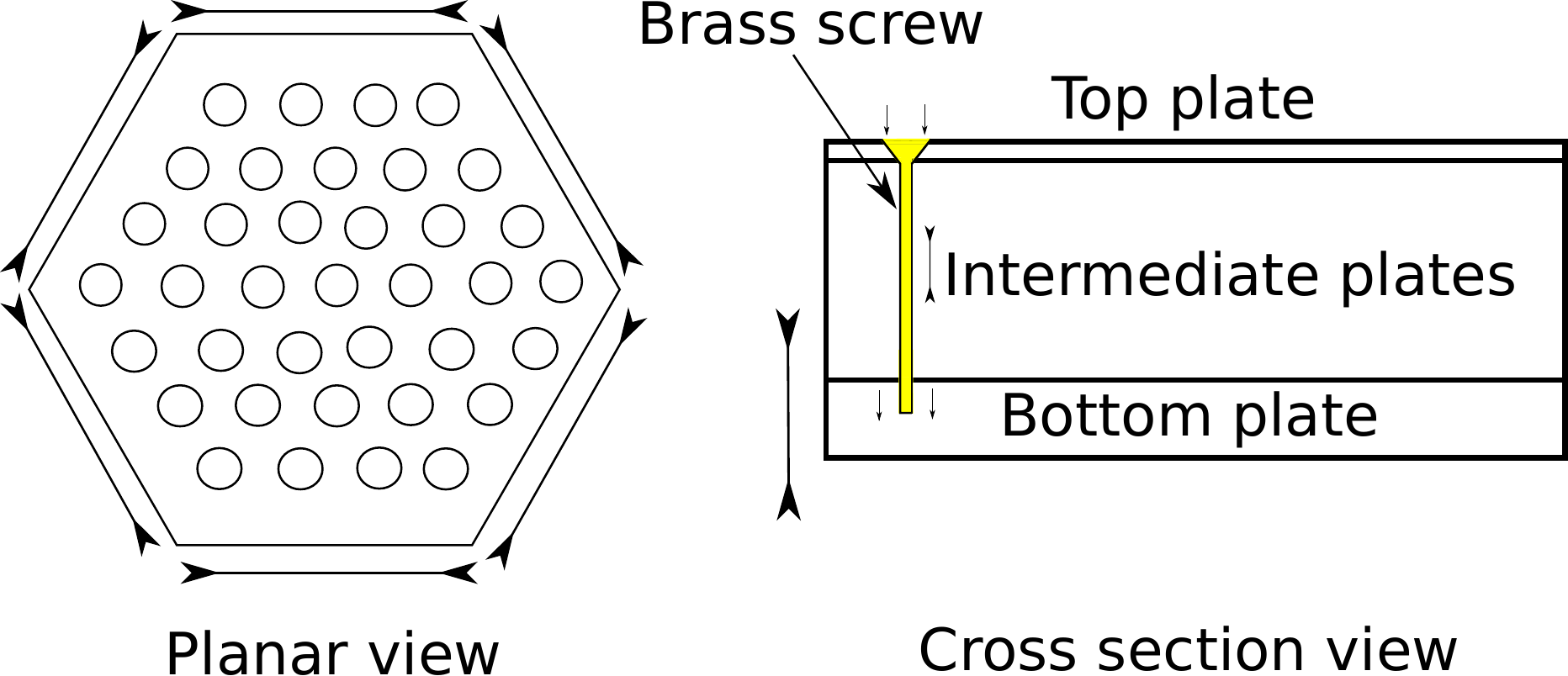}\\
                \end{tabular}
                \caption{\label{thermal:coefs}\emph{Left panel:} shows the two different thermal contractions compared to the reference length at $293$K \cite{ekin2006experimental} \emph{Right panel:} shows the main contraction of the middle plates during cool-down. The contraction affect mostly the surface area. The differential contraction of the bottom plate compared to the brass screws increases the tightening of the array.}
            \end{figure}

            The array is conceived to be used at cryogenic temperature ($\sim 0.1$\,K), where brass screws and aluminum plates are characterized by slightly different expansion coefficients. In Fig.\,\ref{thermal:coefs} are shown in black and red the two different thermal contraction for brass and aluminum, respectively. 
            
            In the right panel of Fig.\,~\ref{thermal:coefs} is shown a sketch representing how different parts in the array contract during cool-down. In particular, the intermediate plates are mostly subjected to a planar contraction while the differential in the contraction of the bottom plate compared to the brass screws result in an increased tightening of the assembly.

            The contraction of a material depends on its mechanical constraints. In the feedhorn array presented in this paper, we can find two different situations. 
            
            The first one concerns the behavior of the brass screws and the inner aluminum plates. The module is held together by the clamping screws that pressing on the top flange create a homogeneous pressure on the various aluminum plates. The plates are mechanically constrained. For this reason, they present mostly a surface contraction reducing their area, but they keep almost unchanged their thickness.

            The second one concerns the contraction the base flange thickness because it is not mechanically constrained. The differential contraction between brass and aluminum is recovered by the significant contraction of the base flange that it ensures to keep constant the pressure on the internal plates avoiding deformations or unscheduled dissasembly.

        \subsubsection{Aperture design}
        \label{sec_aperture_design}
            From the point of view of polarization purity the best design requires that the antenna cavity should be corrugated up to its aperture. This design, however, was difficult to manufacture, because the last plate, 0.3\,mm thick, would have been too thin to ensure the necessary stiffness once the stack was clamped mechanically. For this reason we terminated the aperture with a 2\,mm aluminum plate that was milled with a double-flared aperture to minimize, as much as possible, the induced spurious polarization.

\begin{figure}[htbp]
                \begin{center}
                    \includegraphics[width=0.85\textwidth]{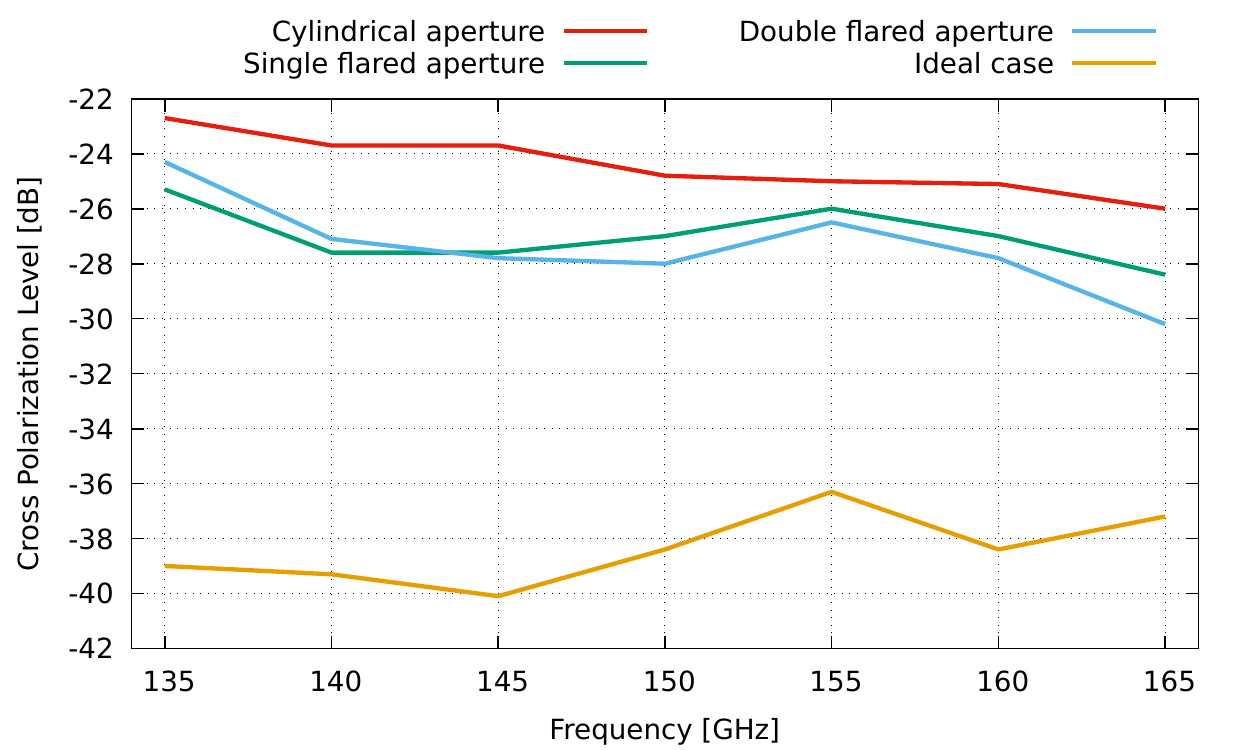}
                    \caption{ \label{arr:cross_flare}Maximum level of cross polarization obtained with various horn apertures. The dual profiled solution is the best compromise between performance and feasibility.}
                \end{center}
            \end{figure}

            Fig.~\ref{arr:cross_flare} compares the maximum level of cross polarization obtained using our solution with the ideal case (corrugations reaching the aperture) and with that obtained using other shapes (cylindrical, single flared). The figure shows that with the dual-flared aperture the maximum level of cross polarization is lower than $-25$\,dB and better than the other shapes on a wider band. The optimized aperture angles are shown in Fig.~\ref{model:cad}

        \subsubsection{Output waveguide design}
        \label{sec_output_waveguide_design}

            The size of the the feedhorn input circular waveguide is 1.7\,mm, and it is shown in Fig.~\ref{model:cad}. The \emph{bottom plate} can be interfaced with two supplementary flanges depending on purpose. The first is a waveguide orthomode transducer (OMT) that splits the radiation into two orthogonal linear polarized components. The OMT description is outside the scope of this paper and will be described in a dedicated work. The second is represented by a flared circular waveguide directly coupled with the KIDs array.

\begin{figure}[htbp]
        \begin{center}

            \begin{tabular}{c | c}
                \includegraphics[width=8.5cm]{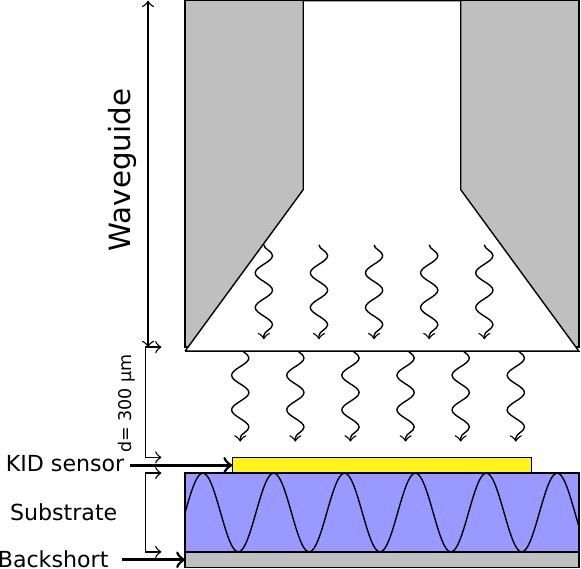} &
 	        \includegraphics[width=6.55cm]{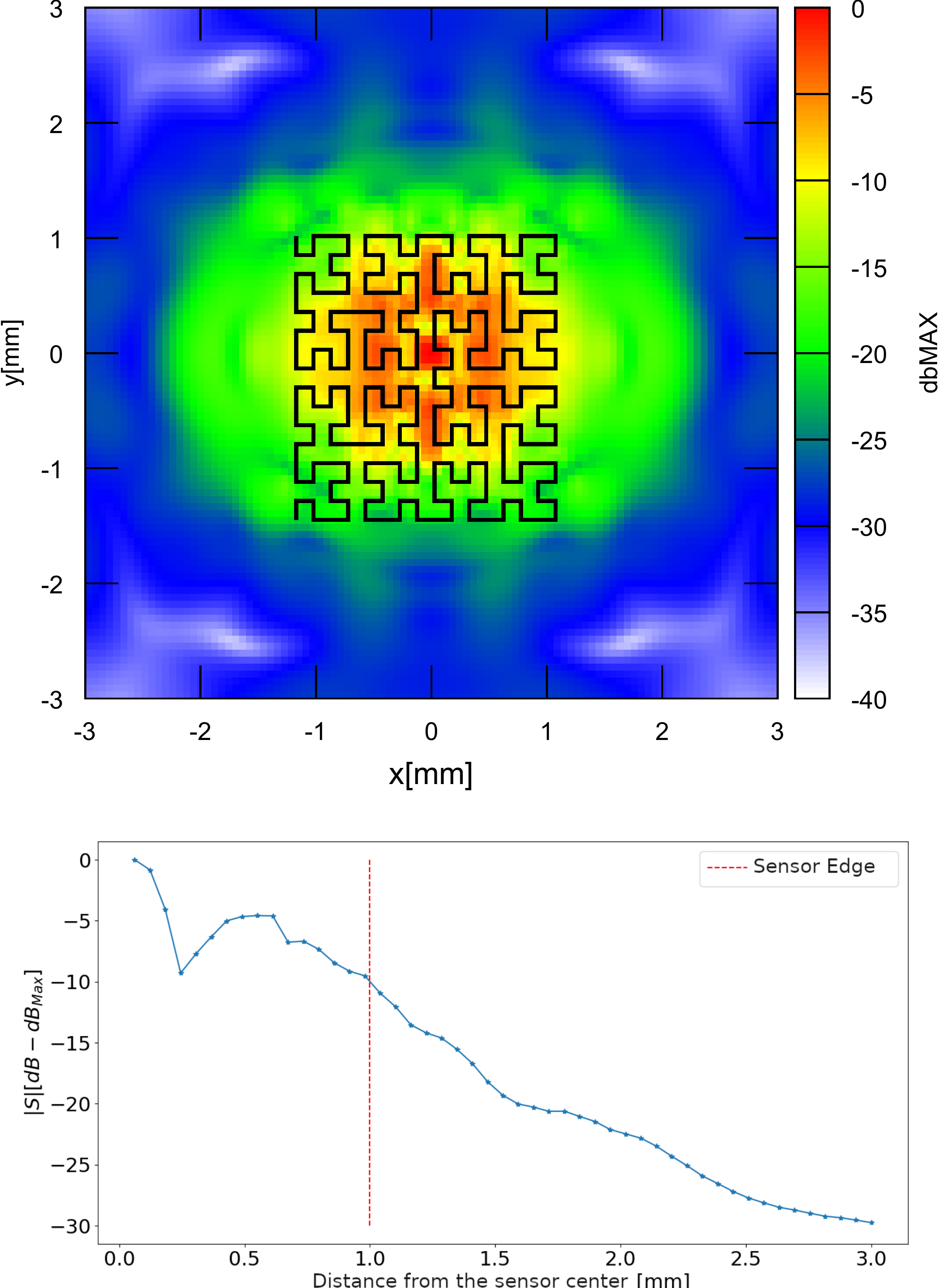}\\
            \end{tabular}

            \caption{\textit{Left panel}: design of the flared waveguide used to illuminate each KID detector. The distance of the flare aperture from the detector is set to $300\,\mu\mathrm{m}$ to maximize the power distribution over the sensor (bottom-right panel). The flare angle of the waveguide is 70$^\circ$.57. \textit{Right panel}: Distribution of the incoming radiation Poynting vector over the sensor.}
            \label{Front:Illuminated}

        \end{center}
        \end{figure}

            We have designed the flare and optimized the distance from the detector to maximize the power distribution over the sensor. The design of the flared waveguide is shown in the left panel of Fig.~\ref{Front:Illuminated}.

        The top-right panel of Fig.~\ref{Front:Illuminated} shows the spatial distribution of the Poynting vector module of the incoming radiation over the surface of the sensor, where the inductive absorber is shaped according to a Hilbert curve. This shape is particularly efficient for total power measurements and is characterized by an excellent level of absorption for both linear polarization components~\cite{d2013nika,Paiella_2019}. At the edge of the detector (1\,mm from the center) the relative power level is $-10$\,dB, falling to $-30$\,dB at 3\,mm, the distance of the closest detector in the array, ensuring that the crosstalk between two detectors is negligible.

        \subsubsection{Simulated beam patterns}
        \label{sec_simulated_beam_patterns}

\begin{figure}[htbp]
                \centering
                \includegraphics[width=0.32\textwidth]{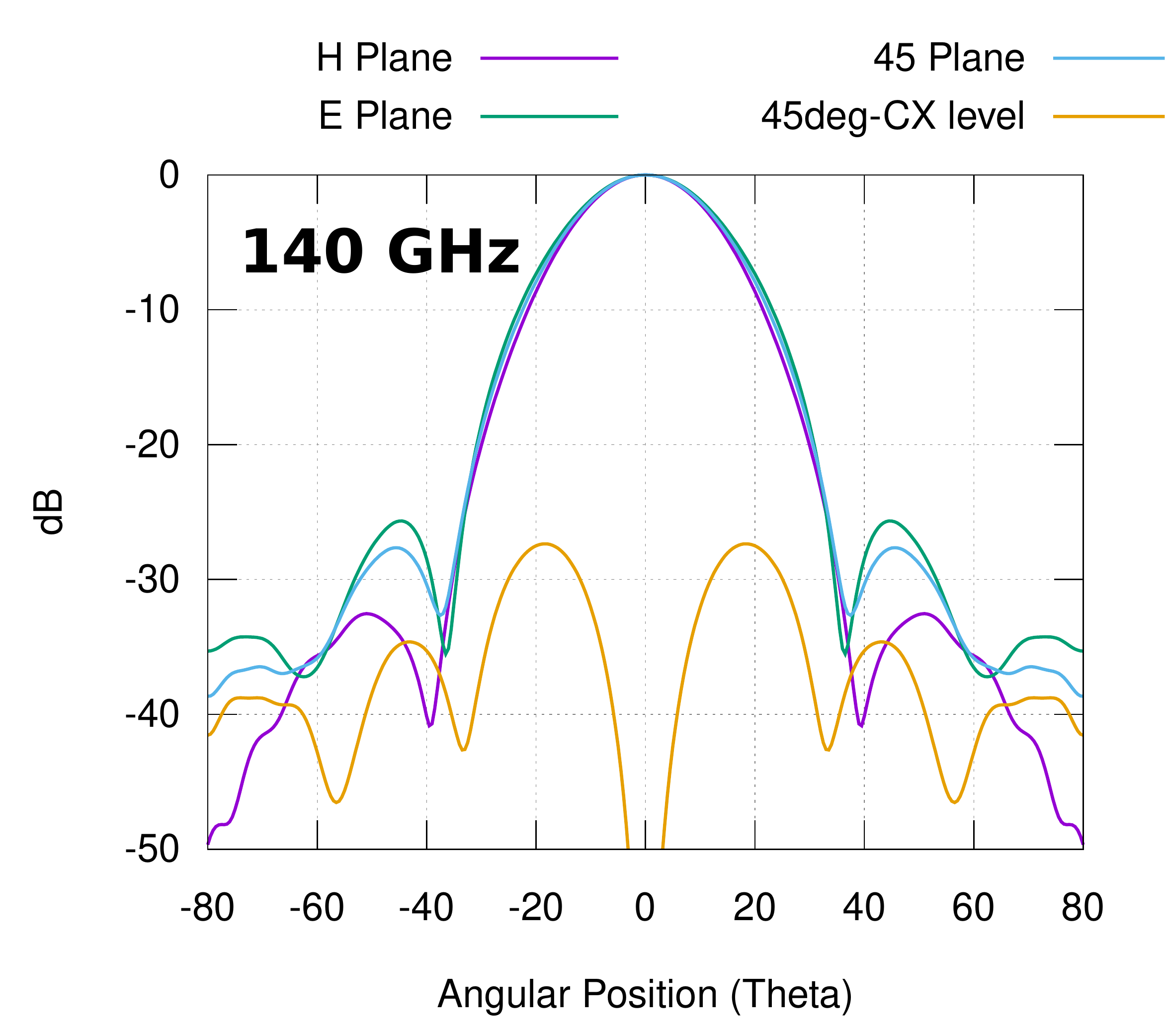}
                \includegraphics[width=0.32\textwidth]{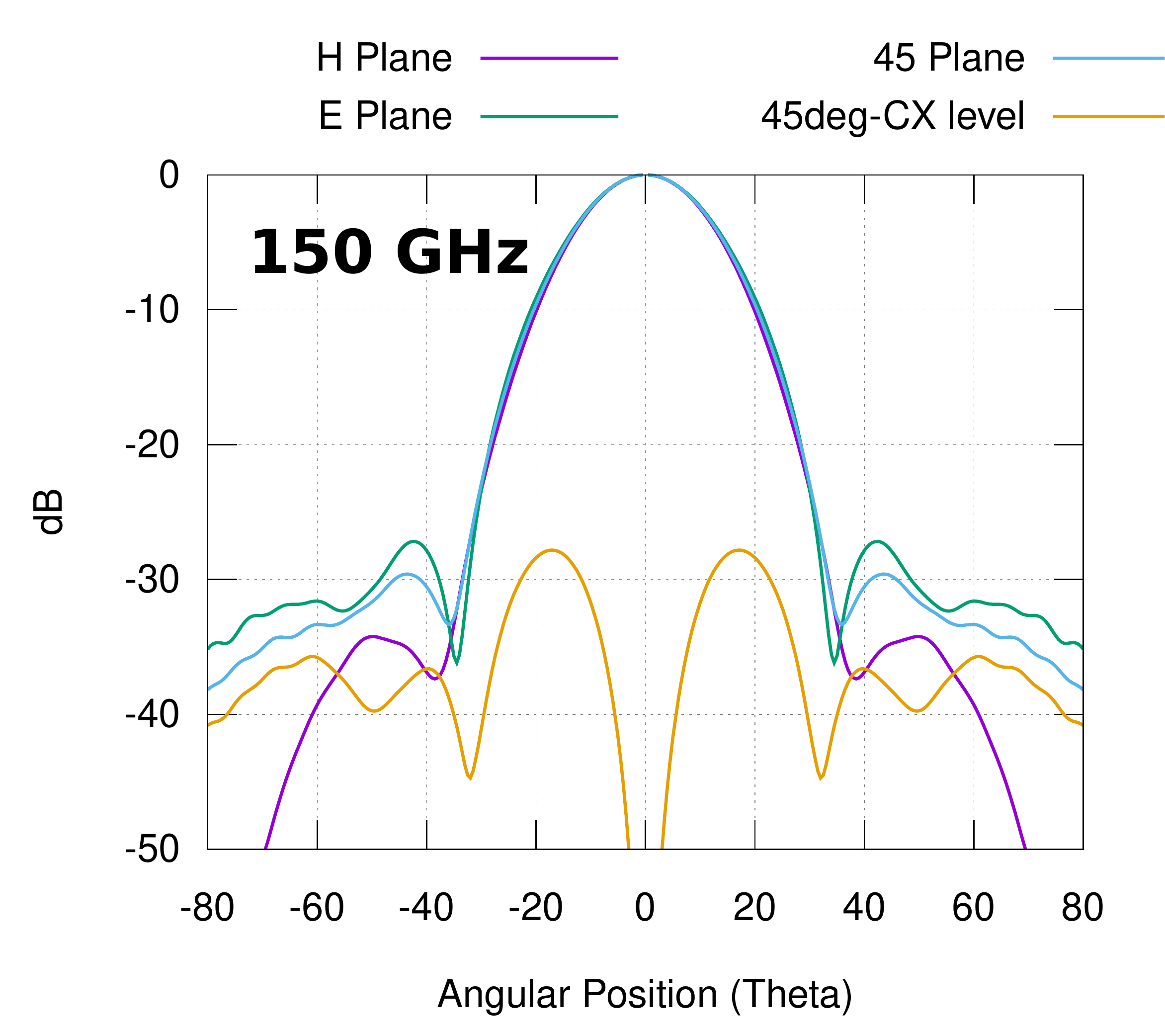}
                \includegraphics[width=0.32\textwidth]{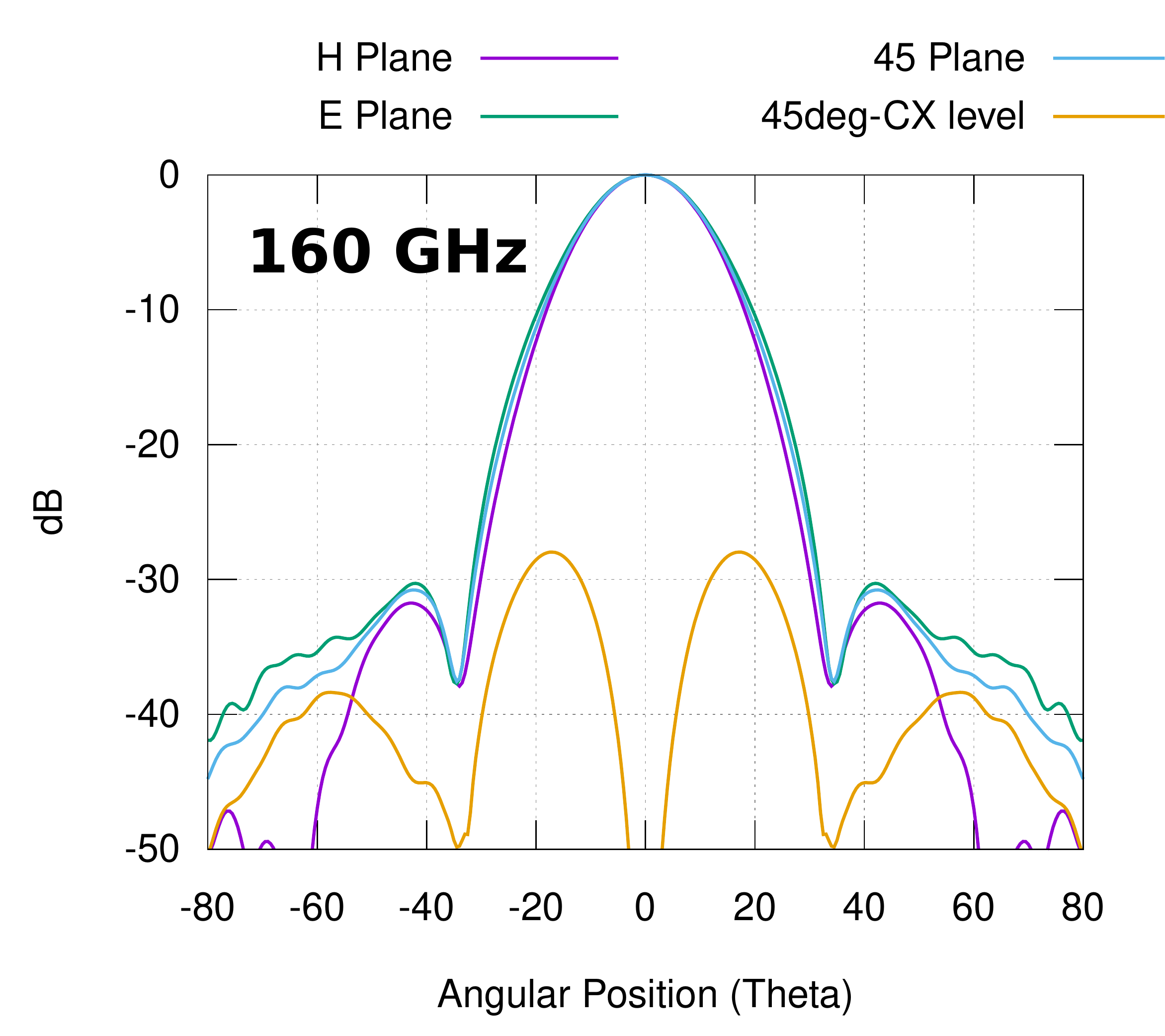}
                \caption{\label{sim:rad:path}Radiation pattern simulations considering the model shown in Fig.~\ref{model:cad} for three co-polar planes (H, E, and 45$^\circ$) and for the 45$^\circ$ cross polarization (CX) plane.}
            \end{figure}

            In Fig.~\ref{sim:rad:path} we show the simulated electromagnetic radiation patterns at 140, 150 and 160\,GHz. The maximum level of cross polarization is always less than $-25$\,dB, and the first and far side-lobes are always less than $-25$\,dB and $-35$\,dB, respectively.

    \subsection{Mechanical Design}
        In the CMB polarization observation, high-density focal planes play a crucial role in reaching the adequate sensitivity for this challenge. For this reason, the mechanical design of the array was focused on the maximization of the number of horns per unit of area.
        
        The final result is a compromise between the ideal horns density where all the tops of the feed horns are placed at the minimal distance to avoid crosstalk and a configuration where the tops of the feed horns have to left space for the assembly screws. 
        
\begin{figure}[htbp]
            \centering
            \includegraphics[width=0.9\textwidth]{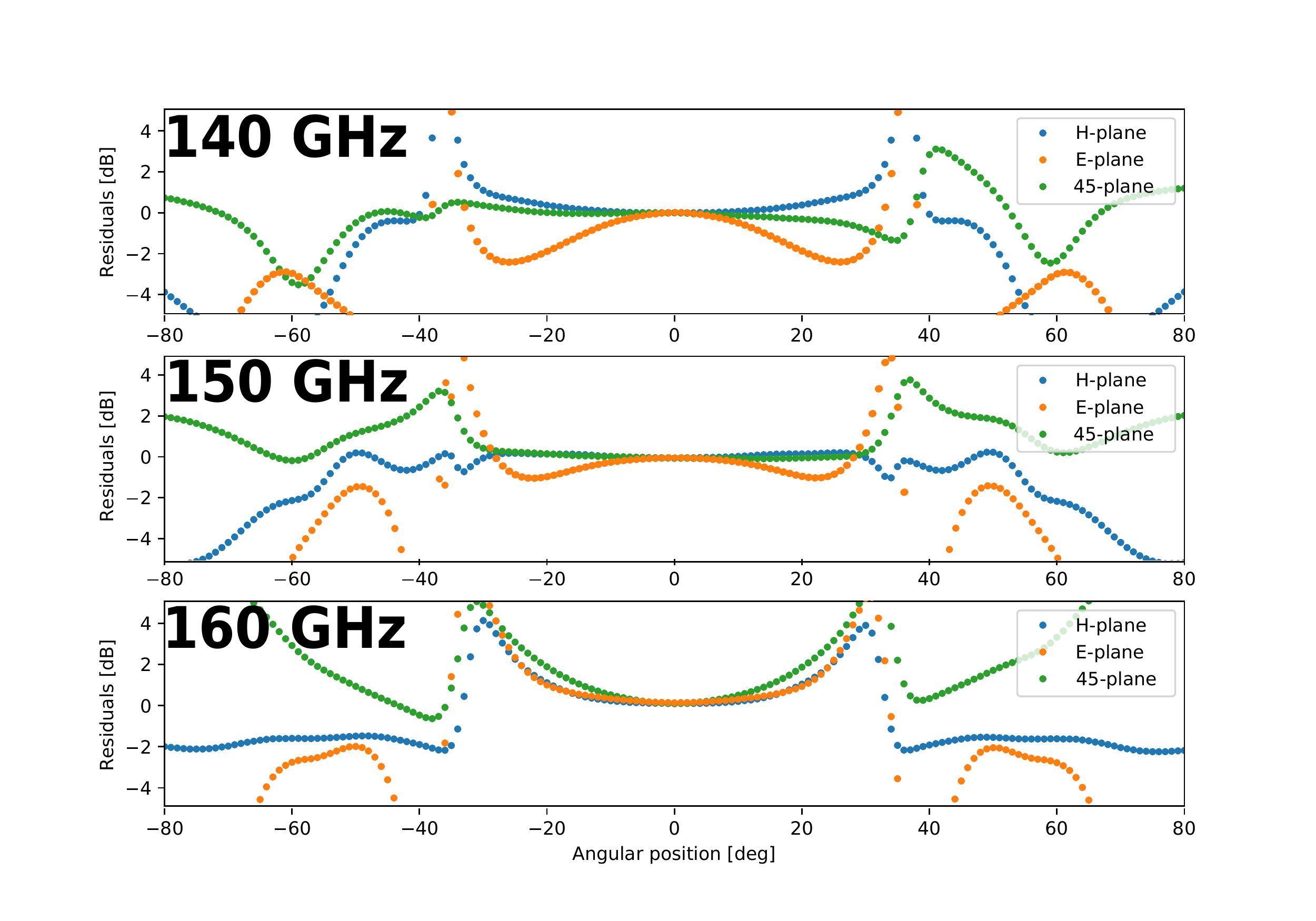}
            \caption{\label{cross:talk:feed}Residuals evaluated comparing the single horn simulation, presented in Fig.\ref{sim:rad:path}, with an improved model characterized by a reference feedhorn ad four apertures of the first-neighborns feedhorns.}
        \end{figure}

        The crosstalk level between close feedhorns was evaluated comparing the radiation patterns simulated with a single-horn model, with the same model integrated with the aperture of the first-neighbors feedhorns. The compatibility level between the two radiation patterns is shown in Fig.\,\ref{cross:talk:feed}. We can appreciate that for the main bea, the residuals are less than 2dB and within 5dB along the whole $\theta$ angular span.
     
        The limit is represented by heads of the tight screws that occupy a significant area. Eventually, we reached a final mechanical design where the tops of the feed horns contact the screws. The degradation of the electromagnetic proprieties are negligible, and we reached a filling factor of $0.6$ $horn/cm^2$. The capability level of the array to collect photons is given by the ratio between the sum of the areas of the horns apertures over of the total array area. We reached a $30\%$ of the working area. The remaining $70\%$ of the area was used for the assembly screws, alignment pins, and the interface with the detectors holder.

        The feedhorn array is fabricated in aluminum, with a hexagonal footprint of 49\,mm side. Between the top and bottom plates there are 38 intermediate plates, each containing 37 antenna holes distributed with a center-to-center distance of 11\,mm, 2 holes for the 2\,mm alignment pins, 17 M2 and 6 M4 screw holes. Half of the intermediate plates are 0.3\,mm thick and contain the corrugations teeth, the other half are 0.4\,mm thick and contain the corrugations grooves.

        The top plate is 2\,mm thick, it hosts the screw heads and the antenna apertures with the double flare described in section~\ref{sec_aperture_design}. The bottom plate is 12\,mm thick, it terminates with circular waveguides and provides the necessary thickness to host the alignment pins, guarantees their orthogonality to the plates and permits the mechanical screwing of the array. We also used this plate to interface the array to the measuring system, to the OMT to illuminate the polarization-sensitive KIDs array, and to the plate containing the flared waveguides described in section~\ref{sec_output_waveguide_design} to illuminate the total power KIDs array.

    \subsection{Manufacturing}
    \label{sec_manufacturing}

        The platelet technique applied to the manufacturing of corrugated feedhorns typically requires drilling circular holes into metal plates that are subsequently stacked, aligned and clamped.  We have used two methods to drill the holes: chemical etching of the thin plates and CNC (Computer Numerical Control) milling applied to the thick top and bottom plates. The chemical etching allowed us a fast and cheap process, potentially scalable to a large number of elements. On the other hand, this method could not be applied to the external plates that required a more substantial thickness for mechanical clamping.

        We checked the achieved mechanical tolerance (hole diameters, center positions, deviation from circularity) against the maximum achievable tolerance of the two techniques: $\pm 0.05$\,mm for chemical etching and $\pm 0.03$\,mm for CNC milling. The chemical etching was performed at Lasertech srl in Milan, CNC milling was carried out at the mechanical workshop of the University of Milan, Department of Physics.

        We measured each antenna and alignment hole of each aluminum plate with our \textit{Werth ScopeCheck 200} metrological machine, we compared the positions and diameters with their nominal values, and we analyzed the form tolerance of the holes.

\begin{figure}[htbp]
            \centering
            \begin{tabular}{c c}
            \includegraphics[width=0.46\textwidth]{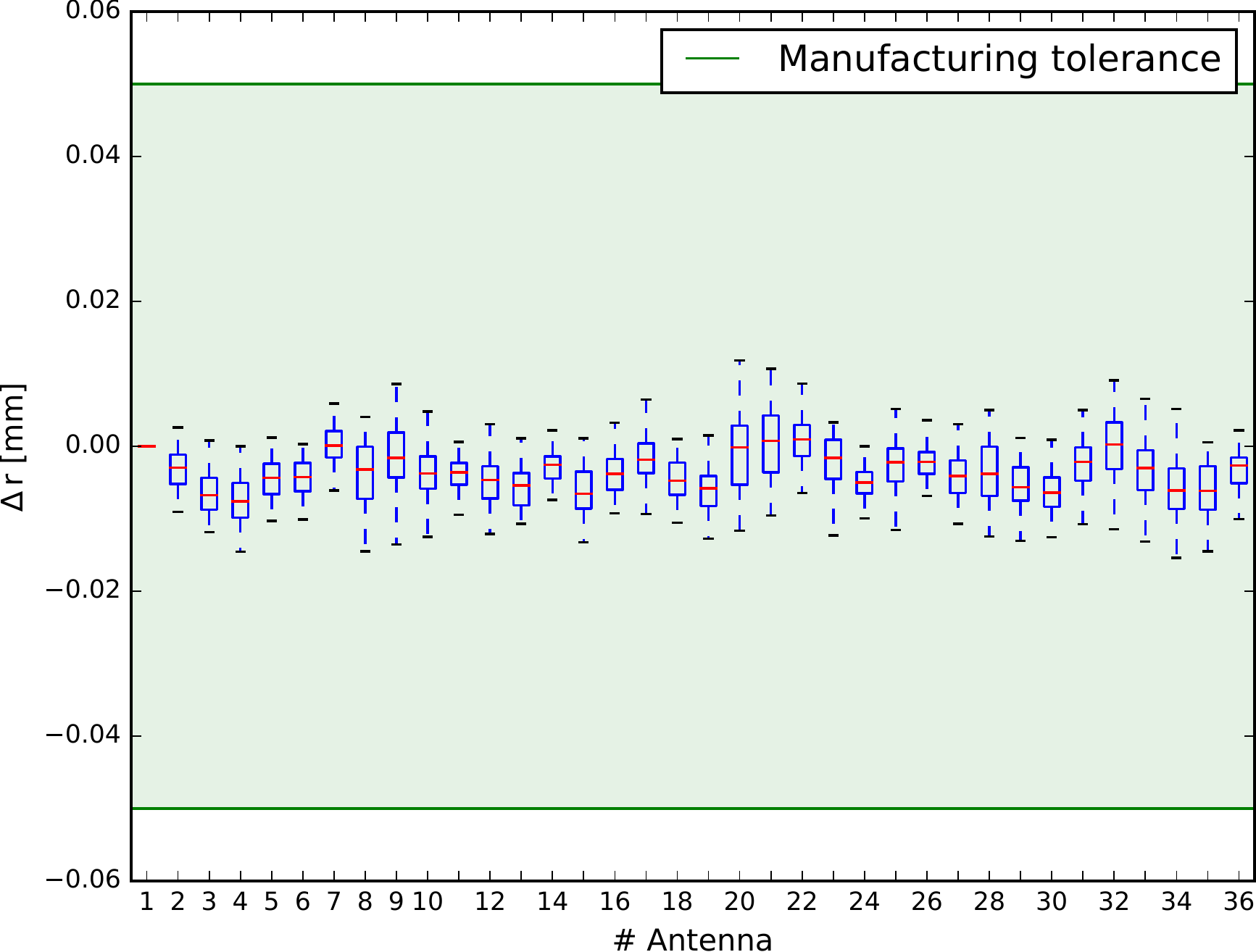} &
            \includegraphics[width=0.46\textwidth]{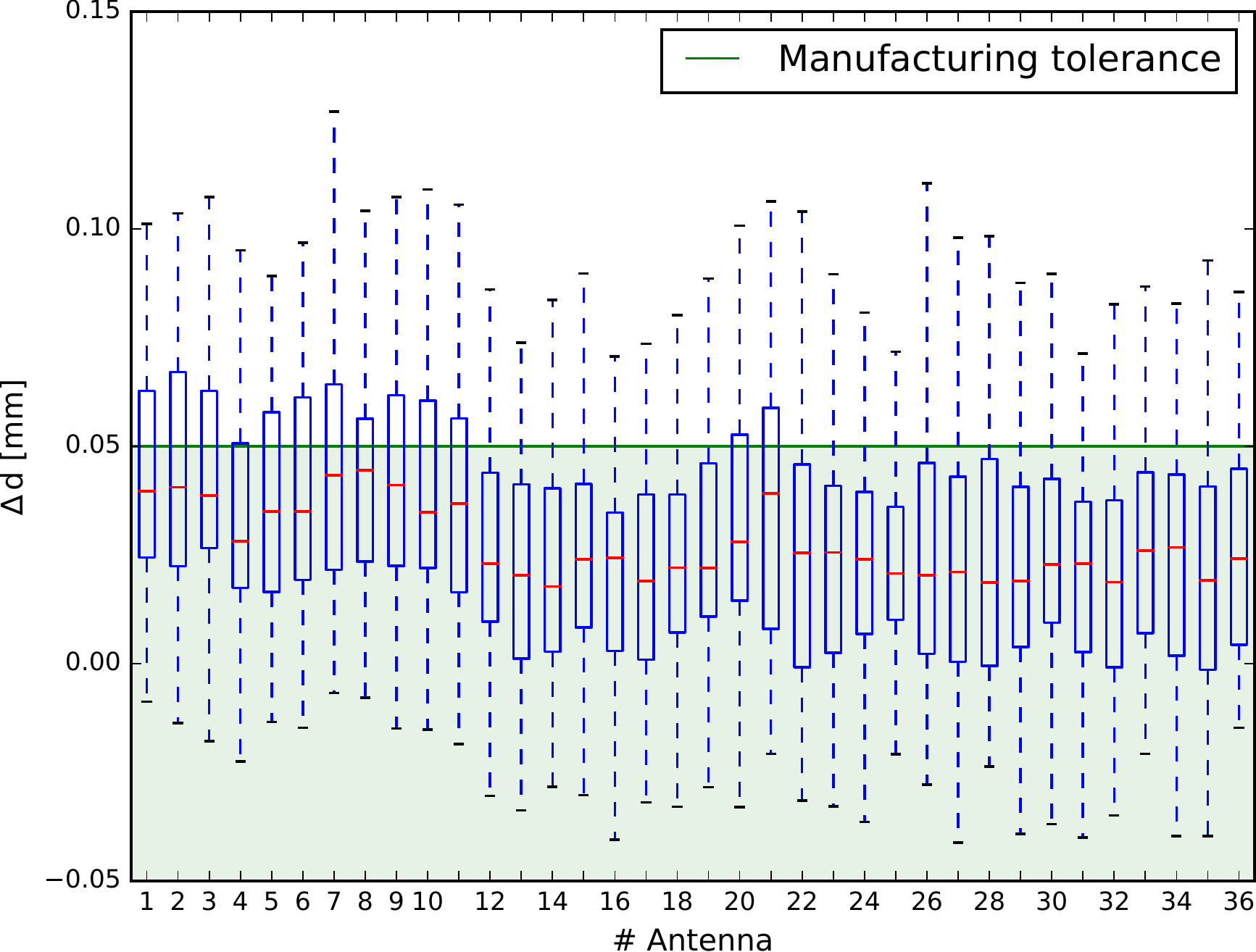}\\
            \end{tabular}
            \caption{\label{KID:MecMeas_position}\emph{Left panel:} Deviation of holes center positions from the nominal value for the chemically etched plates. $\Delta r = \sqrt{(\Delta x)^2+(\Delta y)^2}$, where $\Delta x$ and $\Delta y$ are the deviations of the center coordinates from their nominal value. \emph{Right panel:} Deviation of hole diameters from the nominal value for the chemically etched plates. In the figure are reported 36/37 horns because the measurements of feedhorn number 1 was not recorded by the our metrological machine.}
        \end{figure}

        The metrological measurements of the antennas centers and diameters, for each chemically etched plate, are shown in Fig.\,\ref{KID:MecMeas_position}. The boxplots represent the deviations of the measured quantity from the nominal value for the antenna holes, while the green area highlights the expected tolerance. We can appreciate that the centers of all the antennas holes are within the green-zone of compatibility. Still, the diameters of the holes, on average, were manufactured within the compatibility zone, but we have to notify that are not zero-mean distributed, but they are slightly larger than expected. This means that it is possible to compensate this systematic effect in the design phase, thus improving the final accuracy. Similar results were obtained for the alignment pins of the chemically etched plates.

        The measurements of the top and bottom plates were compliant with the milling mechanical precision of $0.03$\,mm. The holes shape does not significantly deviate from circularity, and this is true both for chemically-etched and milled plates.

        The array was manually aligned using rectified steel dowel pins, and tightened using M2 brass screws and M4 steel screws, for a total weight less than $0.4$\,kg. We decided to align and clamp the plates using brass and steel tools.
        In this way, as presented in section \ref{sec_thermal_assessment}, the alignment and tightness of the plates are guaranteed even at low temperatures.

\section{Electromagnetic characterization}
\label{El:Ch}
\begin{figure}[htbp]
    \centering
    \includegraphics[scale=0.3]{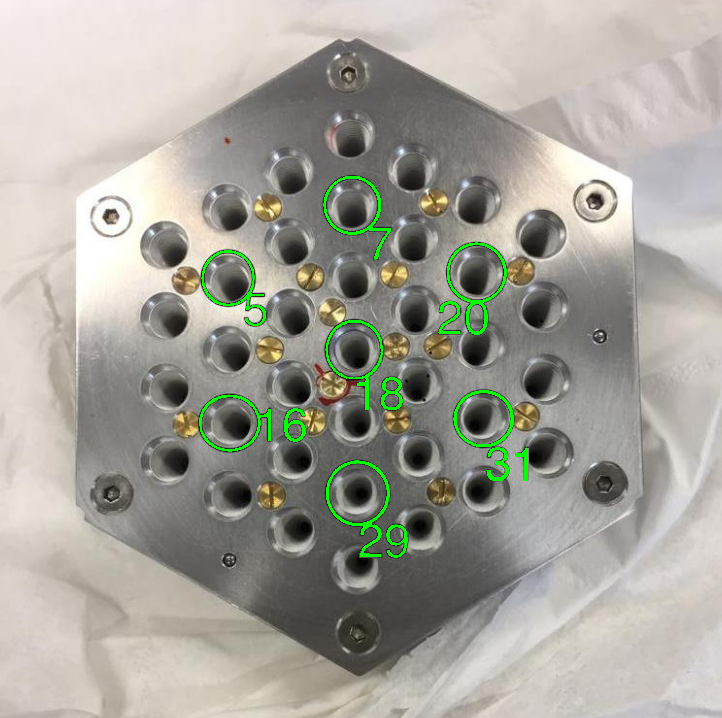}
    \caption{The 7 measured horns (green circles). The number near each horn corresponds to a label used to identify the horn inside the array.}
    \label{HORN:MEAS}
\end{figure}

We have measured the electromagnetic performance of the 7 horns highlighted in Fig.~\ref{HORN:MEAS}. For each horn, we have measured the radiation pattern on four different planes (H, E, $\pm45^\circ$), the maximum level of cross polarization and the return loss. Finally, we have quantified the repeatability of the electromagnetic responses and the compatibility with the electromagnetic simulations.

\subsection{Measurement setup}
\label{sec_measurement_setup}

\begin{figure}[htbp]
        \centering
        \includegraphics[scale=0.65]{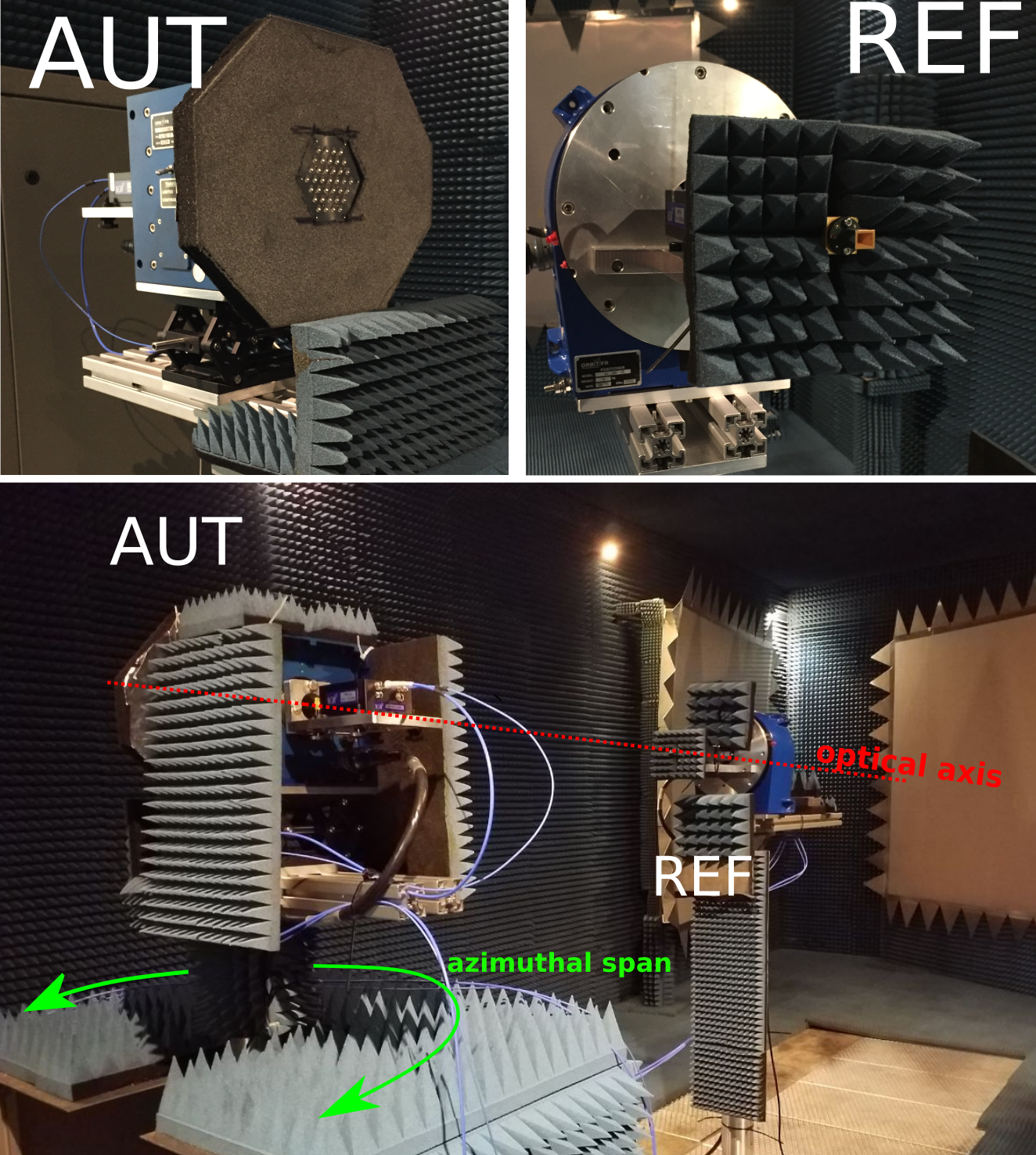}
        \caption{\emph{Top-left panel:} shows a picture of the rotary table mechanically coupled to the feedhorns array. \emph{Top-right panel:} shows a picture of the reference horn and its mechanical interface with its rotary table. \emph{Bottom panel:} An overall view of the measurement setup. The red line represents the horn line-of-sight at an azimuth angle $\theta = 0$, and it is aligned with the main direction of the reference horn.}
        \label{An:Ch}
    \end{figure}

    The antenna-under-test (AUT) is connected to the VDI 110-170\,GHz receiver (RX) frequency module extender. The reference launcher (REF) is a linearly polarized pyramidal standard gain horn (STG 29240-20) made by Flann coupled to a VDI 110-170\,GHz transmitter (TX) frequency module extender. The whole setup is shown in is shown Fig.~\ref{An:Ch} where all the parts are mounted and ready for measurements.

    The distance between the REF and AUT is 1.5\,m, which ensures a distance  of at least two times the far-field distance between the source and the receiver. The TX and RX frequency extenders are connected to a Vector Network Analyzer (VNA, Anritsu Vector-Star MS6467B), equipped with a broad-band test-set Anritsu 3739C.

    The AUT and the REF are both mounted on a rotary table. The two tables can rotate around their polarization axis to select the co-polar and cross-polar planes. Furthermore, the AUT table can rotate around its azimuth axis with an angle span of $\pm90$\,deg.

\subsection{Radiation Patterns and Cross Polarization}
\label{sec_beam_patterns}

    In this section, we show the measurements of the electromagnetic response of the 7 horns and their cross polarization patterns. For clarity, in this section we show only the H-plane and the cross polarization measurements. All the other measurements are reported in Appendix\,\ref{horn:spec}.

\begin{figure}[htbp]
        \centering
        \includegraphics[width=0.4\textwidth]{keys}\\
        \includegraphics[width=0.32\textwidth]{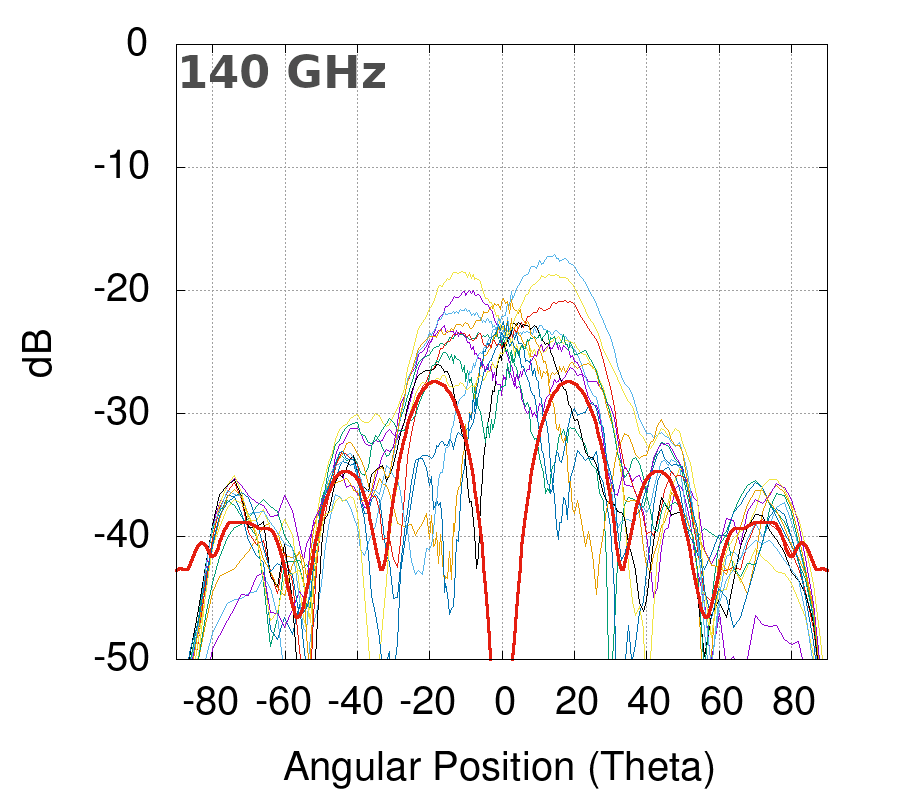}
        \includegraphics[width=0.32\textwidth]{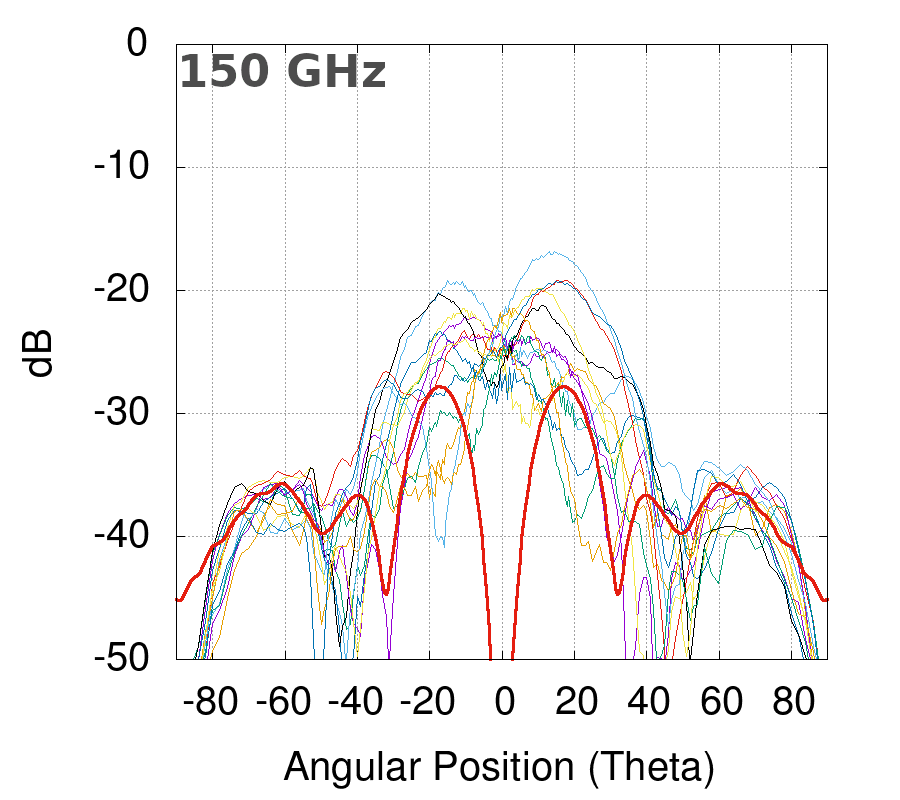}
        \includegraphics[width=0.32\textwidth]{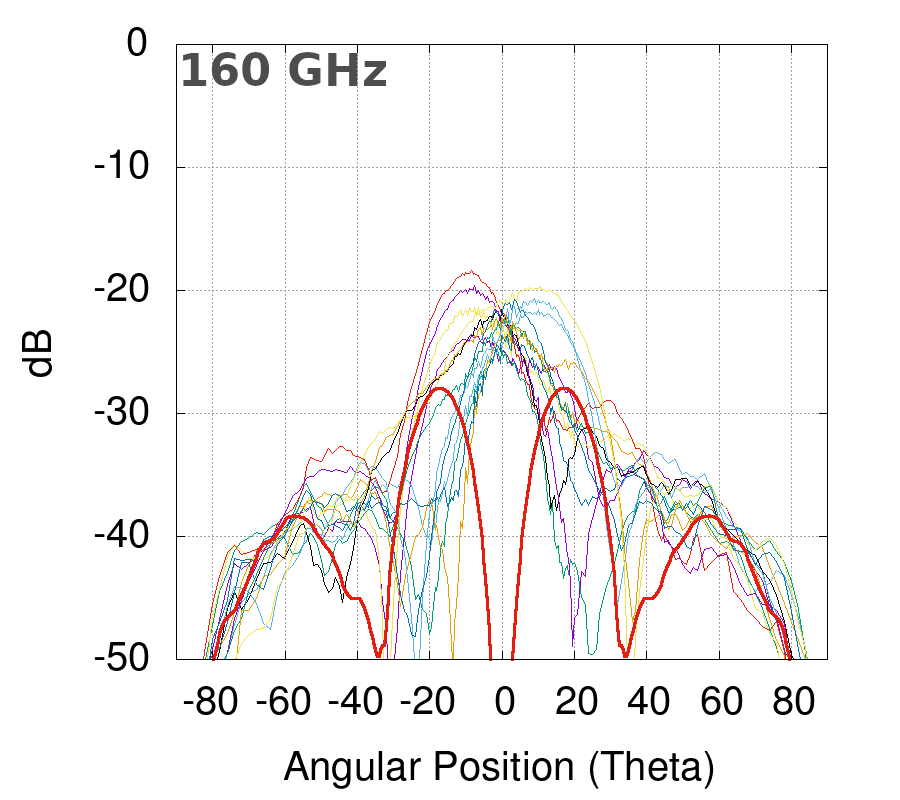}
        \caption{The cross-polar radiation patterns on $45$ deg and $-45$ deg planes. The red lines represent the simulated cross-polar pattern.}
        \label{fig:three:graphs:cross}
    \end{figure}

    Figure~\ref{fig:three:graphs_Hplane} shows the H-plane for three different frequencies. In each plot we show the seven measured H patterns, together with the corresponding simulation. The simulation was performed for each feedhorn with the CST Microwave Studio software starting from the mechanical measurement shown in section \ref{proto}. In the plot, we present the average response of the whole set of simulations to evaluate both the compatibility level with the simulation and the repeatability of the measurements. The deviations of the measurements from the simulation are typically within $\pm 2.0$~dB in the whole angular range and $\pm 1.0$~dB within the main beam angular span ($\pm 20$ $deg$), which indicates an excellent repeatability of the manufacturing process. For each feedhorn we measured the cross polarization radiation pattern for both the $45^\circ$ and the $-45^\circ$ planes to evaluate the absolute maximum level of cross polarization (see Fig.~\ref{fig:three:graphs:cross}). The measured maximum level of cross polarization is about $-20$~dB, about $\sim 7$ dB higher than the simulated value. This is still comparable to the best performance of the current state-of-art of the CMB experiments that use silicon lenslets instead of feedhorns \cite{arnold2012bolometric}.

    In an attempt to prepare for future improvements, we have identified three possible sources of this cross polarization excess: ($i$) uncertainties in the alignment of the experimental setup, ($ii$) possible misalignment of the antenna plates caused by the mechanical accuracy of the alignment pin holes, and ($iii$) the cross polarization of the transition from circular to rectangular waveguide used to connect the AUT to the RX frequency extender. We discuss in detail the first two points in Appendix~\ref{sec_crosspol_verification}, where we show that part of the discrepancy is consistent with the mechanical accuracy of the alignment pin holes. To characterize the effect of the transition cross polarization we are planning measurements using an OMT being developed in the framework of the same project in place of the transition. We will report these measurements in a forthcoming paper dedicated to the OMT development and performance.

    \begin{figure}[htbp]
        \centering
        \includegraphics[width=0.4\textwidth]{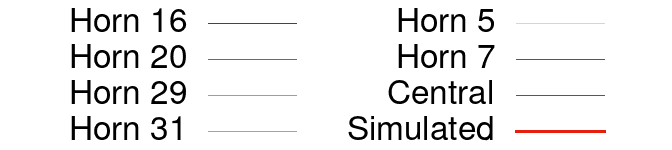}\\
        \includegraphics[width=0.32\textwidth]{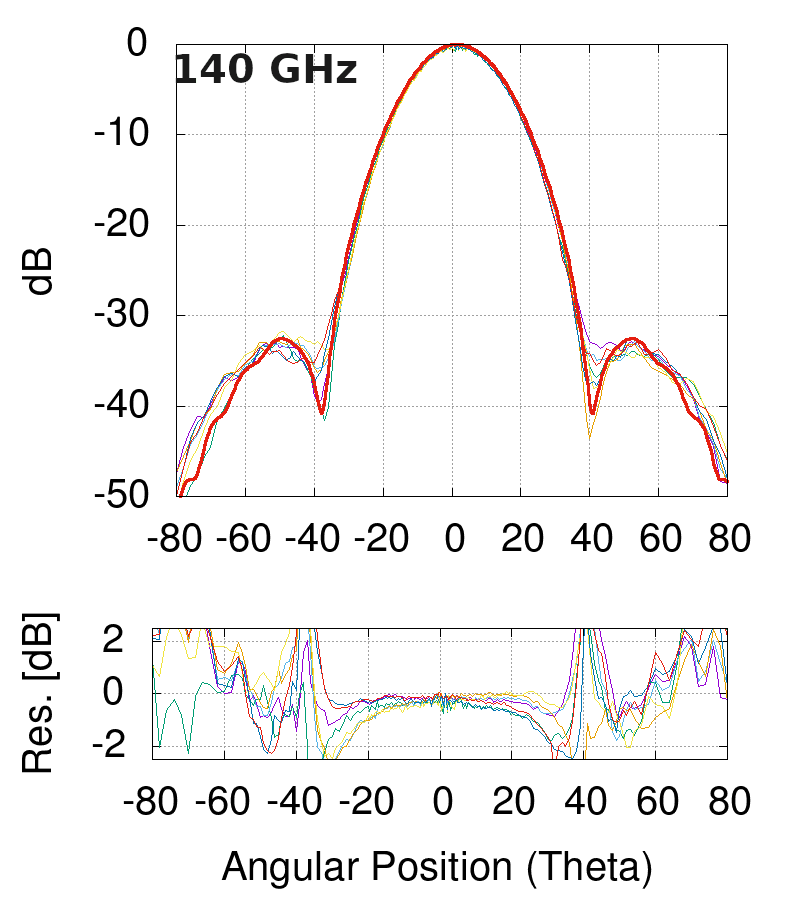}
        \includegraphics[width=0.32\textwidth]{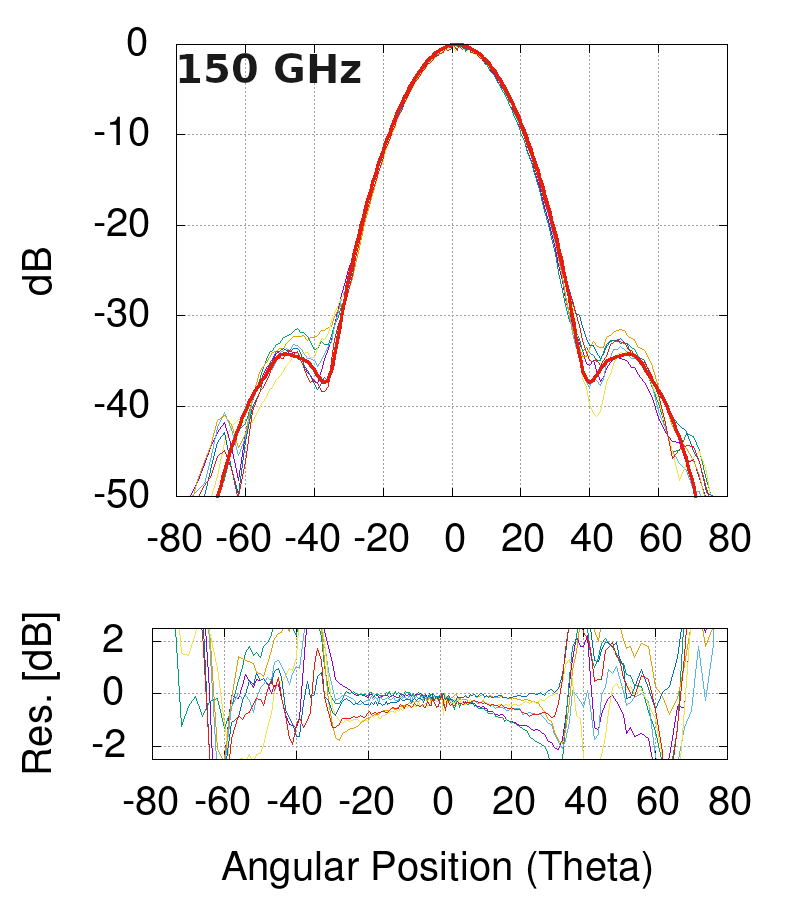}
        \includegraphics[width=0.32\textwidth]{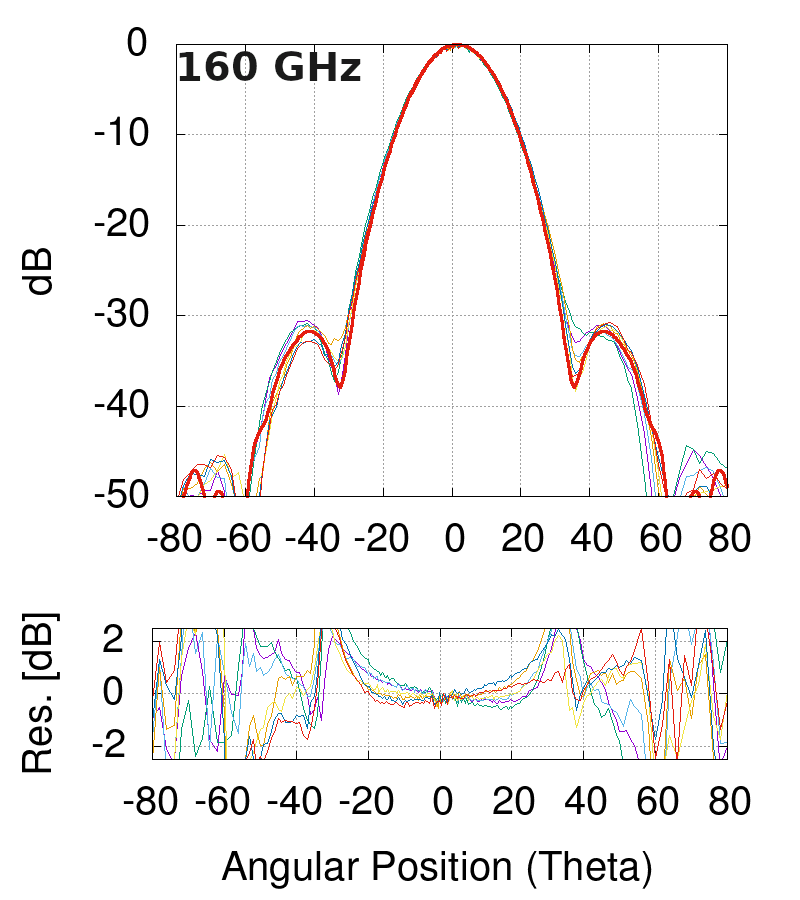}
        \caption{The H-plane radiation patterns of the measured feedhorns at three different frequencies. The bottom panels represent the residual between the measurements and the simulation.}
        \label{fig:three:graphs_Hplane}
    \end{figure}

\subsection{Return loss}

\begin{figure}[ht]
        \centering
        \includegraphics[width=0.96\textwidth]{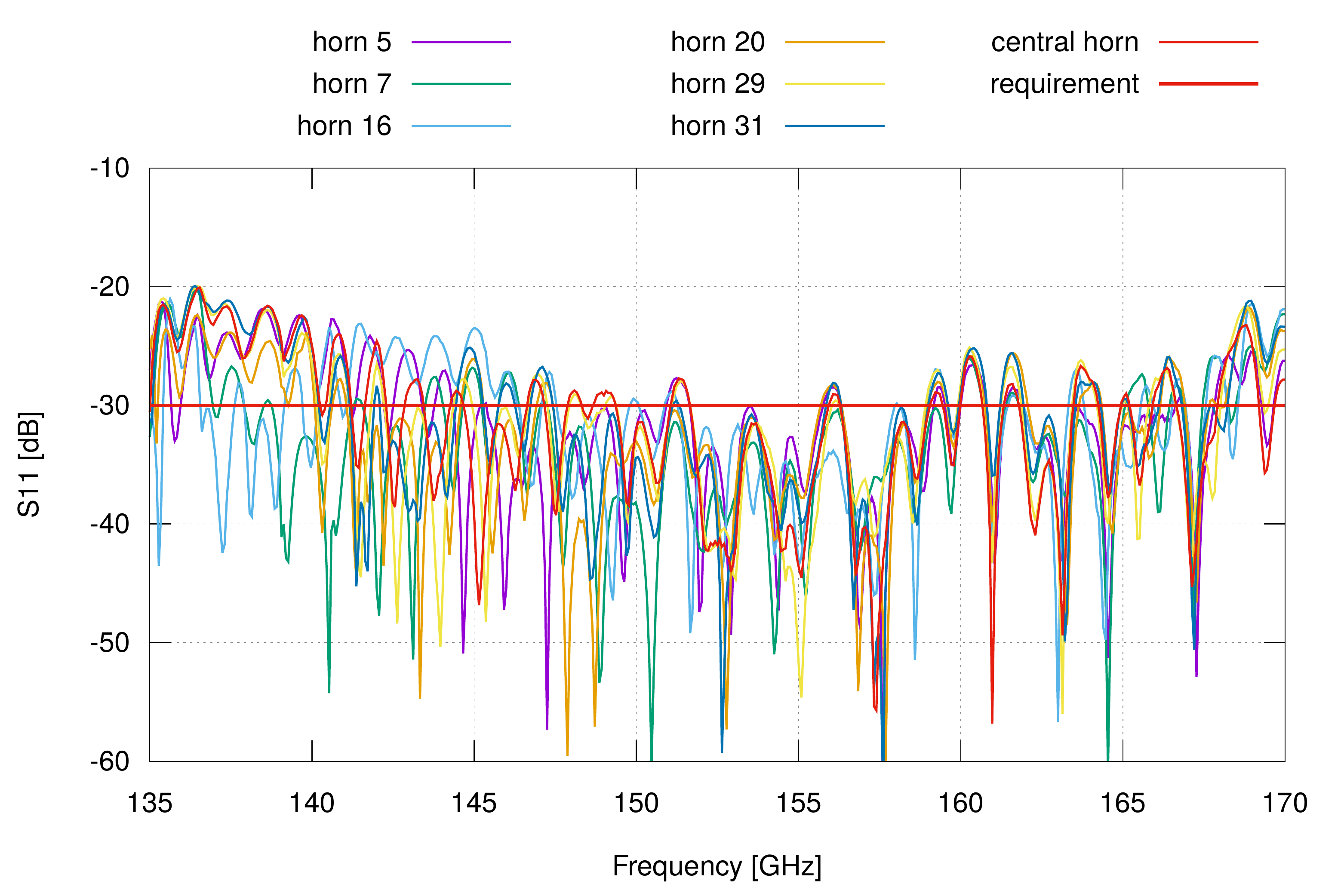}
        \caption{Measured return loss for the 7 horns compared with the $-$30\,dB requirement.}
        \label{RL:KID}
    \end{figure}

    Figure~\ref{RL:KID} shows the return loss of the 7 measured horns on the extended bandwidth of $135-170$\,GHz. We see that the return loss is compatible, on average, with the $-$30\,dB requirement ($\sim-$25\,dB at the edge of the band and $\sim-$30\,dB at the center of the band). We also notice the very good repeatability of the measurement for the various horns.

\section{Discussion and conclusions}
\label{discuss}
In this work we have described a D-band feedhorn array manufactured with chemically etched aluminum platelets that were subsequently mechanically clamped together. We have shown that their electromagnetic performance in terms of repeatability, return loss ($<-25$\,dB), cross-polarization ($<-20$\,dB) and sidelobe levels is competitive with similar arrays developed for CMB polarization measurements. In Table~\ref{lit:feed} We compare our work with similar studies published in the recent literature.

\begin{table*}[htbp]
    \renewcommand{\arraystretch}{1.25}
  \begin{center}
    \caption{\label{tab_comparison_horns}Performance comparison of corrugated feedhorn arrays published in the literature.}
    \label{lit:feed}
    \begin{tabular}{lccccc}
        \hline
        & Freq. band & \multirow{2}*{Bandwidth} & CX-Pol & RL               & \multirow{2}*{$N_\mathrm{feeds}$}\\
        & [GHz]      &            & [dB]   & [dB]             &            \\
        \hline
        \hline
        J. P. Nibarger \cite{nibarger201284}\dotfill       & 120\,-\,170           & 1:1.42 & $< -23$\dg& $< -20$\ddg& \zero84\\
        R. Datta \cite{datta2014horn}\dotfill              & \zero 70\,-\,175      & 1:2.33 & $< -20$\dg& N.P. & 255 \\
        F. Del Torto \cite{del2011w}\dotfill               & \zero 75\,-\,110      & 1:1.47 & $< -25$\dg& $< -30$\ddg& \zero\zero4\\
        L. Lucci \cite{lucci2012stackable}\dotfill         & \zero 33\,-\,\zero 50 & 1:1.51 & $< -27$\dg& $< -30$\ddg& \zero\zero7\\
        Y. Beniguel \cite{beniguel2005design}\dotfill      & \zero 70\,-\,110      & 1:1.57 & $< -30$\dg& $< -20$\ddg& \zero\zero1 \\
        V. Tapia \cite{tapia2016ultra}\dotfill             & \zero 67\,-\,116      & 1:1.73 & $< -17$\dg& N.P. & \zero\zero1\\
        S. Sekiguchi \cite{sekiguchi2016broadband}\dotfill & 120\,-\,270           & 1:2.25 & $< -20$\dg& $< -15$\ddg& \zero\zero1\\
        S. Sekiguchi \cite{sekiguchi2016broadband}\dotfill &  \zero 80\,-\,180     & 1:2.25 & $< -20$\dg& $< -15$\ddg& \zero\zero4\\
        \textbf{This Work}\dotfill                         & 140\,-\,170           & 1:1.21 & $< -20^\dagger$& $< -25^\ddagger$& \zero\zero37\\
        \hline
        \multicolumn{6}{l}{\small{$^\dagger$ The design value was $-$25\,dB}}\\
        \multicolumn{6}{l}{\small{$^\ddagger$ The design value was $-$27\,dB}}\\
    \end{tabular}
    \end{center}
\end{table*}

Nibager~\cite{nibarger201284} and Datta~\cite{datta2014horn}, for example, produced arrays of feedhorns built in silicon platelets that were subsequently metal-plated (with copper or gold) and bonded. Del Torto~\cite{del2011w} and Lucci~\cite{lucci2012stackable}, instead, built the horns directly in aluminum from platelets and rings, respectively, by direct milling. Binguel~\cite{beniguel2005design} produced only one feedhorn by electroforming, achieving and excellent level of cross polarization ($<-30$\,dB).

This comparison shows that our approach is competitive, in terms of performance, with all the platelet feedhorns. It also shows advantages in terms of cost and manufacturing time. Indeed, we can work directly the aluminum without being limited by the longer processing times of direct milling. Finally, mechanical clamping offers one more advantage with respect to bonding, allowing one to dismount the array and change individual platelets if needed.

The excellent agreement between measurements and simulations shows that we have achieved the necessary mechanical accuracy in the manufacturing process and in the platelets alignment allowing us to produce feedhorns with predictable performance.

Our design has still margins for improvements. In particular we may be able to improve the cross polarization by realizing proper corrugations into the thick top plate used for the horn apertures. In the future we will explore possible techniques in this direction to improve the polarization response.

\begin{acknowledgements}
\label{akwn}
This project has been carried out in the framework of the project \emph{Kinetic Inductance Detector for Space} funded by the \emph{Agenzia Spaziale Italiana} (ASI). The authors wish to thank the entire staff of the machine shop of Physics Department of UNIMI, \emph{L'Officina}, for their contribution and key discussions about the realization and assembly process.

\end{acknowledgements}

\bibliographystyle{spphys}
\bibliography{Ref}

\appendix
\newpage
\section{Appendix: Complete set of measured beam patterns}
\label{Appendix:A}

In this section we show the remaining set of radiation patterns not shown in the paper main body. For every feedhorn we display three different planes: $E$, $+45^\circ$ and $-45^\circ$. In Table~\ref{tab:all:result} we list the numerical values of the beam FWHM and the maximum cross polarization level.

\label{horn:spec}
\begin{figure}[htbp]
     \centering
     \includegraphics[width=0.32\textwidth]{keys}\\
     \includegraphics[width=0.32\textwidth]{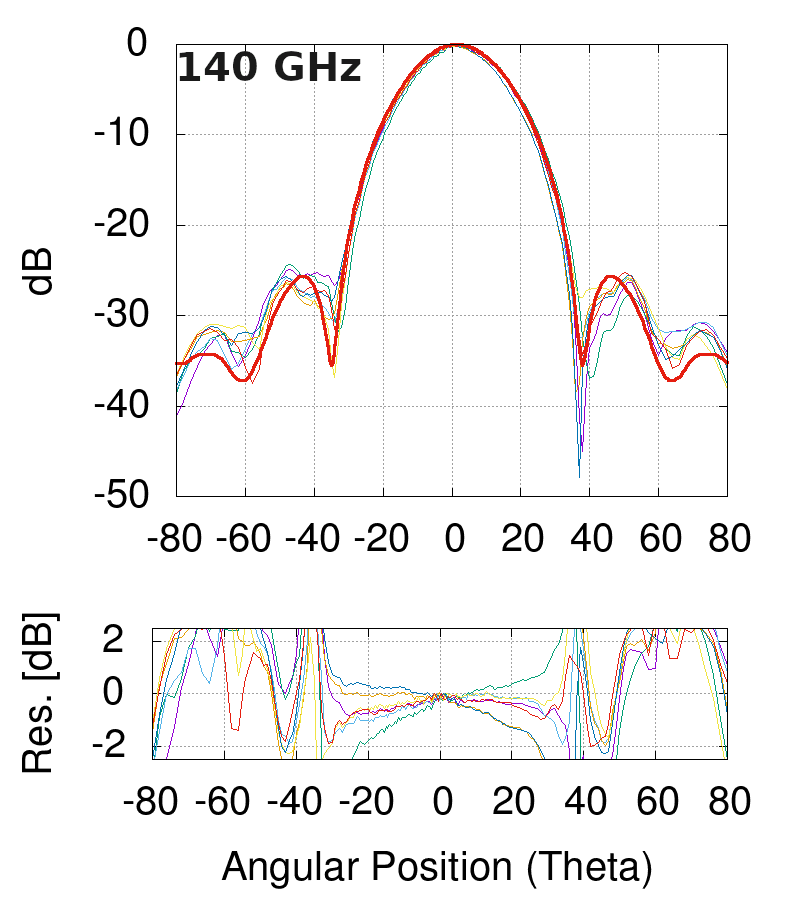}
     \includegraphics[width=0.32\textwidth]{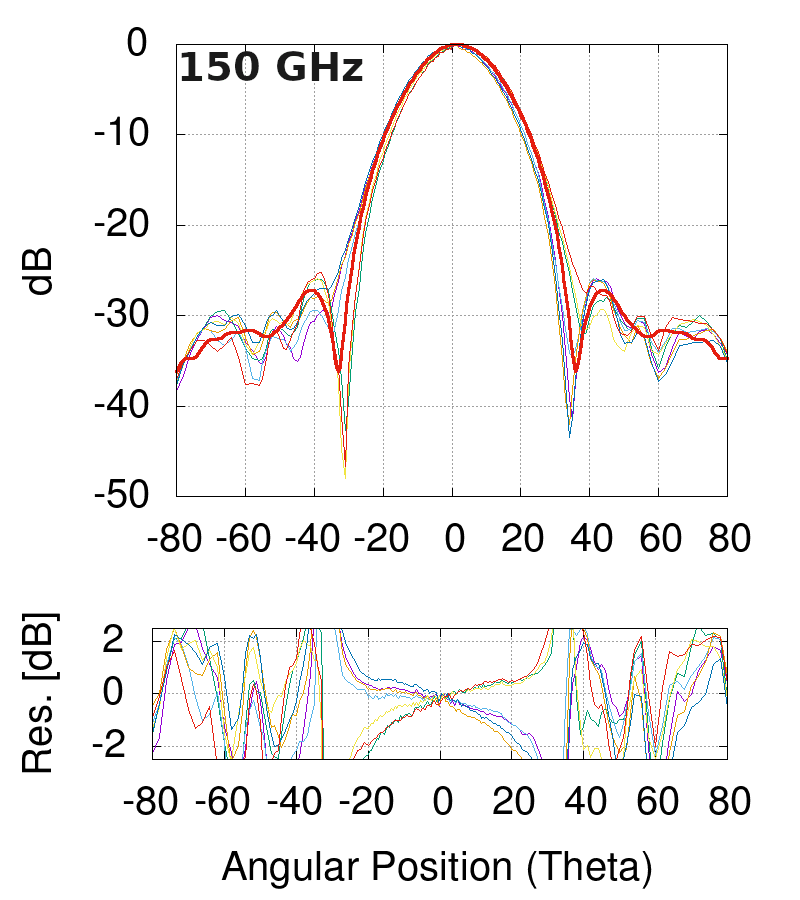}
     \includegraphics[width=0.32\textwidth]{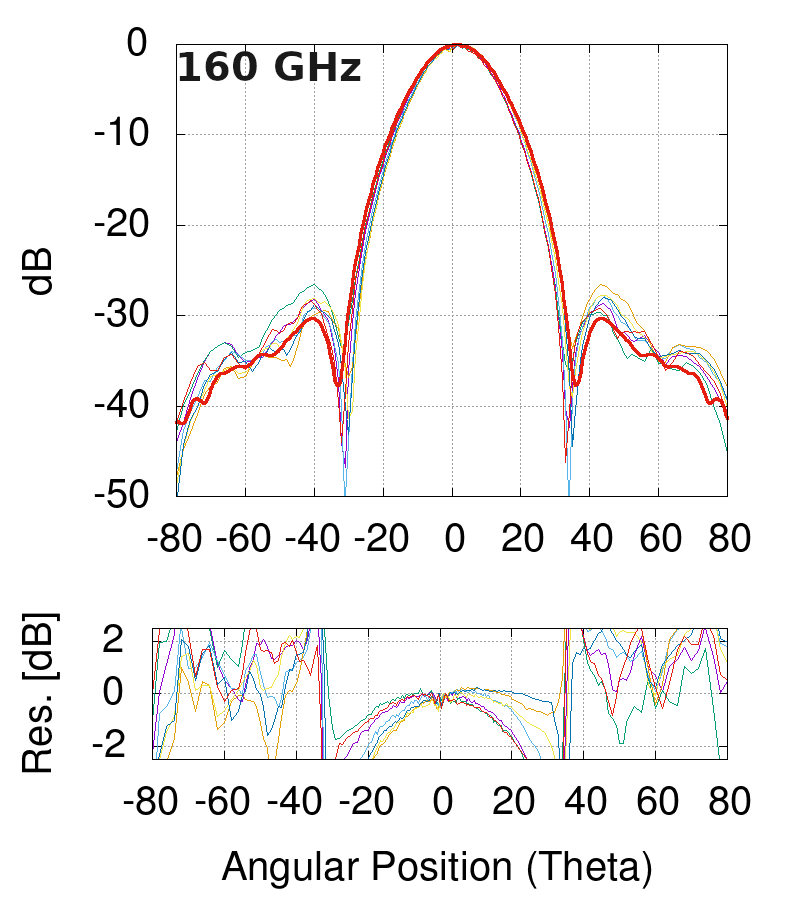}
     \caption{E-plane measurements.}
     \label{fig:three graphs_Eplane}
\end{figure}
\begin{figure}[htbp]
     \centering
     \includegraphics[width=0.32\textwidth]{keys}\\
     \includegraphics[width=0.32\textwidth]{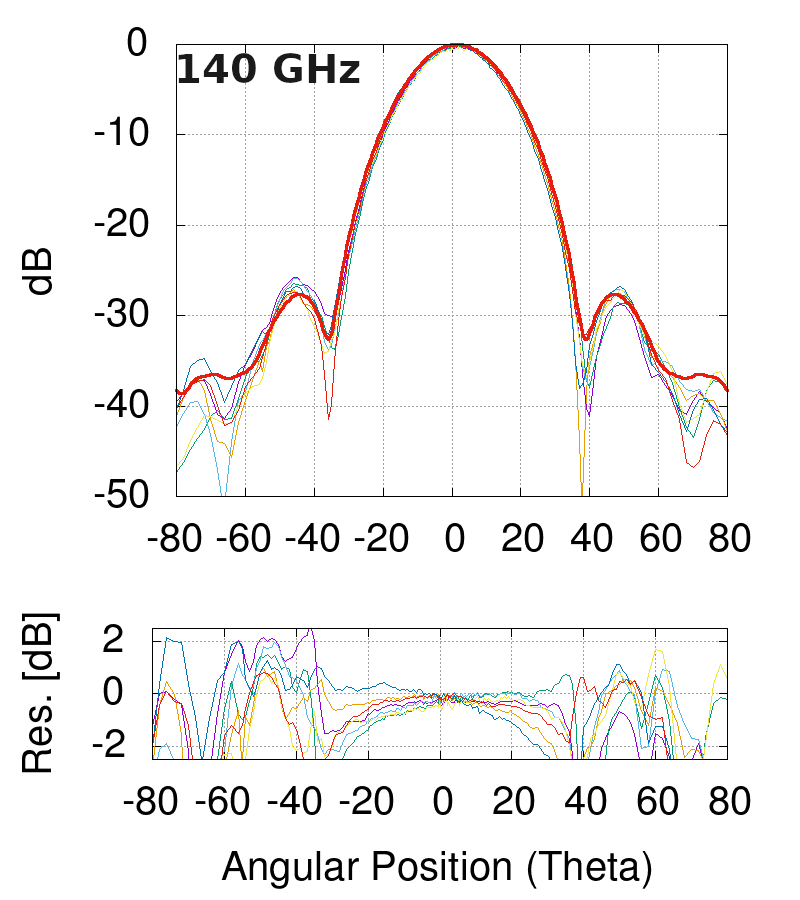}
     \includegraphics[width=0.32\textwidth]{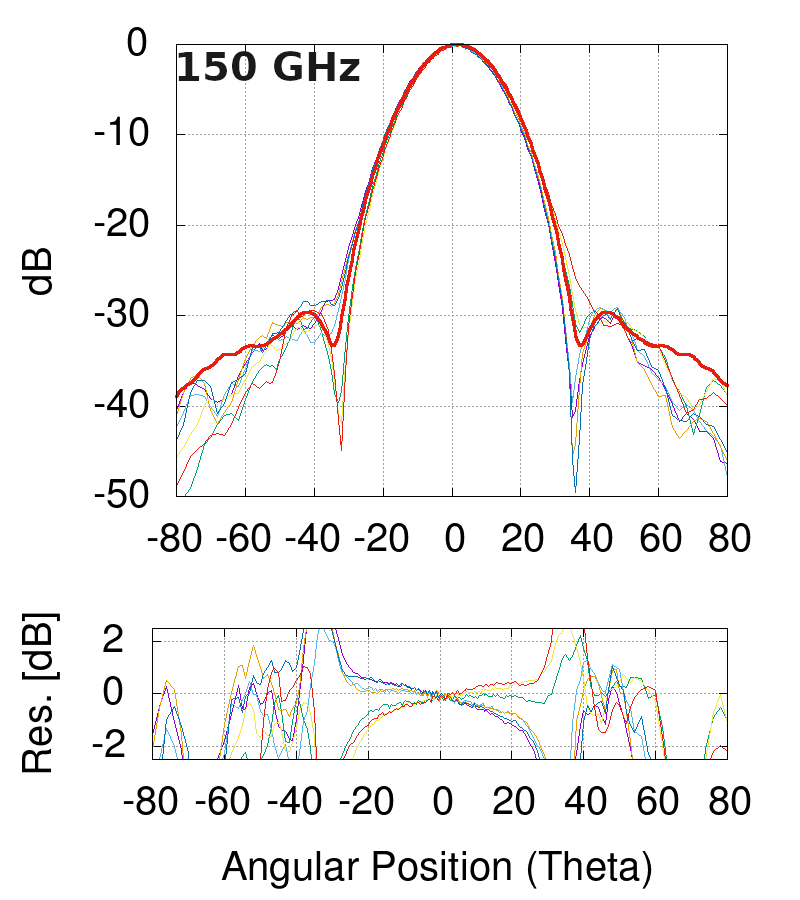}
     \includegraphics[width=0.32\textwidth]{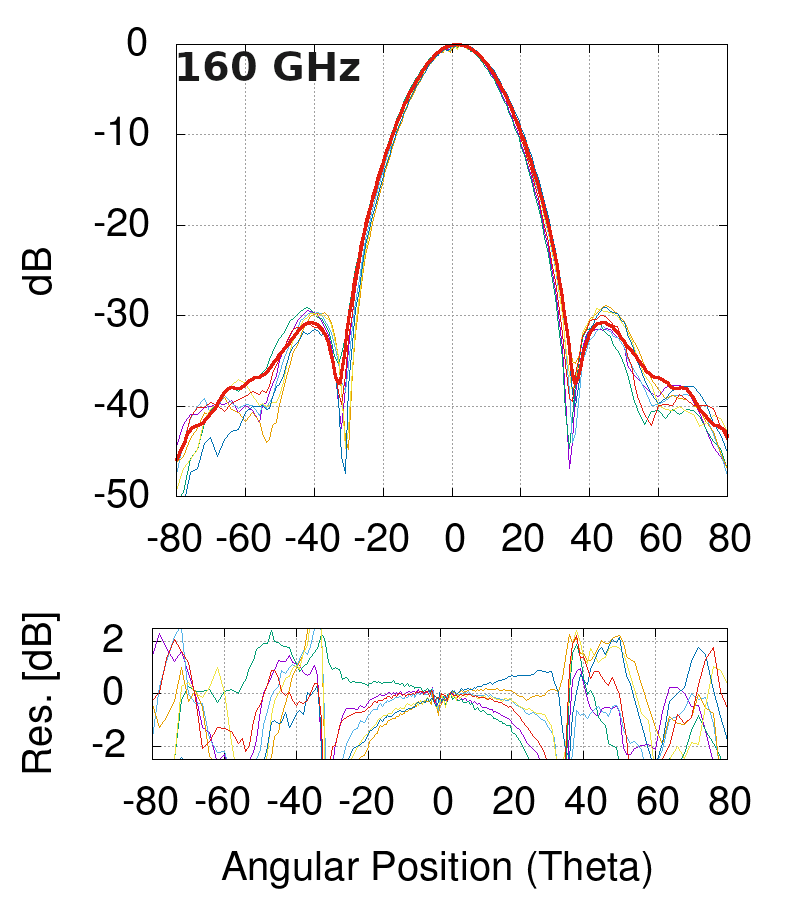}
     \caption{45$^\circ$-plane measurements.}
     \label{fig:three graphs_45plane}
\end{figure}
\begin{figure}[htbp]
     \centering
     \includegraphics[width=0.32\textwidth]{keys.png}\\
     \includegraphics[width=0.32\textwidth]{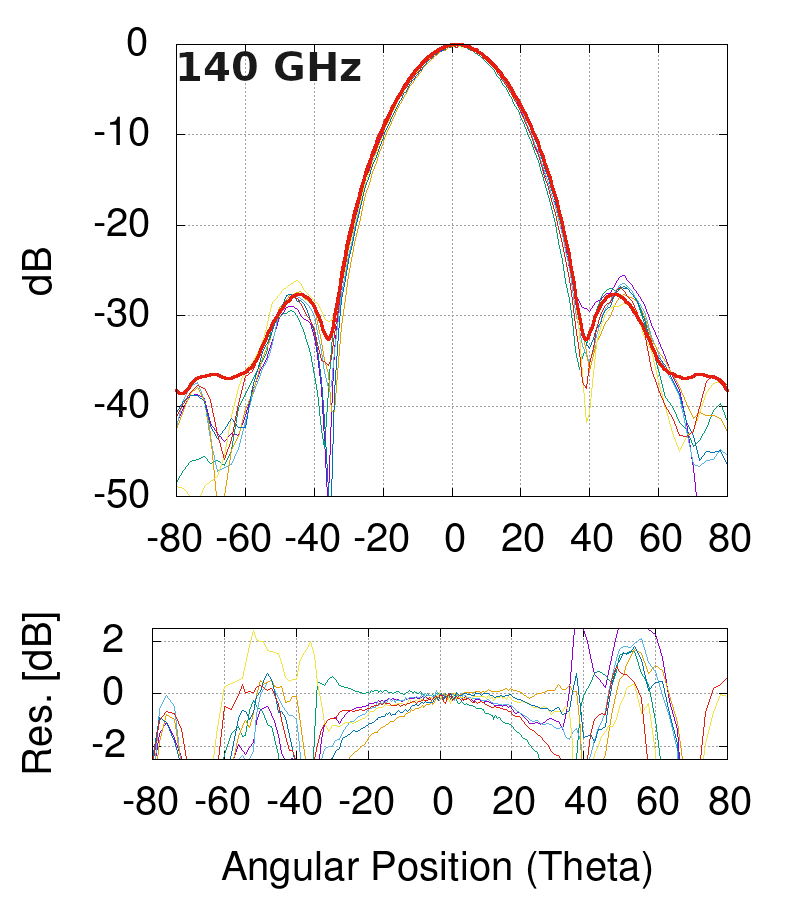}
     \includegraphics[width=0.32\textwidth]{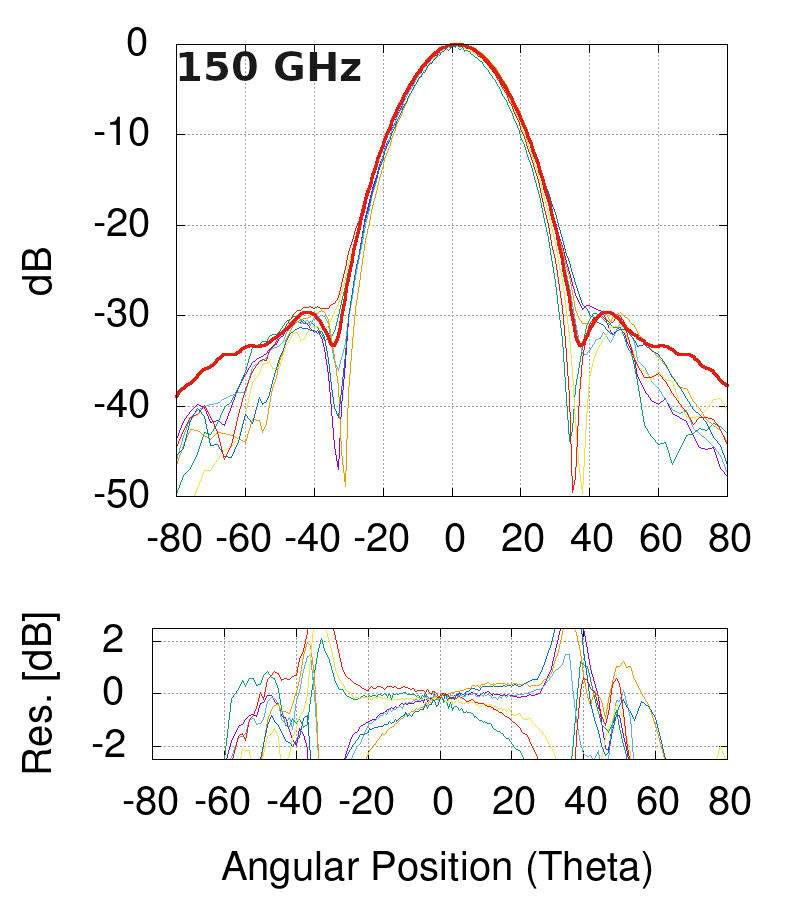}
     \includegraphics[width=0.32\textwidth]{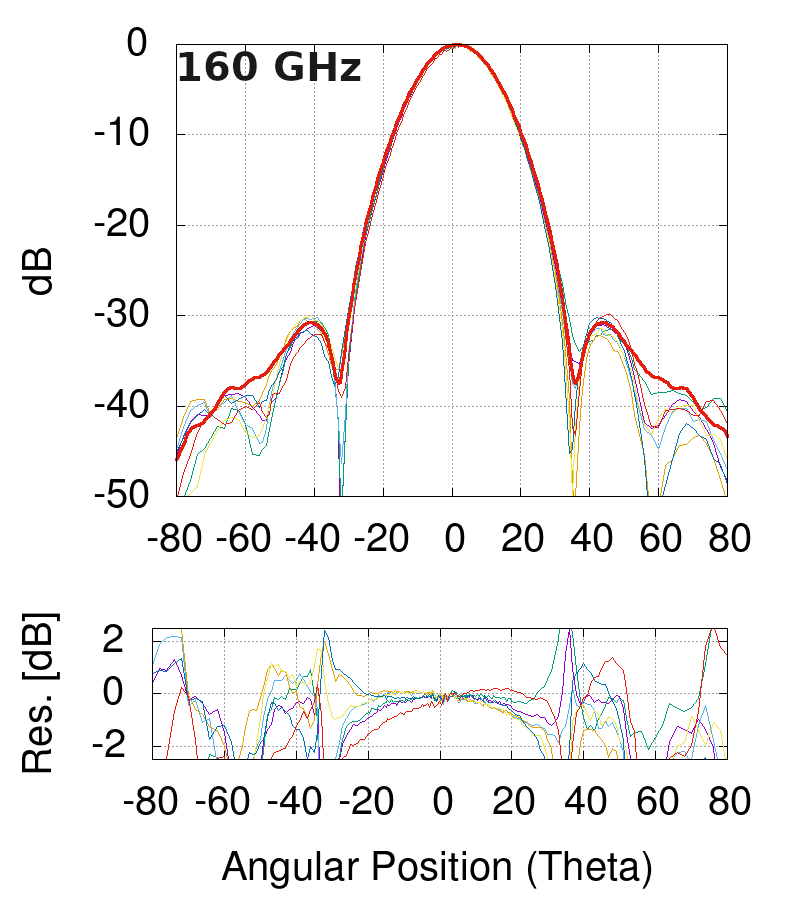}
     \caption{-45$^\circ$-plane measurements.}
     \label{fig:three graphs_m45plane}
\end{figure}

\begin{table}[htbp]
  \caption{\label{tab:all:result}Numerical values of the beam FWHM and maximum cross polarization derived from the radiation patterns.}
  \renewcommand{\arraystretch}{1.15}
    \begin{center}
    \begin{tabular}{@{}|c|c|c|c|c|c|c|@{}}
        \hline
        \multirow{2}*{Horn ID} & \multirow{2}*{Freq. [GHz]} & \multicolumn{4}{c|}{FWHM [deg]} & \multirow{2}*{CX-level [dB]} \\
        \cline{3-6}
                &             & H-plane & E-plane & $+45^\circ$-plane & $-45^\circ$-plane &      \\
        \hline
        \multirow{3}*{$5$} & \multirow{1}*{$140$} &  22.0 & 24.5 & 22.0 & 23.0  & \multirow{1}*{$-$22.8}  \\
            & \multirow{1}*{$150$} &  21.0 & 20.5 & 20.5 & 21.5   & \multirow{1}*{$-$23.6}  \\
            & \multirow{1}*{$160$} &  19.0 & 19.0 & 19.5 & 19.0   & \multirow{1}*{$-$19.6}  \\
        \hline
        \multirow{3}*{$7$} & \multirow{1}*{$140$} &  23.0 & 21.5 & 22.5 & 24.0   & \multirow{1}*{$-$17.1}  \\
            & \multirow{1}*{$150$} &  21.5 & 21.0 & 20.5 & 21.5   & \multirow{1}*{$-$16.8}  \\
            & \multirow{1}*{$160$} &  19.5 & 19.5 & 19.0 & 19.5   & \multirow{1}*{$-$20.6}  \\
        \hline
        \multirow{3}*{$16$} & \multirow{1}*{$140$} &  22.5 & 22.5 & 23.0 & 23.5   & \multirow{1}*{$-$20.0}  \\
            & \multirow{1}*{$150$} &  21.0 & 20.0 & 21.0 & 21.0   & \multirow{1}*{$-$22.2}  \\
            & \multirow{1}*{$160$} &  19.5 & 19.5 & 19.0 & 19.5   & \multirow{1}*{$-$23.6}  \\
        \hline

        \multirow{3}*{$20$} & \multirow{1}*{$140$} &  22.5 & 21.5 & 22.5 & 23.5   & \multirow{1}*{$-$21.4}  \\
            & \multirow{1}*{$150$} &  21.5 & 20.5 & 20.0 & 21.0   & \multirow{1}*{$-$19.2}  \\
            & \multirow{1}*{$160$} &  19.0 & 19.5 & 19.5 & 19.5   & \multirow{1}*{$-$21.5}  \\
        \hline

        \multirow{3}*{$29$} & \multirow{1}*{$140$} &  22.0 & 24.0 & 23.0 & 23.0   & \multirow{1}*{$-$18.5}  \\
            & \multirow{1}*{$150$} &  20.5 & 21.0 & 21.0 & 21.0   & \multirow{1}*{$-$21.5}  \\
            & \multirow{1}*{$160$} &  20.0 & 19.5 & 19.5 & 19.5   & \multirow{1}*{$-$21.3}  \\
        \hline
        \multirow{3}*{$31$} & \multirow{1}*{$140$} &  22.5 & 23.5 & 22.5 & 23.0   & \multirow{1}*{$-$20.8}  \\
            & \multirow{1}*{$150$} &  20.5 & 20.0 & 21.0 & 21.0   & \multirow{1}*{$-$19.2}  \\
            & \multirow{1}*{$160$} &  19.0 & 19.5 & 19.0 & 19.5   & \multirow{1}*{$-$18.4}  \\
        \hline
        \multirow{3}*{C} & \multirow{1}*{$140$} &  22.0 & 23.5 & 22.5 & 22.0   & \multirow{1}*{$-$18.7}  \\
            & \multirow{1}*{$150$} &  20.0 & 21.5 & 21.5 & 21.5   & \multirow{1}*{$-$19.3}  \\
            & \multirow{1}*{$160$} &  19.0 & 19.0 & 19.5 & 19.5   & \multirow{1}*{$-$19.7}  \\
        \hline
    \end{tabular}
    \end{center}
\end{table}

\newpage
\section{Appendix: Analysis of the cross polarization mismatch}
\label{sec_crosspol_verification}

    In this appendix we discuss possible sources of the discrepancy between the measured and simulated cross polarization highlighted in Fig.~\ref{fig:three:graphs:cross}. We first address briefly the uncertainty of the alignment between the source and receiver antennas during the beam pattern measurement and then we study the impact of plates misalignment induced by the mechanical accuracy in the alignment pin holes.

\subsection{Alignment in the experimental setup}
\label{sec_setup_alignment}

    We had aligned the transmitting and receiving antennas with a laser pointer with the setup in the H-plane scan configuration. When we measured the other planes we did not repeat the alignment procedure, but we simply rotated the antennas to set the other measurement planes. Therefore we cannot exclude that this movement may have introduced a slight misalignment with respect to the starting position.

    If we look at the beam patterns in Fig.~\ref{fig:three:graphs_Hplane} we see that the residuals between measured and simulated patters confirm the quality of our alignment. Indeed, if we look at the residuals in the other planes (see Figs.~\ref{fig:three graphs_Eplane}, \ref{fig:three graphs_45plane}, and \ref{fig:three graphs_m45plane}) we can appreciate slightly larger residuals compared with those of the H-plane, which suggests that some misalignment could have occurred during the motors movement.

    We believe, however, that this possible misalignment is not enough to explain a difference of $\sim7$ dB in the cross polarization patterns.

\subsection{Analysis of misalignment among antenna plates}
\label{sec_plates_misalignment_analysis}

    We have studied the impact of plates misalignment by associating at every plate a random displacement extracted from a normal distribution. We describe this effect as:

    \begin{equation}
        d\vec{r}_n = (\sigma \cdot |d\vec{r}|_n + \mu) \cdot \hat{\vec{n}}_n,
        \label{eq_displacement_model}
    \end{equation}
    where $|d\vec{r}|_n$ is the length of the $n$-th plate displacement, $\hat{\vec{n}}_n$ is the displacement direction, $\sigma$ and $\mu$ are the standard deviation and mean of the normal distribution, respectively.

    We performed the analysis following two complementary approaches: (i) we tested various values of $\sigma$ and found the value that best reproduces the measured cross polarization, and (ii) we estimated $\sigma$ from the metrological analysis and compared the beam pattern simulated in this condition with the measured one for two horns in the array.

    \subsubsection{Estimate of the misalignment necessary to reproduce measurements}
    \label{sec_esimate_misalignment}

        In this part of the work we considered the fact that during the plates alignment we could check for any displacement of the tooth plates near the aperture. All the throats and the teeth towards the output waveguide, instead, could not be checked. This allowed us to assume that the first 9 teeth from the aperture were perfectly aligned while we applied the displacement model of Eq.~\ref{eq_displacement_model}. In Table~\ref{mis:model} we list the parameters that best reproduce the measured data.

        \begin{table}[h!]
            \renewcommand{\arraystretch}{1.5}
            \centering
            \caption{\label{mis:model}Parameters in the misalignment model that reproduce the measured cross polarization.}
            \begin{tabular}{p{5cm} c r }
                \hline
                Plate ID & $\sigma$ [$\mu$m] & $\mu$ [$\mu$m]   \\
                \hline
                \hline
                1-10\dotfill & 30.0 & 0.0 \\
                teeth-plate 10-end\dotfill& 0.0 & 0.0    \\
                throat-plate \dotfill & 30.0 & 0.0    \\
                \noalign{\smallskip}\hline
            \end{tabular}
        \end{table}

\begin{figure}[htbp]
            \centering
            \includegraphics[width=0.32\textwidth]{keys}\\
            \includegraphics[width=0.32\textwidth]{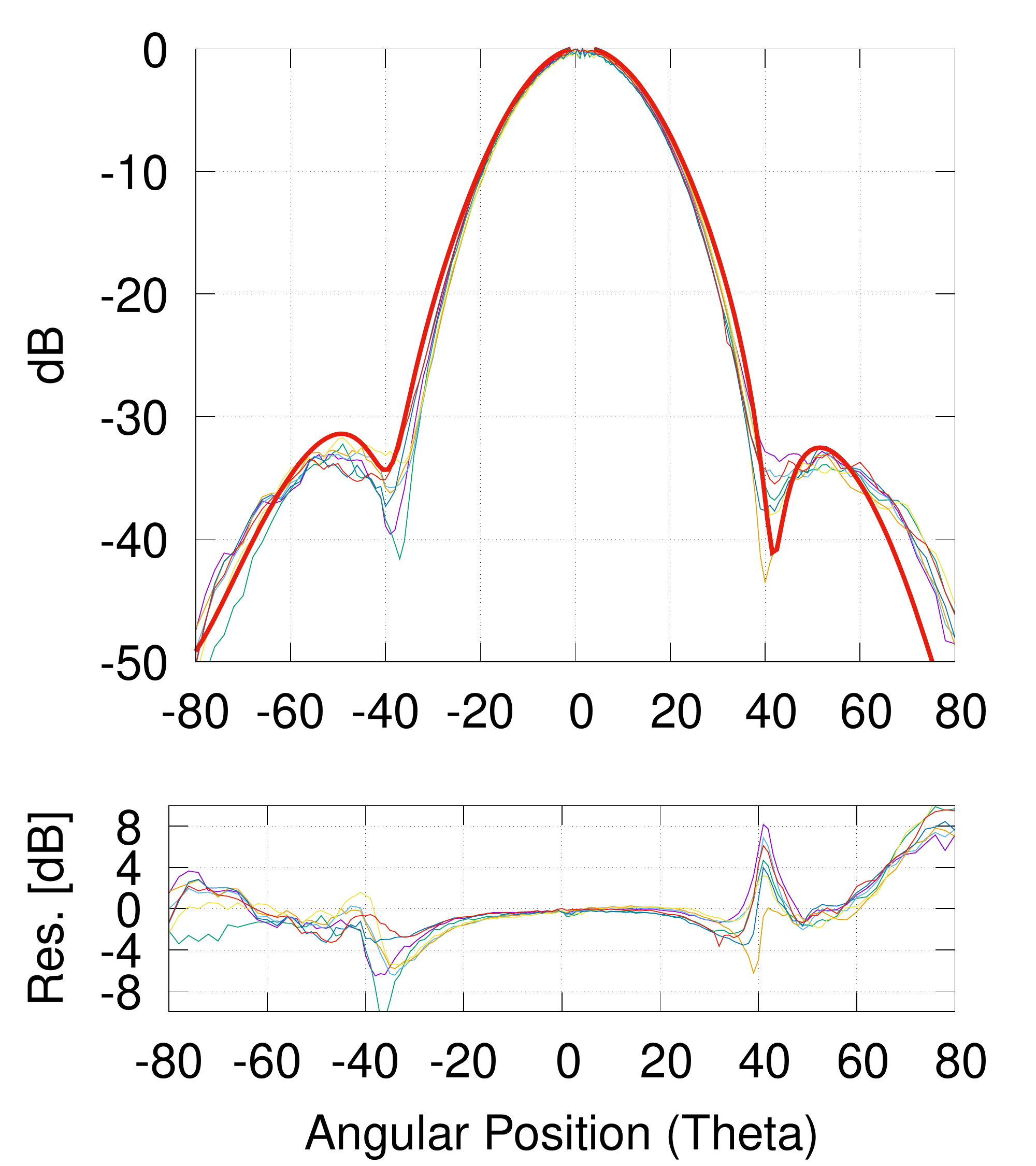}
            \includegraphics[width=0.32\textwidth]{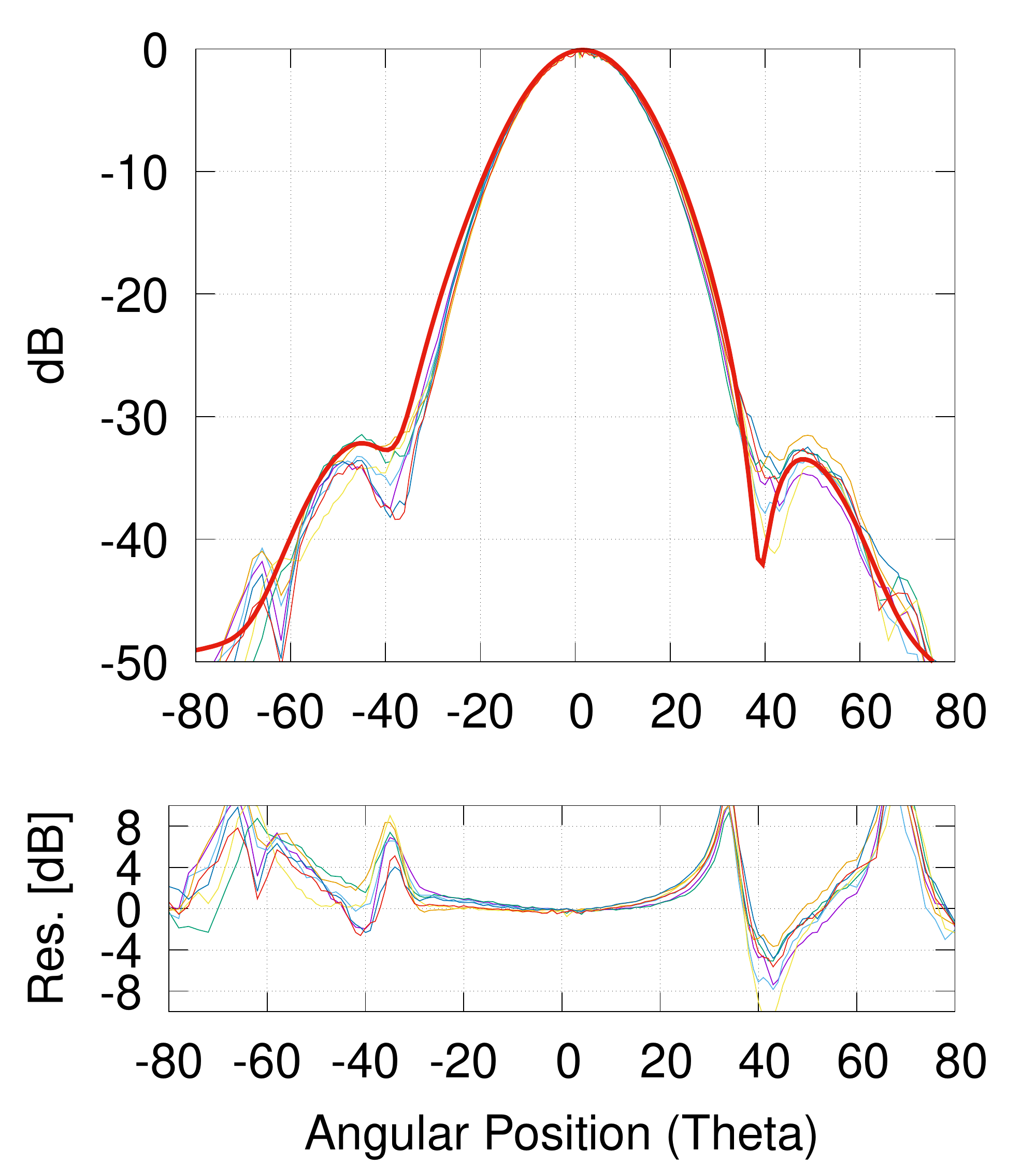}
            \includegraphics[width=0.32\textwidth]{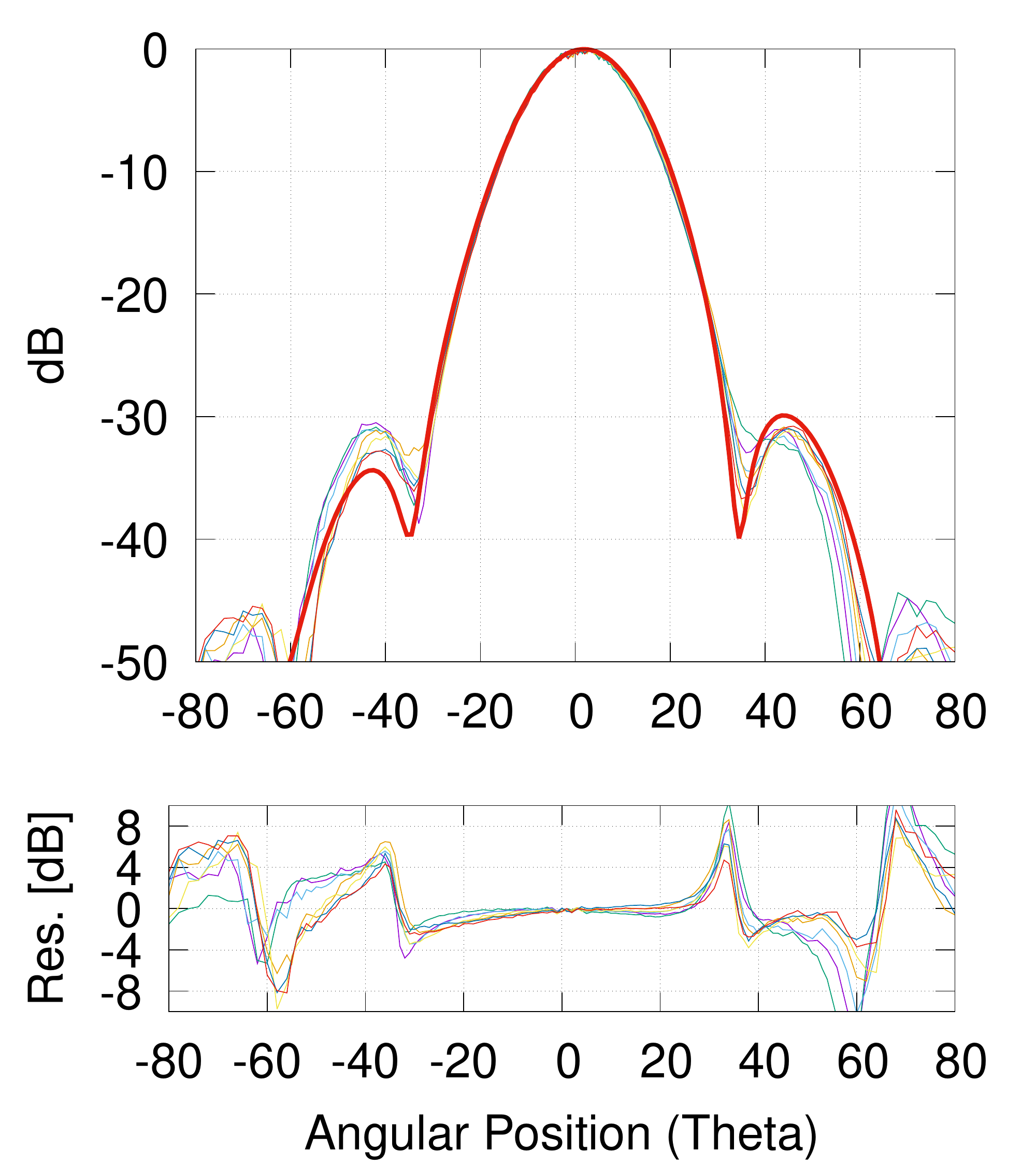}
            \includegraphics[width=0.32\textwidth]{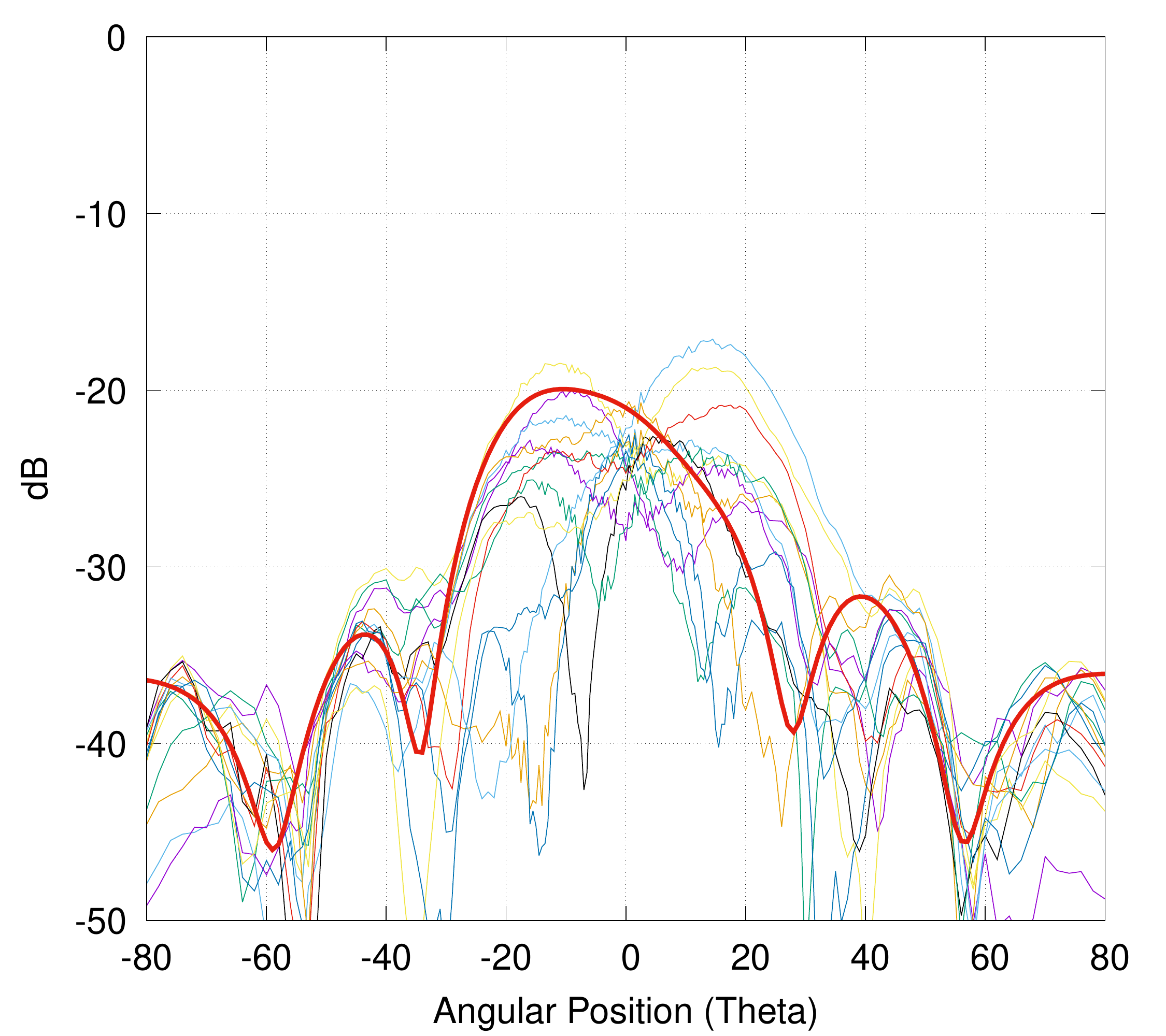}
            \includegraphics[width=0.32\textwidth]{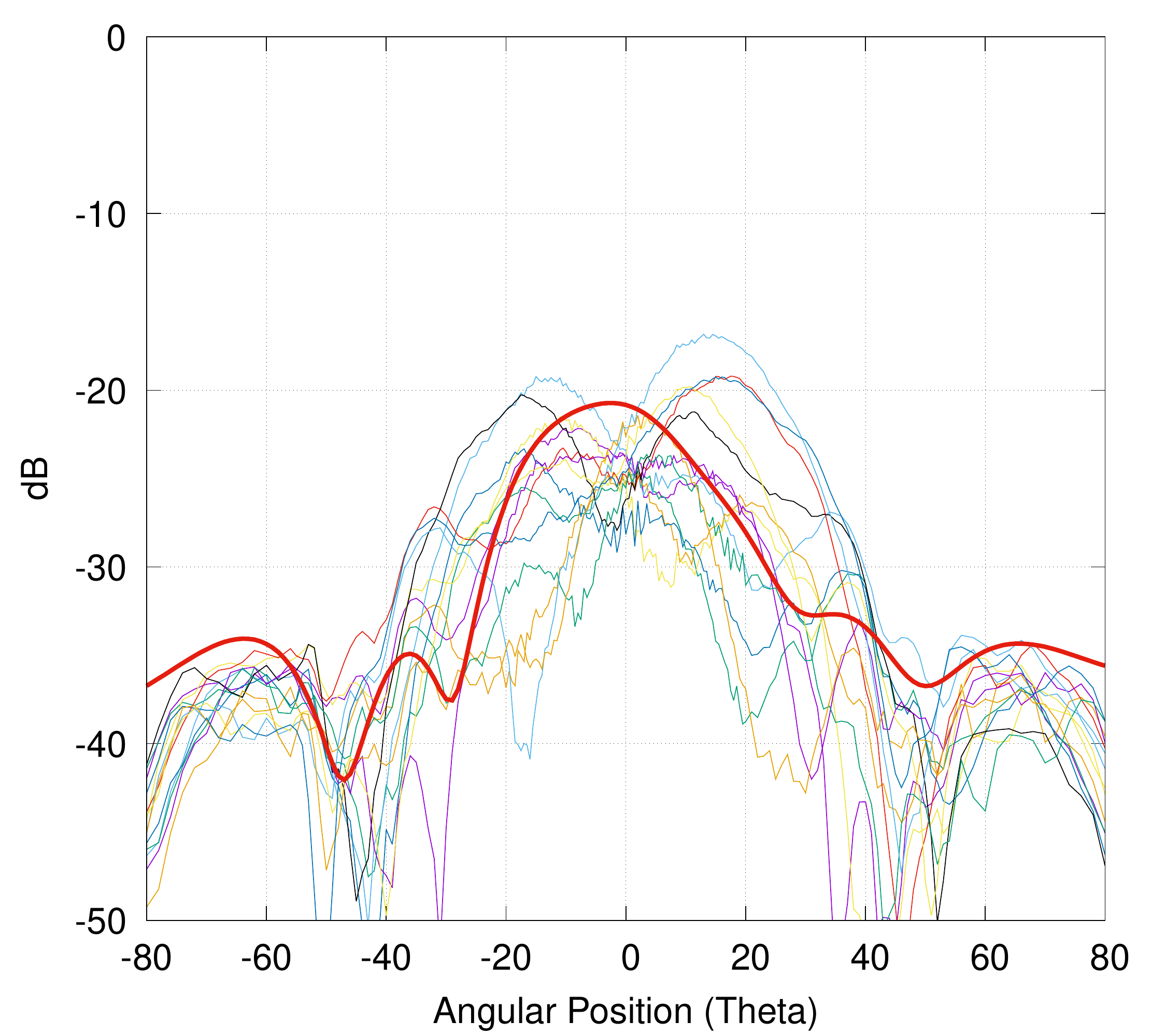}
            \includegraphics[width=0.32\textwidth]{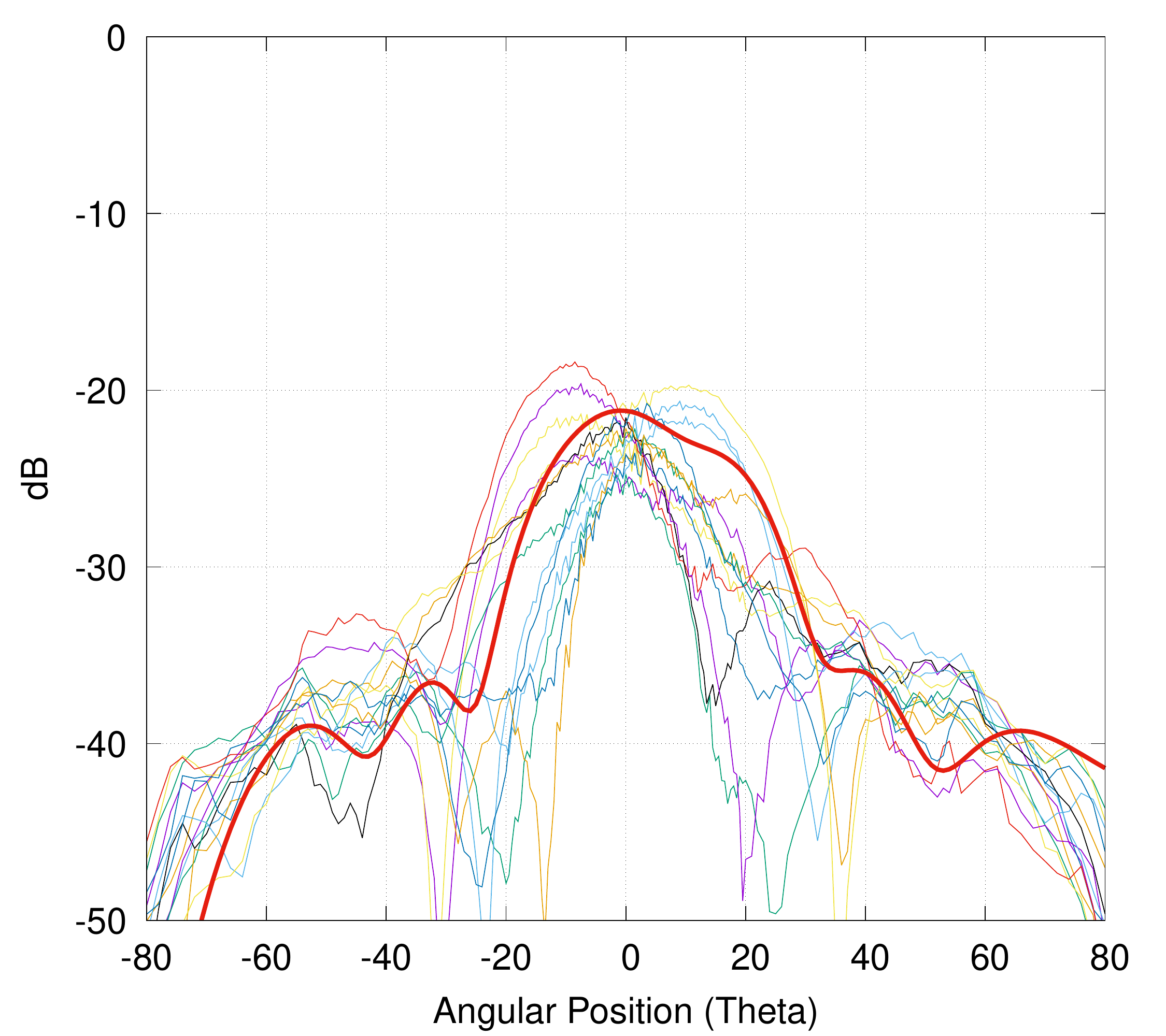}
            \caption{Comparison of measured beam patterns with the simulation performed considering the displacement model of Eq.~\ref{eq_displacement_model} with the parameters in Table~\ref{mis:model} (140\,GHz -left-, 150\,GHz -middle-, 160\,GHz -right-). The top row represents the H-plane pattern, the middle row is the difference between the measurement and the simulation and the bottom row is the cross polarization pattern.}
            \label{fig_check_misalignment_1}
        \end{figure}

        In Fig.~\ref{fig_check_misalignment_1} we show the results of this analysis by comparing the measured co-polar (H-plane) and cross-polar patterns with the beams simulated considering the parameters in Table~\ref{mis:model}. We see that the simulated beam is compatible with the data, although the asymmetry in the co-polar pattern is probably somewhat overestimated compared to the measured data.

    \subsubsection{Simulations considering the results of metrological measurements}
    \label{sec_simulations_with_metrological_measurements}

\begin{figure}[htbp]
            \centering
            \includegraphics[width=0.5\textwidth]{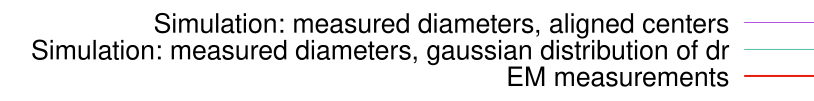}\\
            \includegraphics[width=0.32\textwidth]{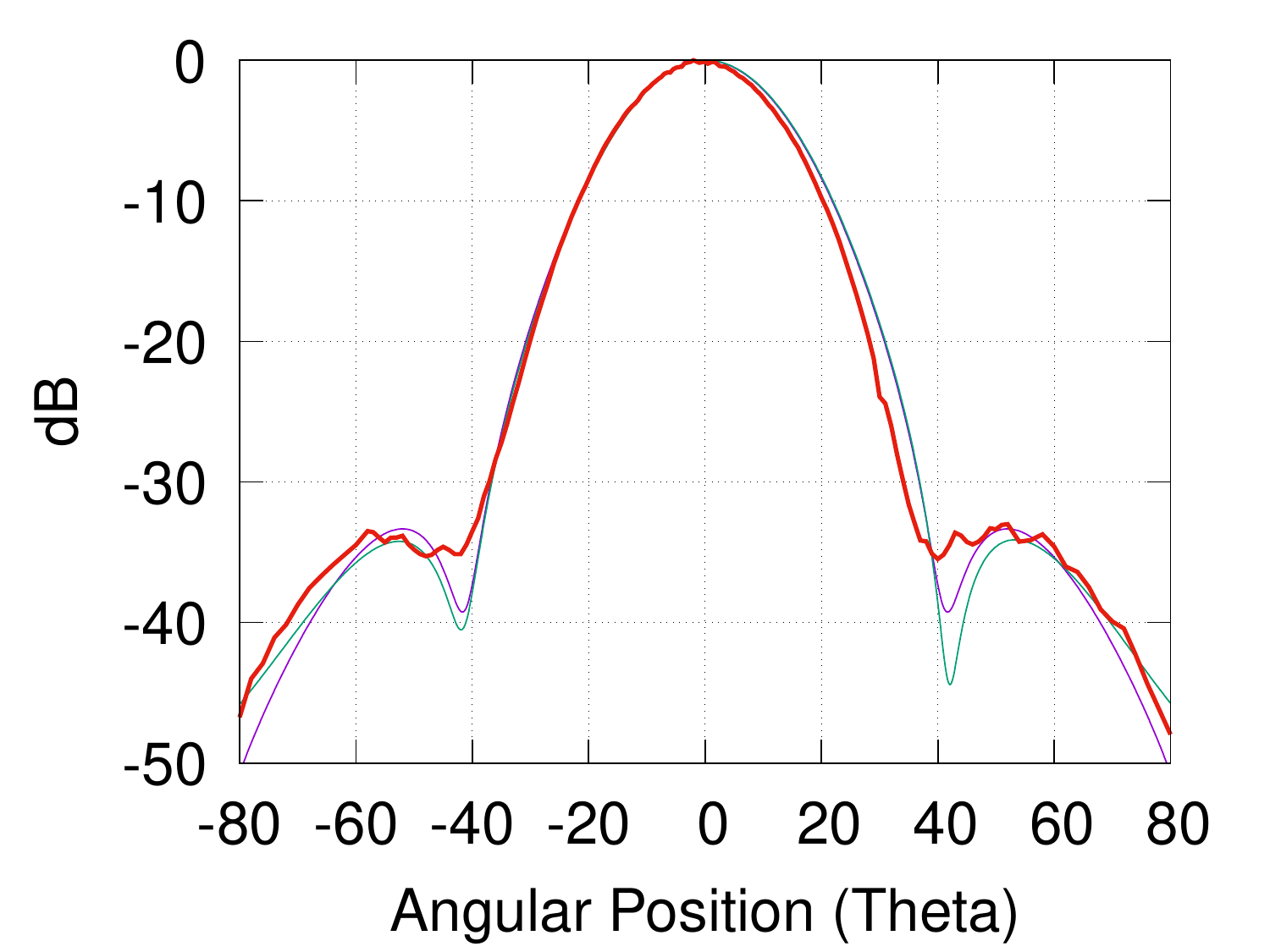}
            \includegraphics[width=0.32\textwidth]{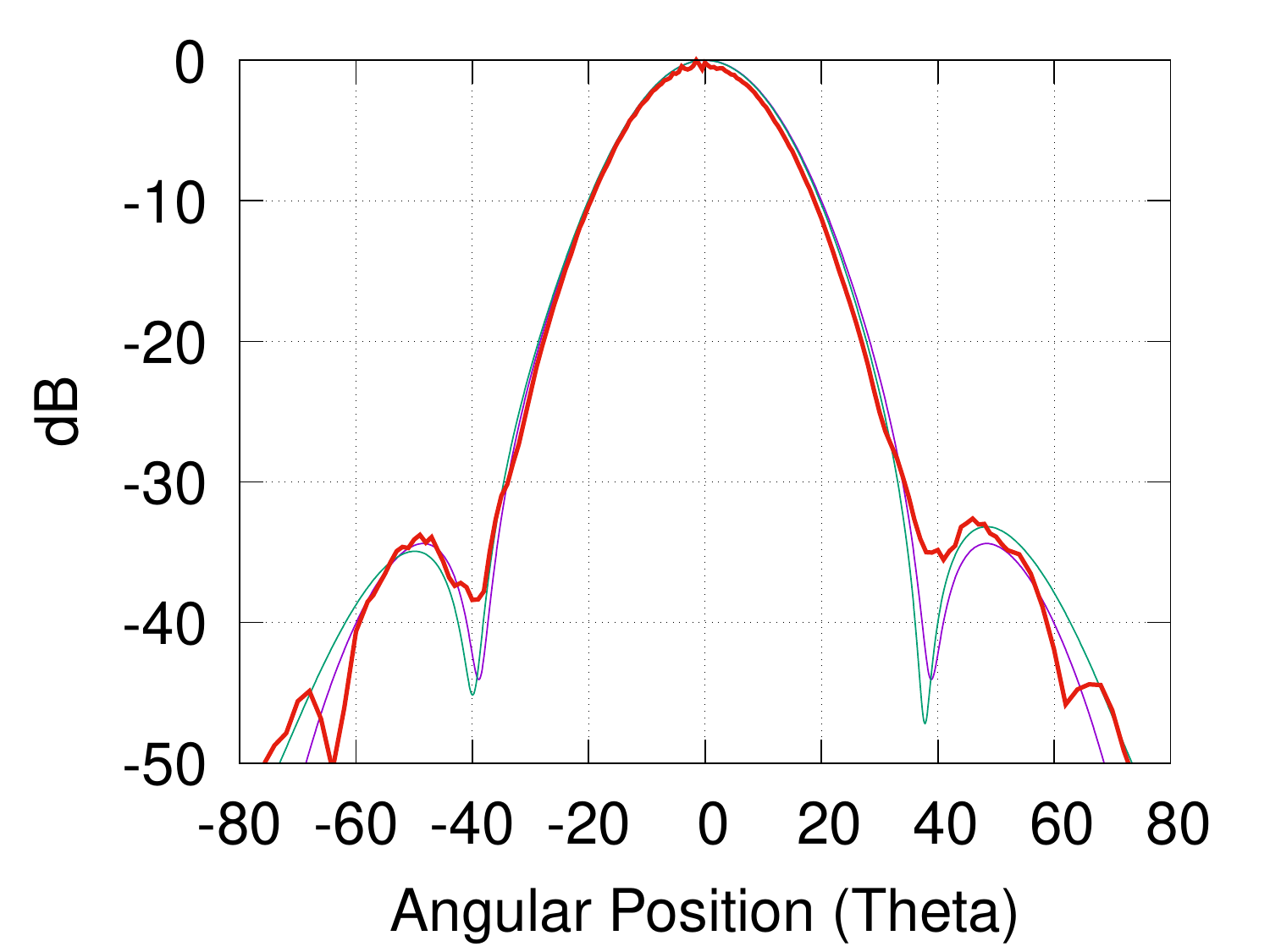}
            \includegraphics[width=0.32\textwidth]{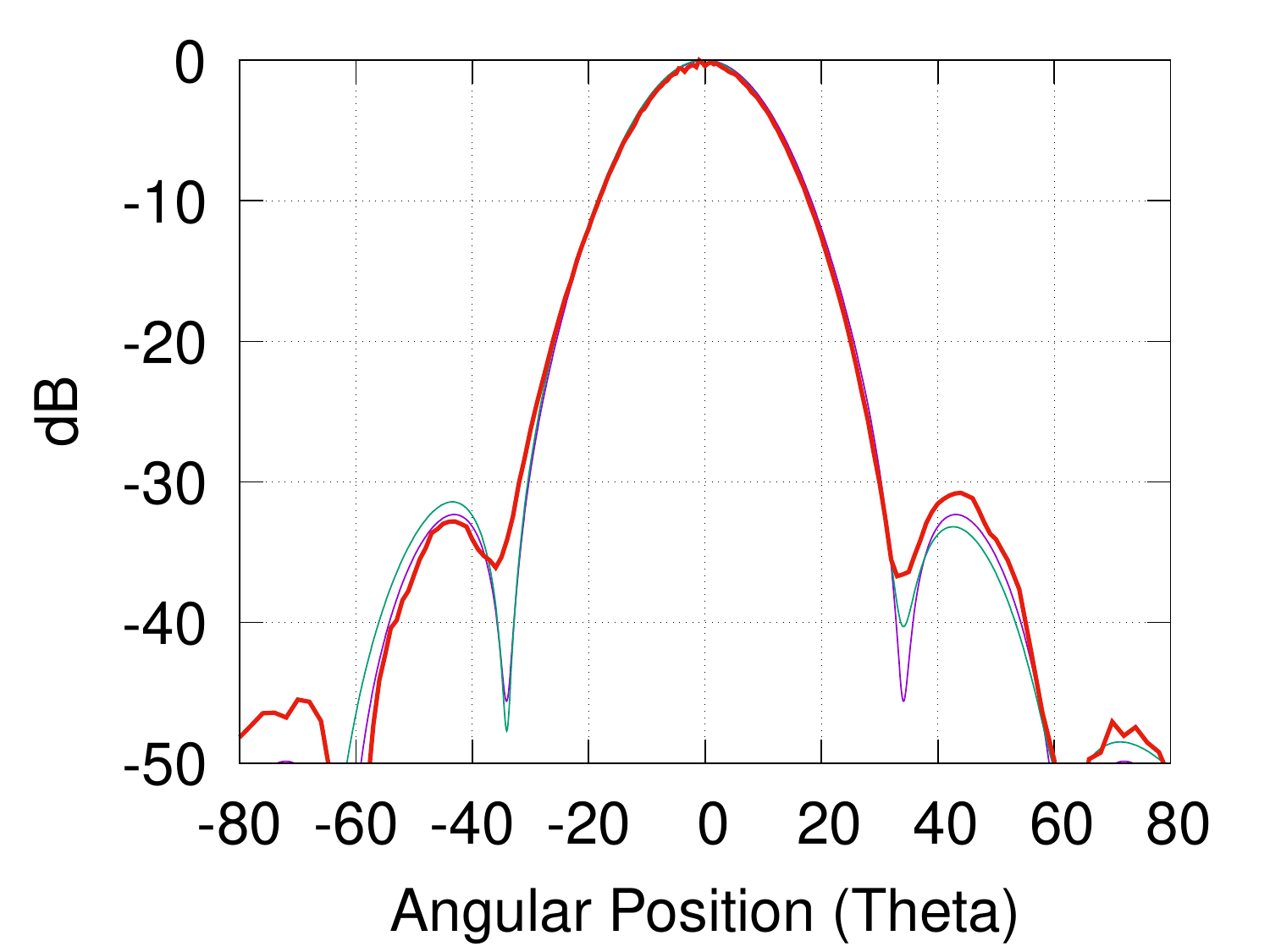}
            \includegraphics[width=0.32\textwidth]{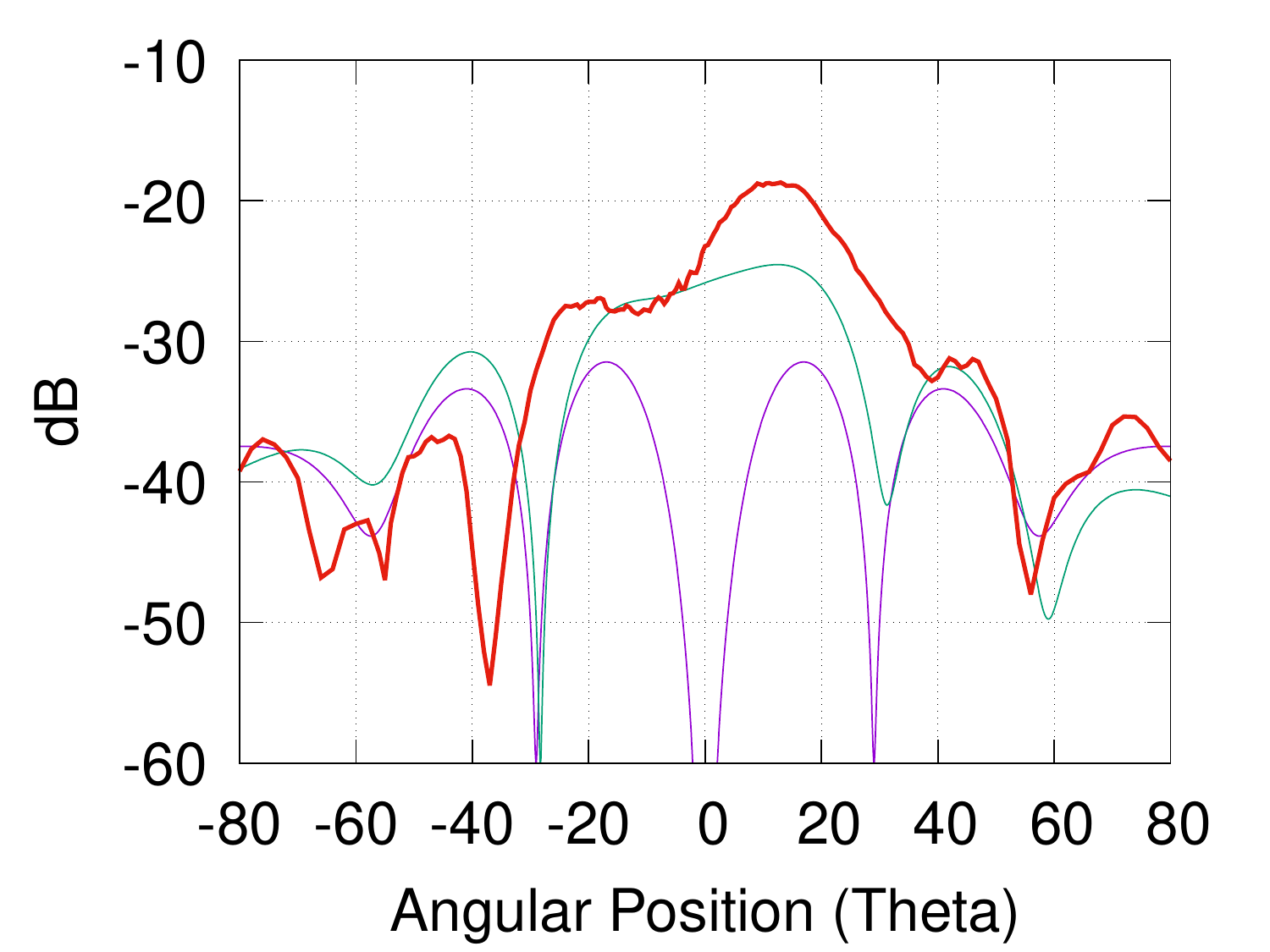}
            \includegraphics[width=0.32\textwidth]{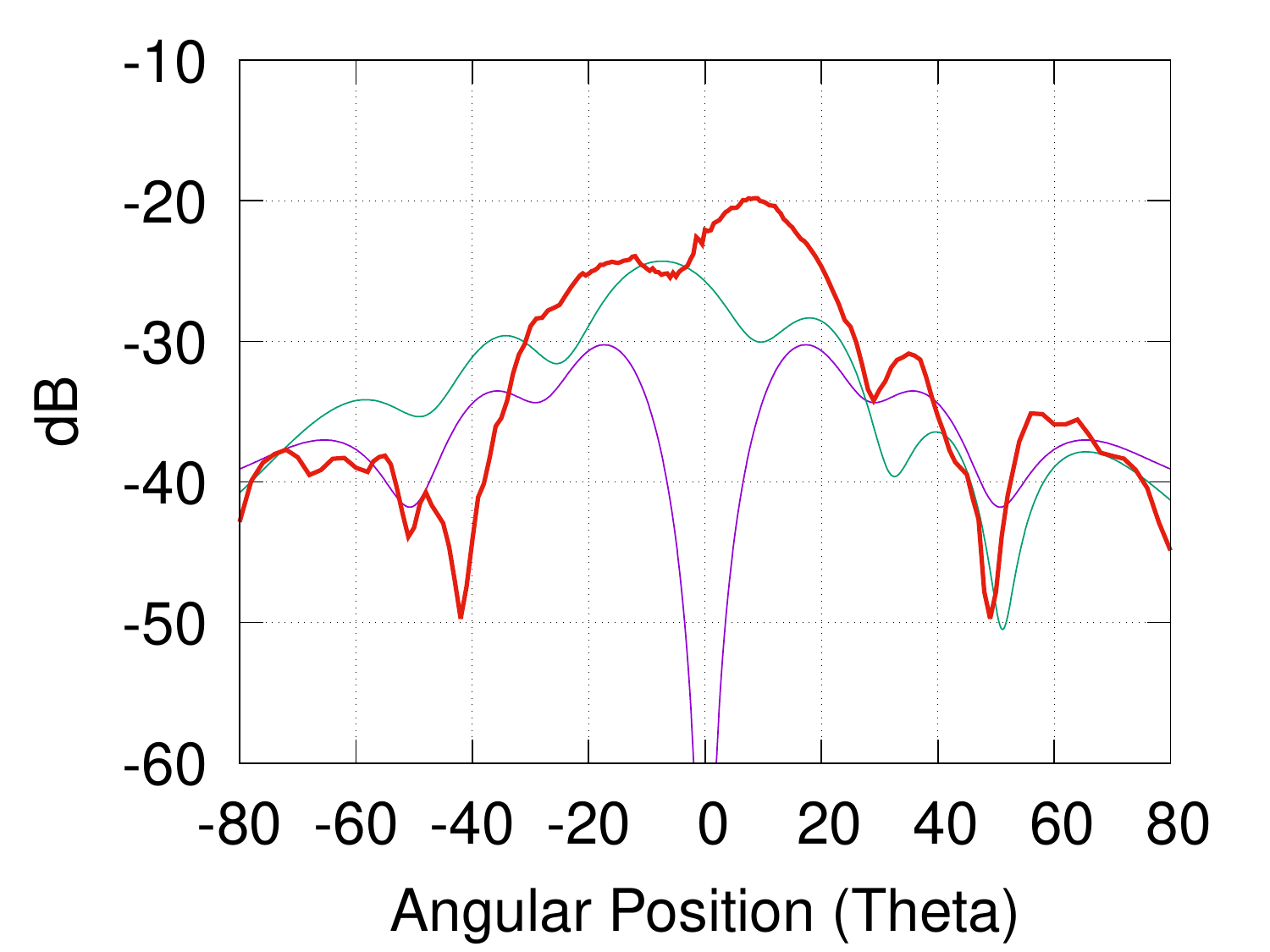}
            \includegraphics[width=0.32\textwidth]{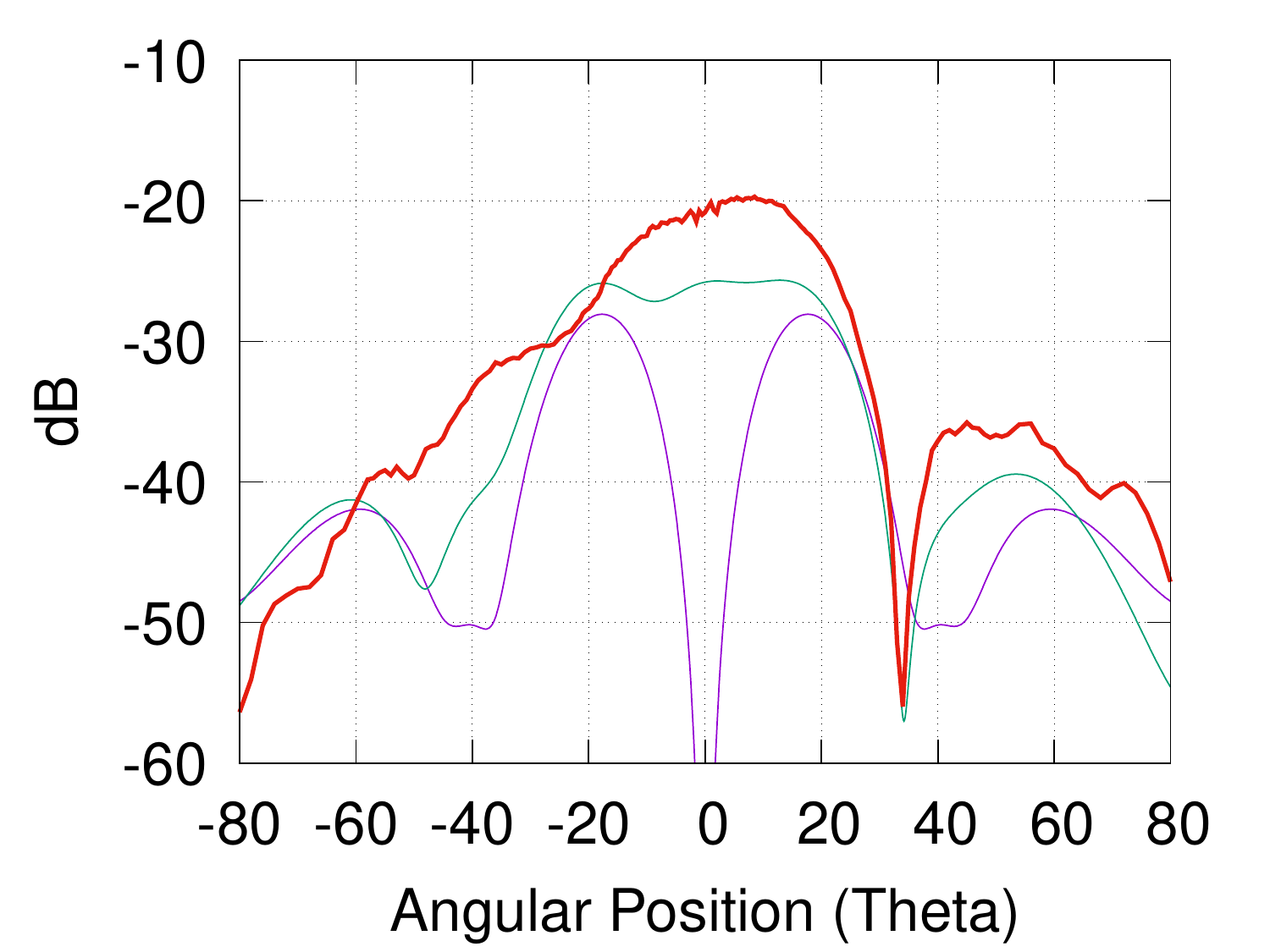}
            \includegraphics[width=0.32\textwidth]{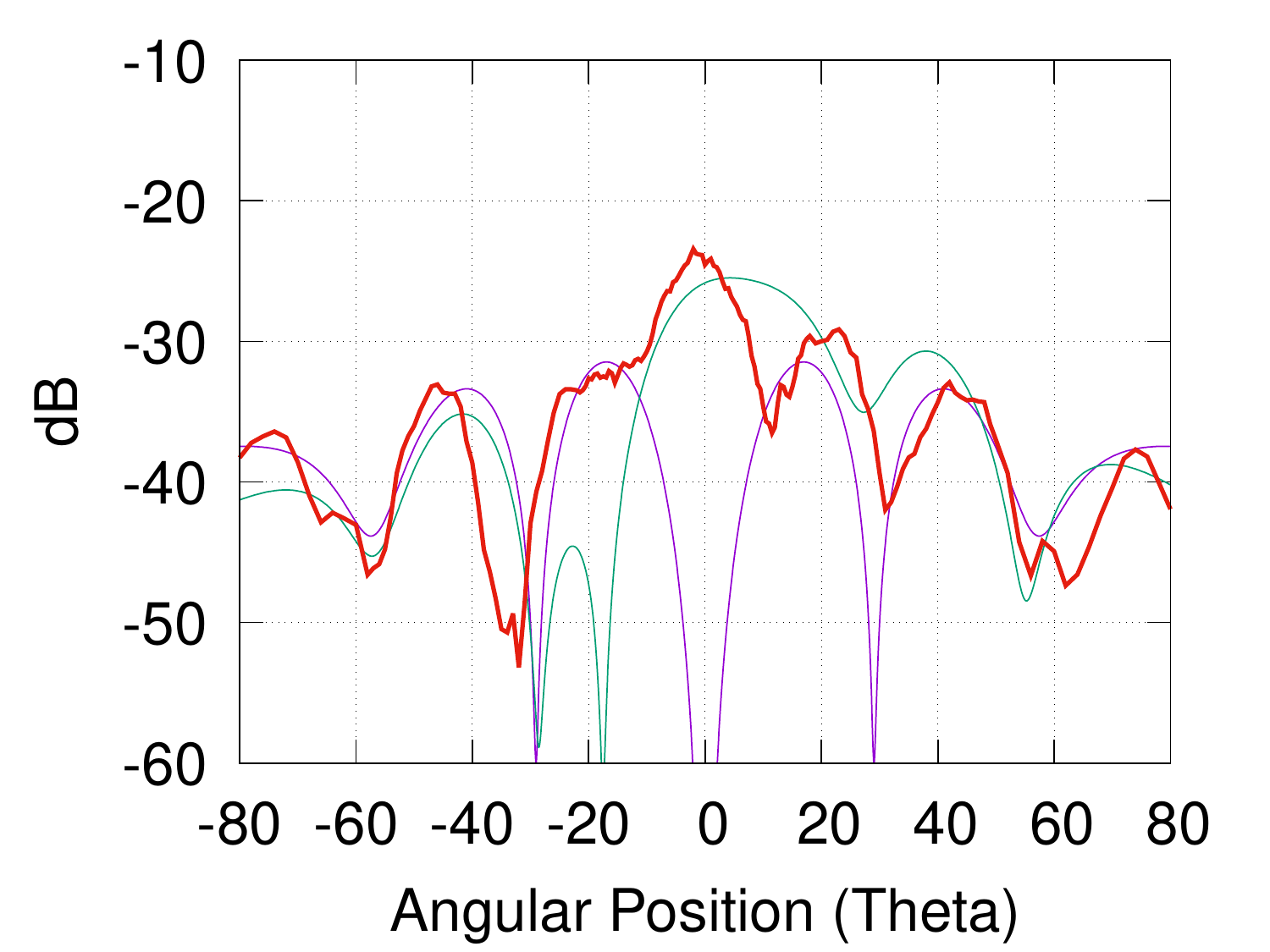}
            \includegraphics[width=0.32\textwidth]{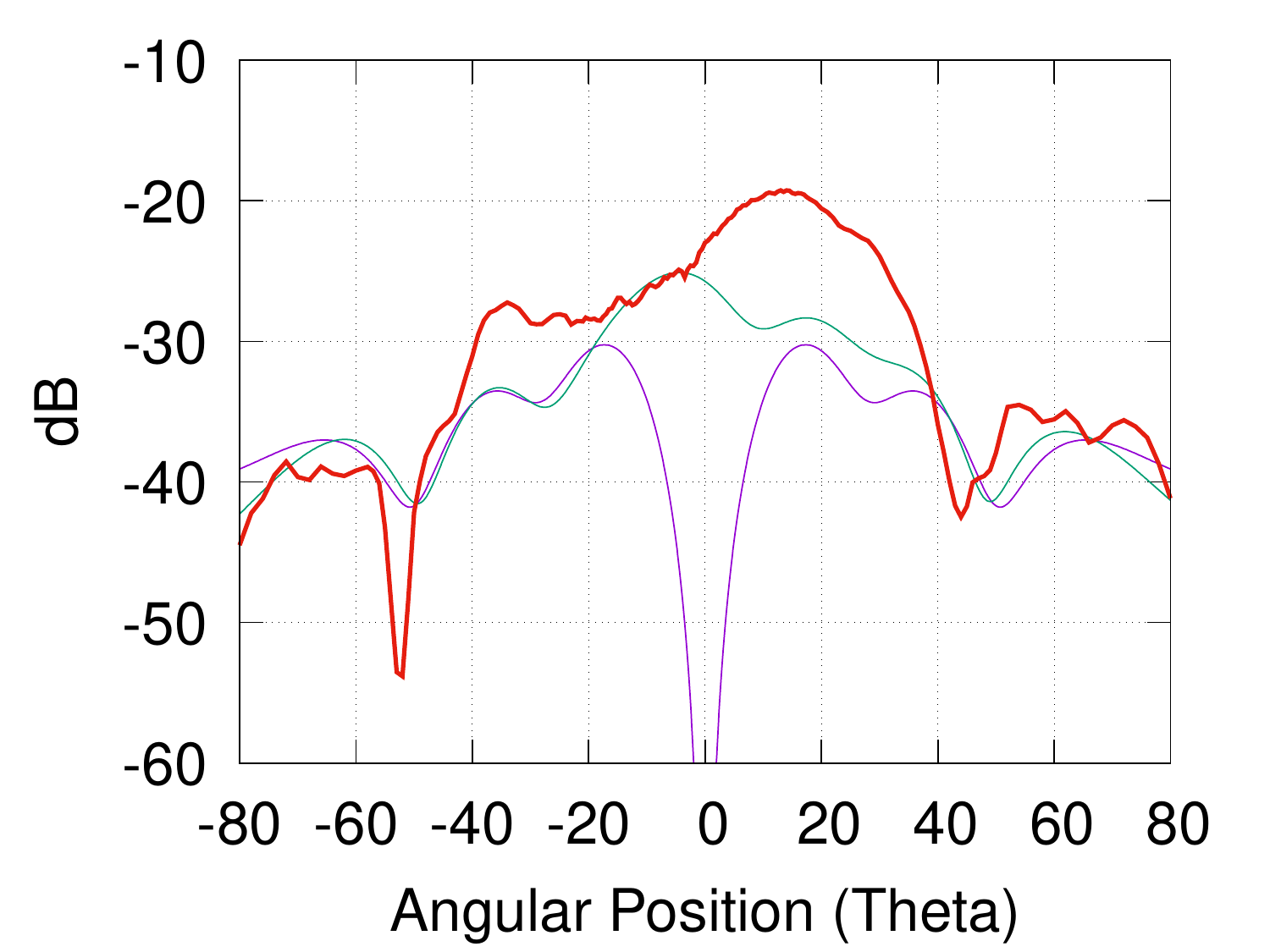}
            \includegraphics[width=0.32\textwidth]{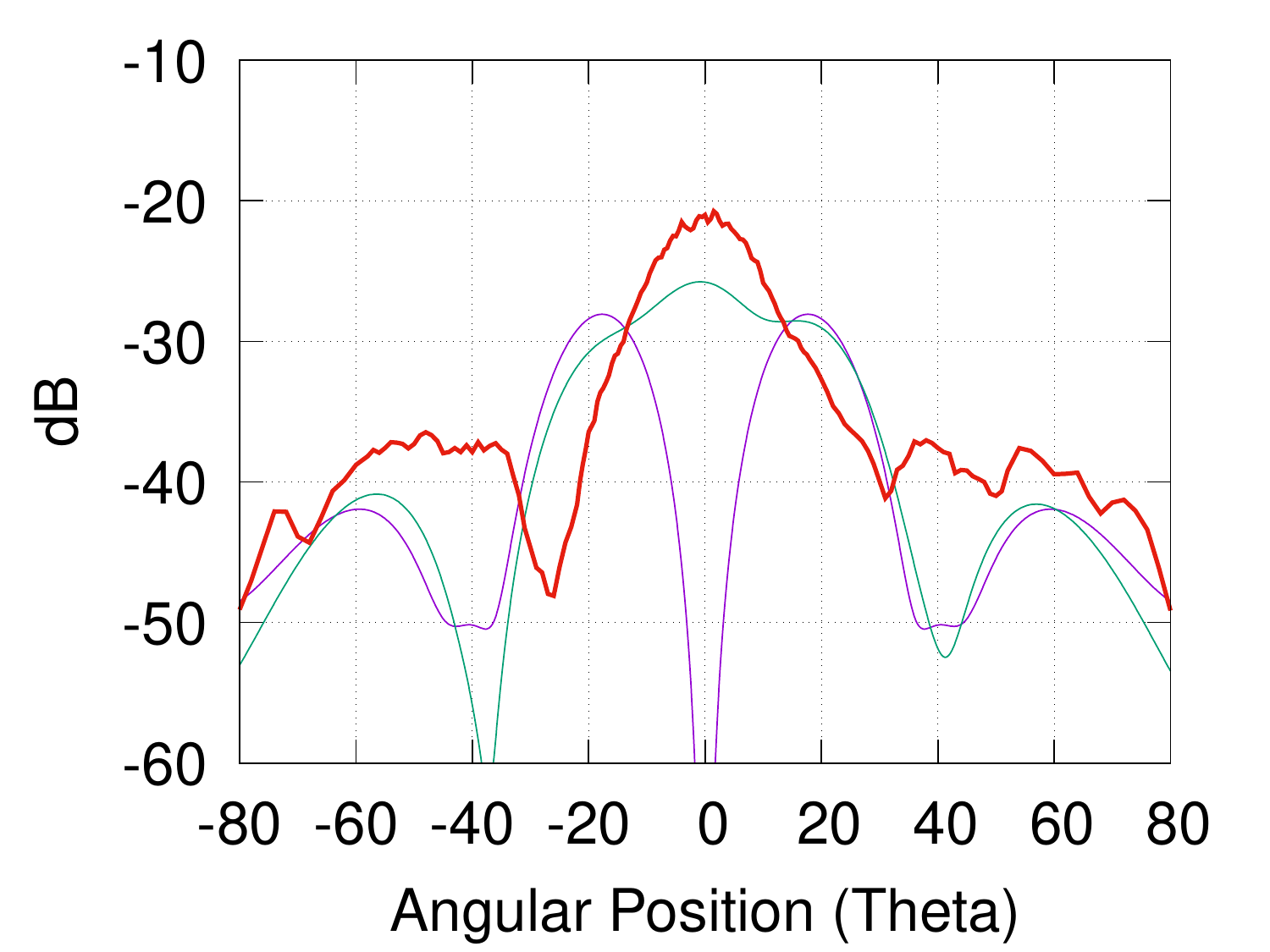}
            \caption{Simulated patterns and comparison with the electromagnetic measurements for the central antenna (140\,GHz -left-, 150\,GHz -middle-, 160\,GHz -right-). The top row is the co-polar H-plane pattern, the middle and bottow rows are the $\pm 45^\circ$ cross polarization patterns, respectively.}
            \label{fig:Antenna_18}
        \end{figure}

\begin{figure}[htbp]
            \centering
            \includegraphics[width=0.5\textwidth]{keys2.png}\\
            \includegraphics[width=0.32\textwidth]{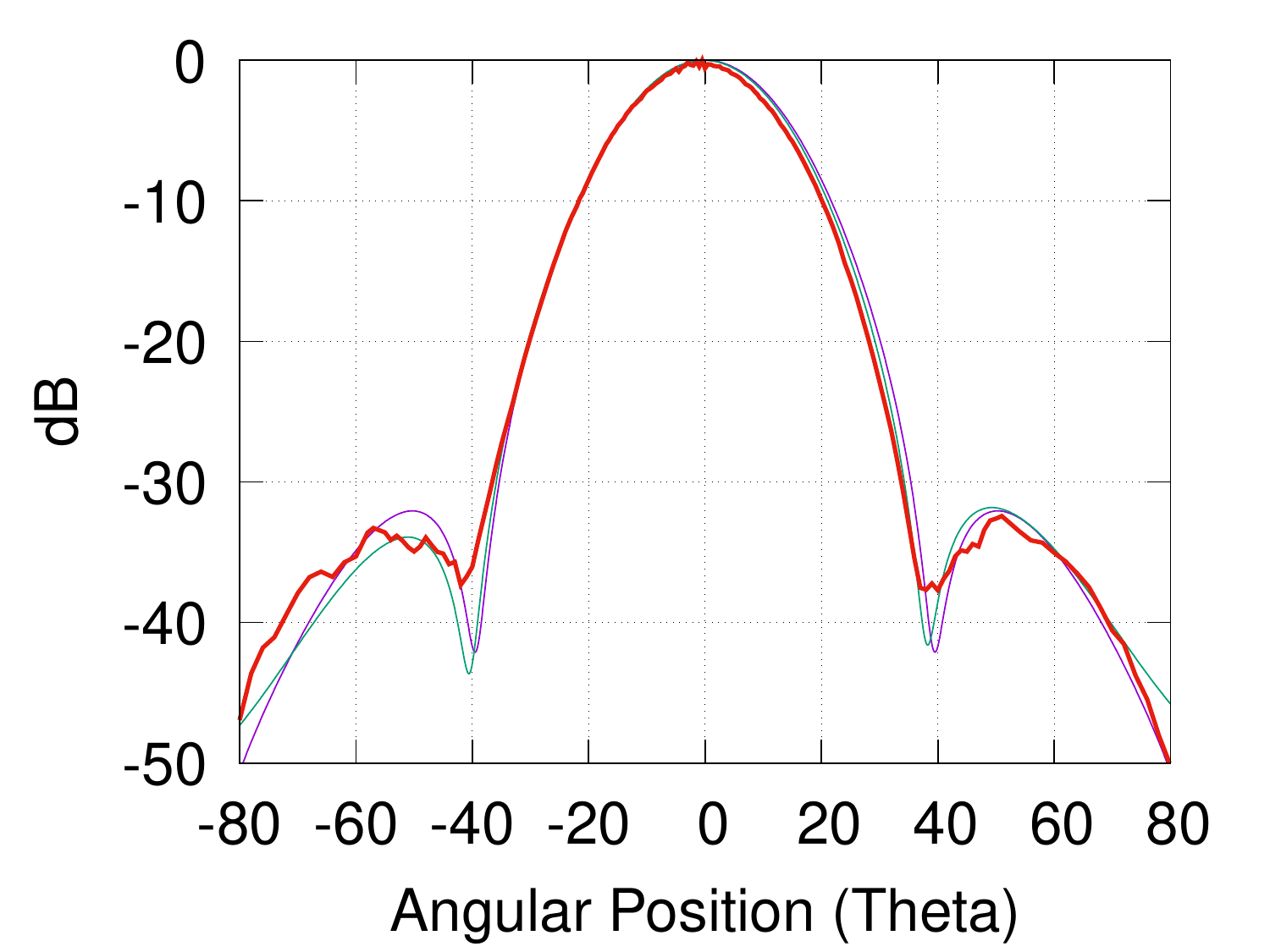}
            \includegraphics[width=0.32\textwidth]{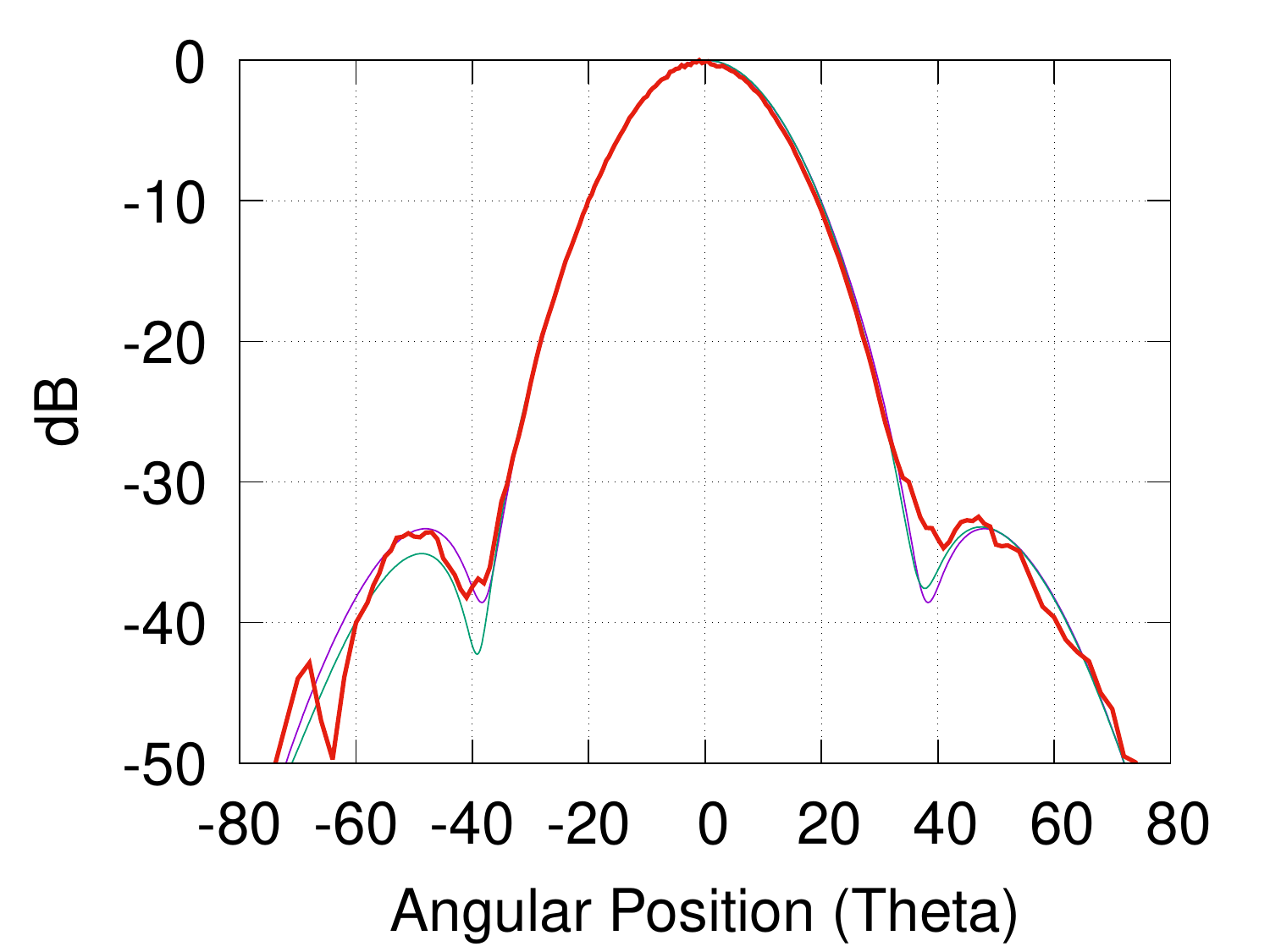}
            \includegraphics[width=0.32\textwidth]{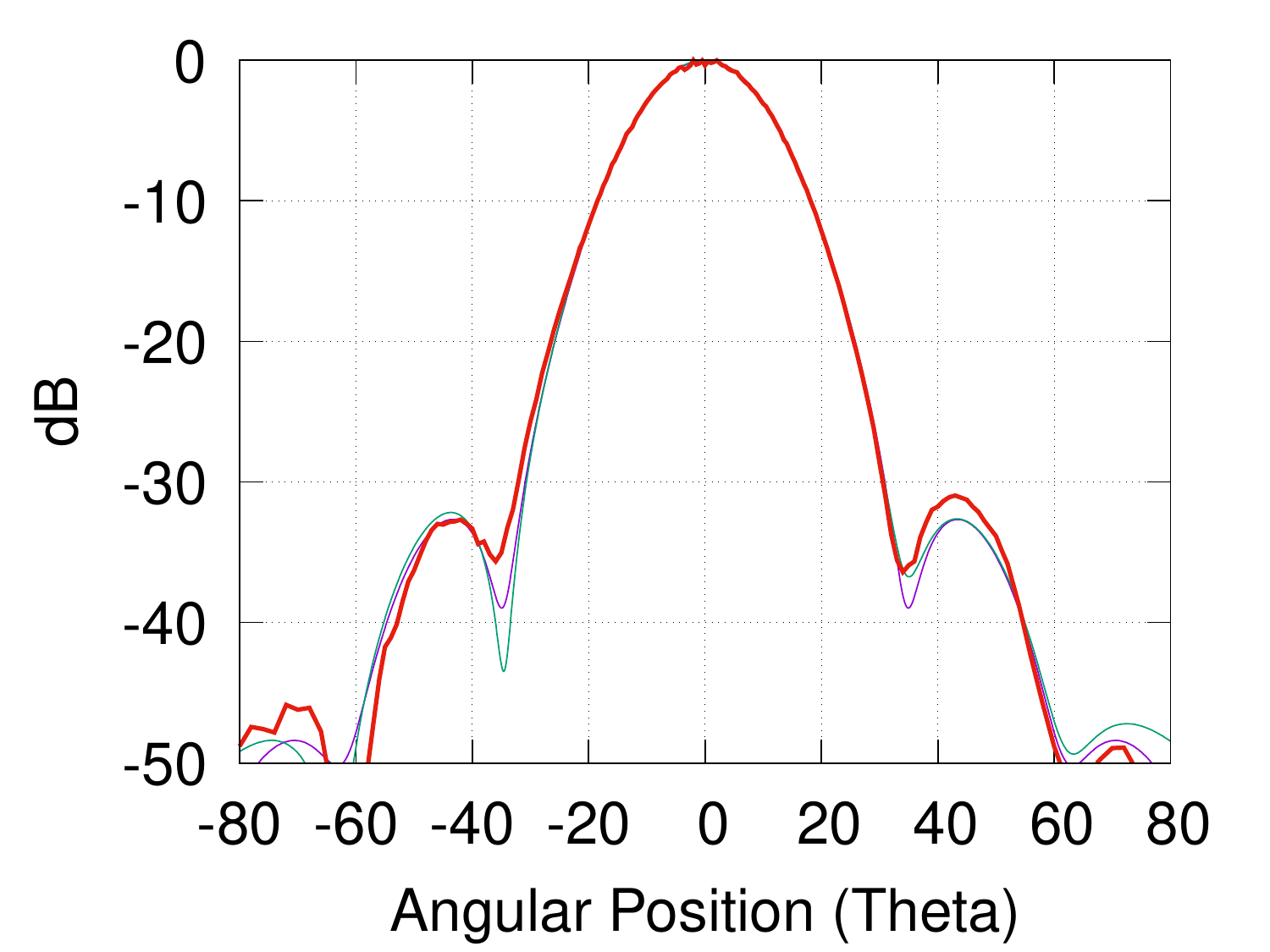}
            \includegraphics[width=0.32\textwidth]{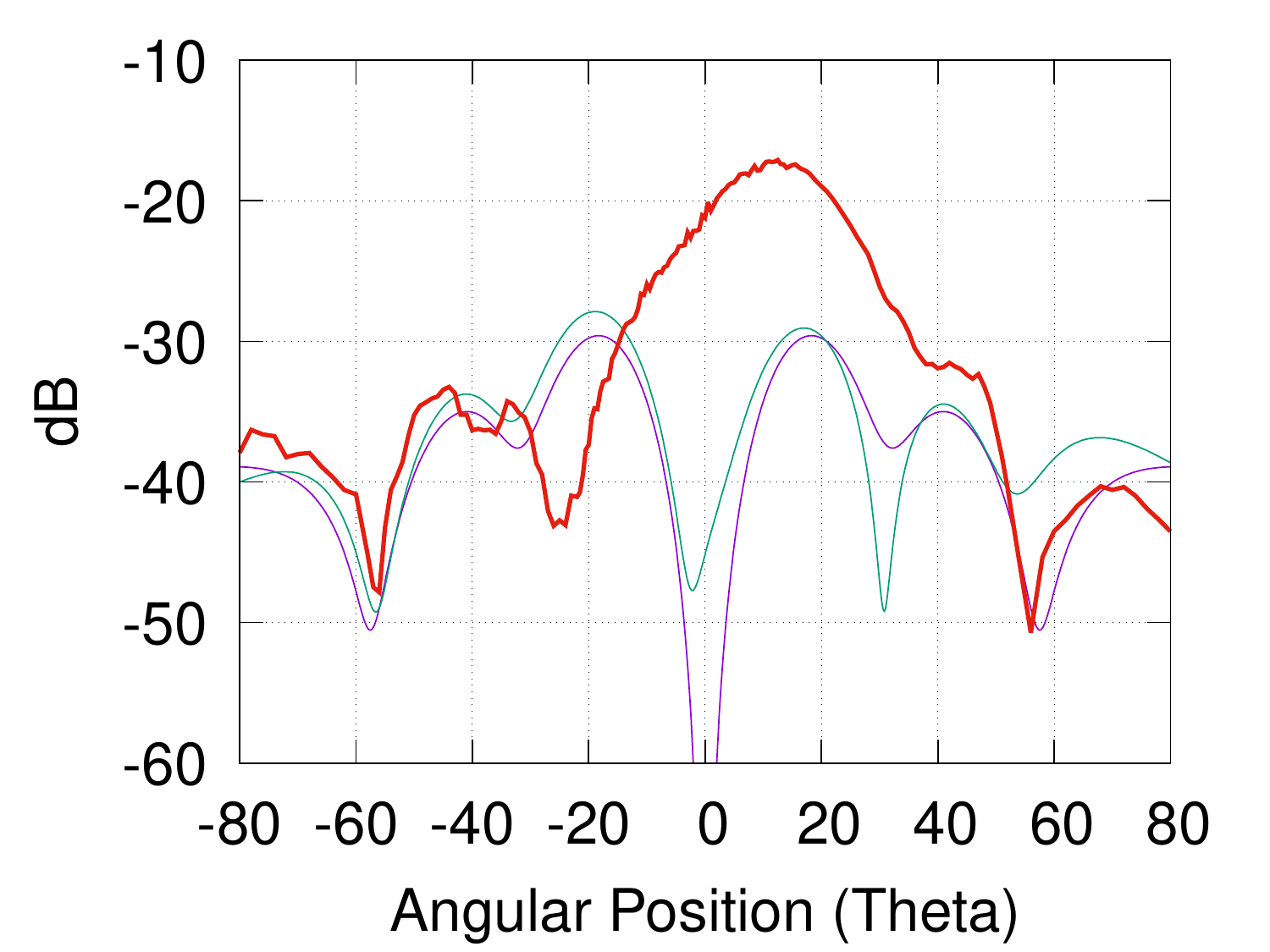}
            \includegraphics[width=0.32\textwidth]{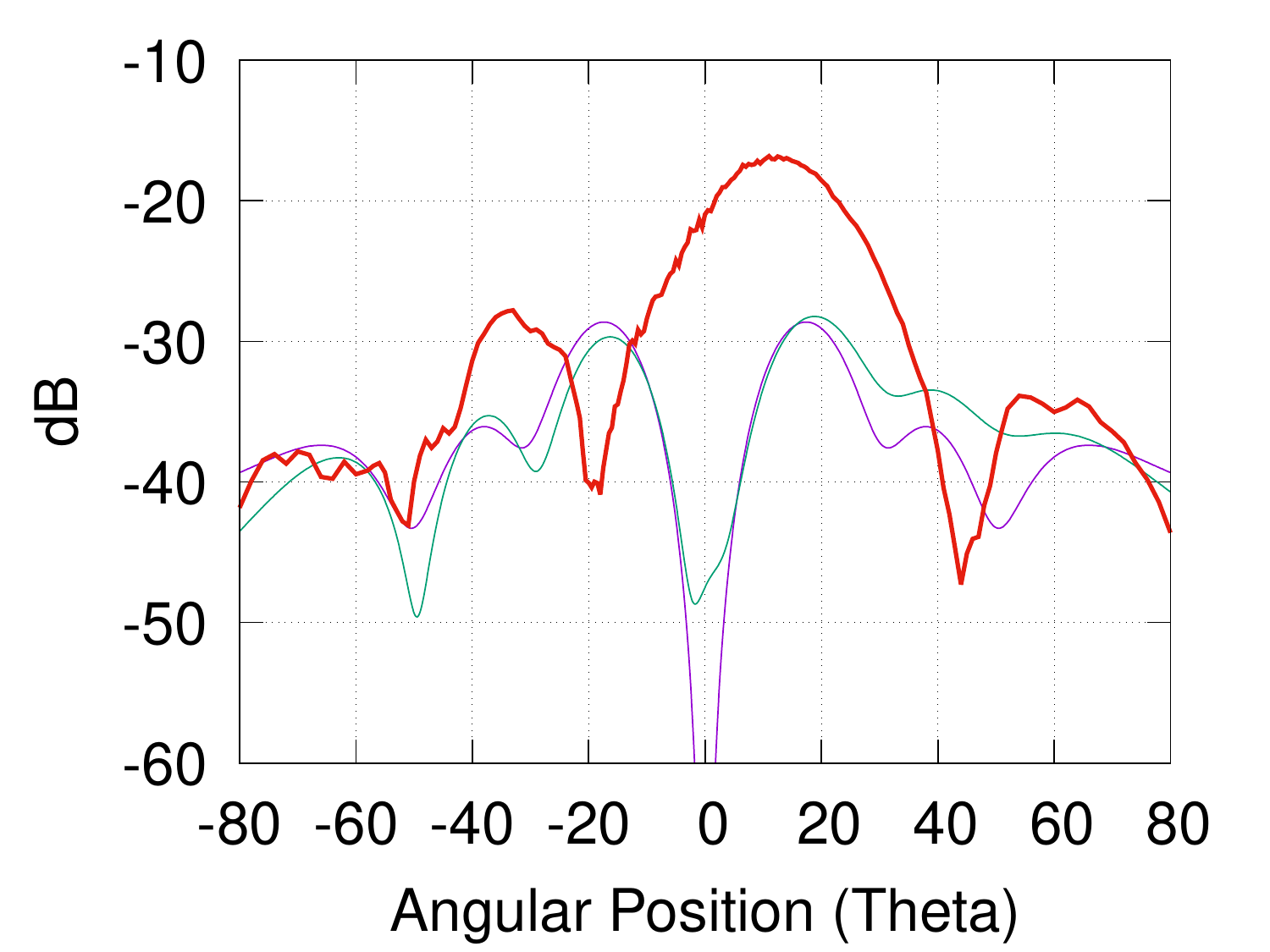}
            \includegraphics[width=0.32\textwidth]{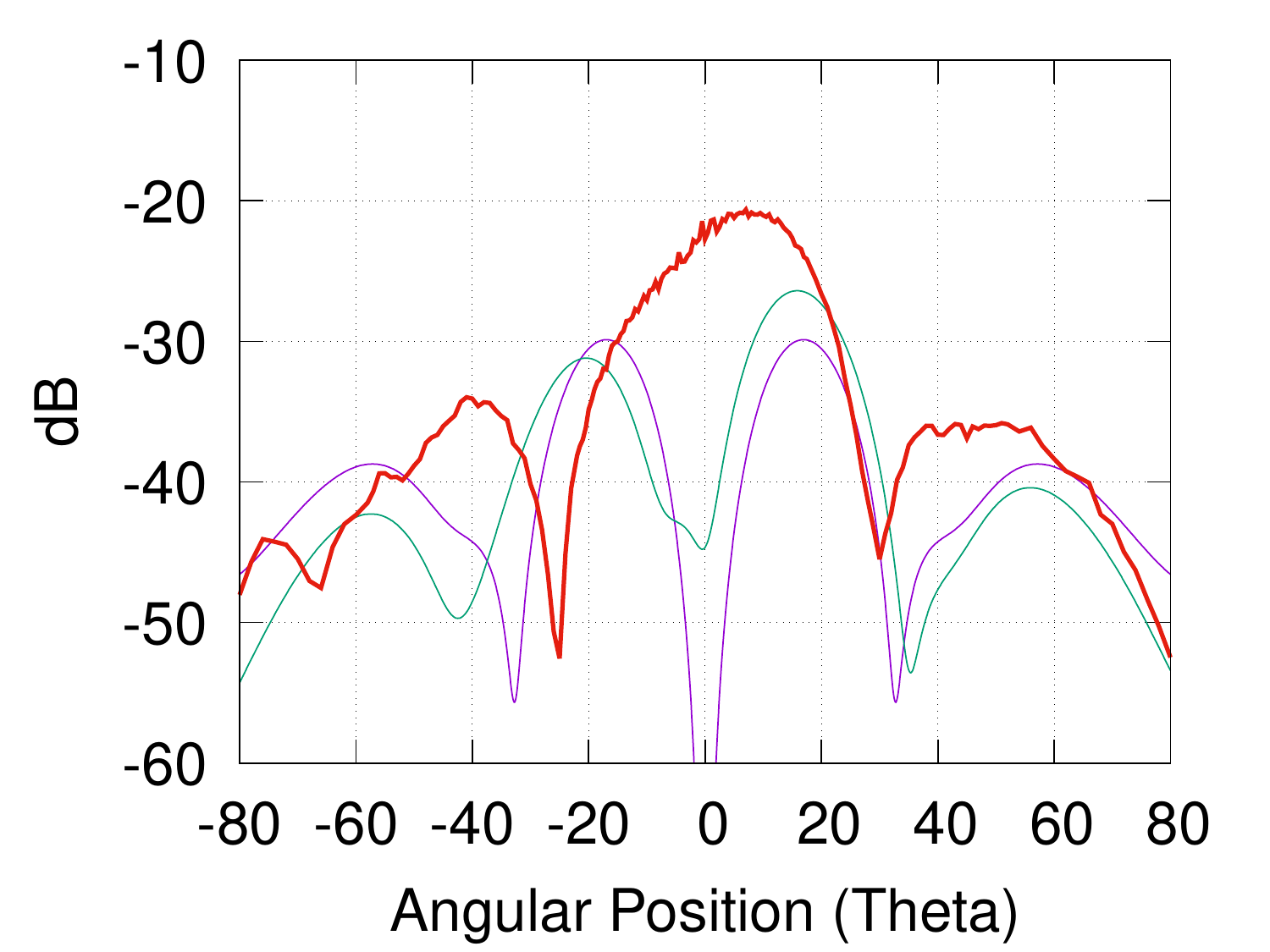}
            \includegraphics[width=0.32\textwidth]{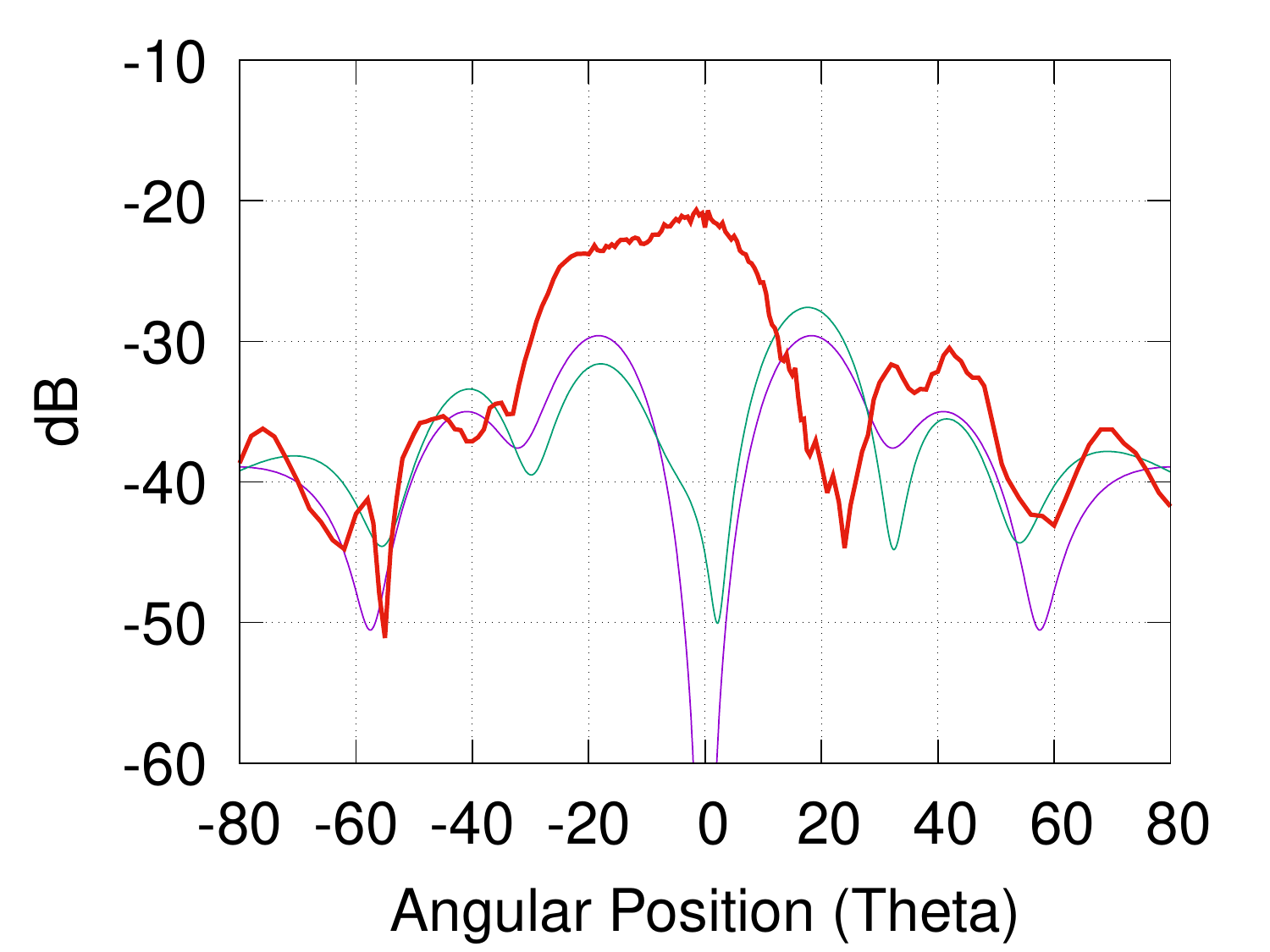}
            \includegraphics[width=0.32\textwidth]{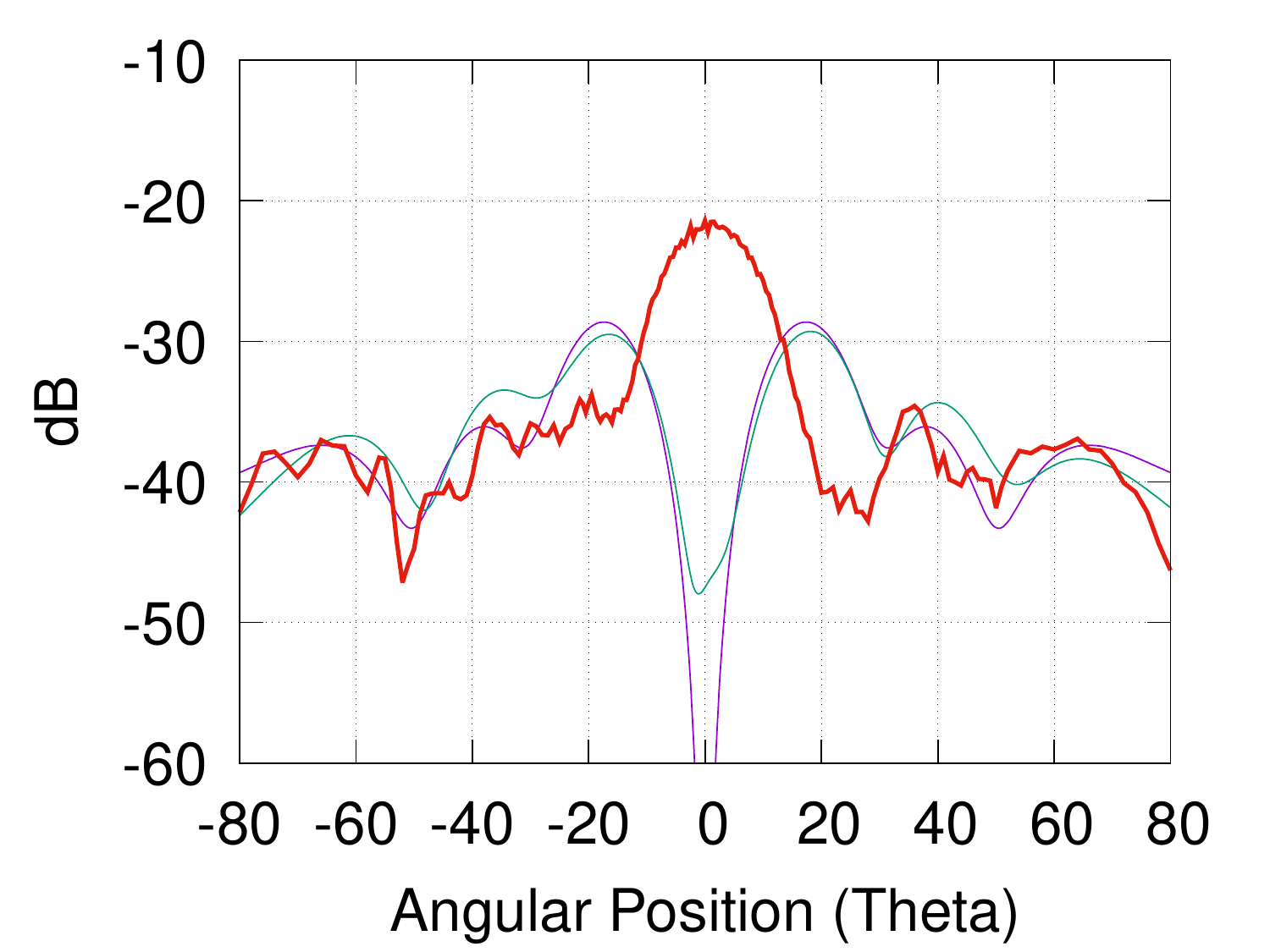}
            \includegraphics[width=0.32\textwidth]{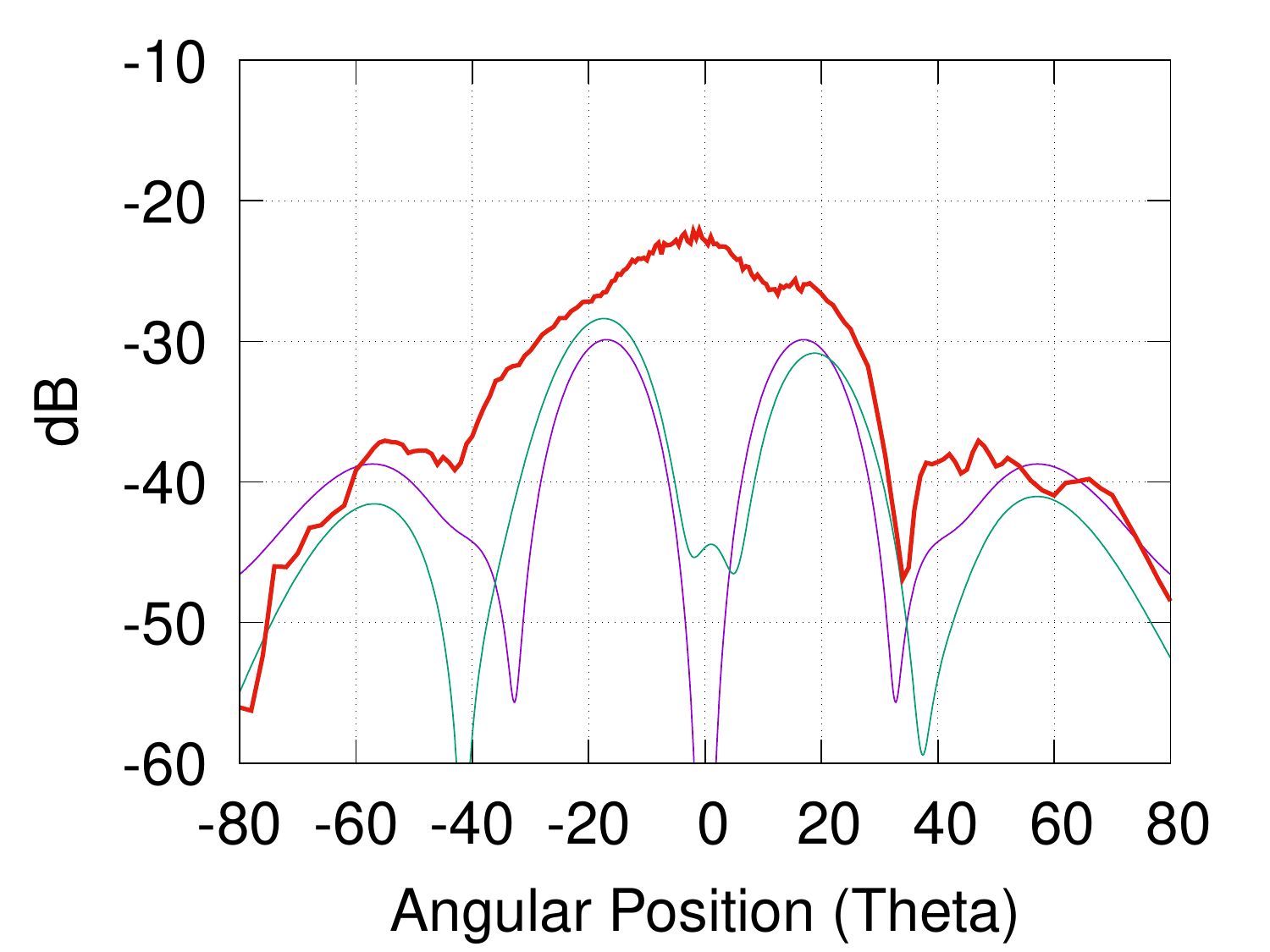}
            \caption{Simulated patterns and comparison with the electromagnetic measurements for the antenna n. 7 (140\,GHz -left-, 150\,GHz -middle-, 160\,GHz -right-). The top row is the co-polar H-plane pattern, the middle and bottom rows are the $\pm 45^\circ$ cross polarization patterns, respectively.}
            \label{fig:Antenna_7}
        \end{figure}

        In this case we performed electromagnetic simulations of two specific antennas: the central one and the antenna number 7, which is one of the peripheral units and the one with the worst measured cross polarization. For each antenna we considered their mechanically measured diameters and a misalignment based on the pin mechanical measurements, which showed that the mean deviation between the pin hole radius and its nominal value is 0.015\,mm. We considered the vector position $\vec{r}$ of each antenna hole in the polar coordinate system, $\vec{r}=(r\cos\theta,r\sin\theta)$, and we assumed $r$ normally distributed around $\mu=\ $0.0\,mm with $\sigma=\ $0.015\,mm and $\theta$ randomly distributed between $[0,2\pi)$.
        We then compared the results with the simulation of a perfectly aligned profile based on the measured antenna diameters and with the electromagnetic measurement. We show the results in Figures~\ref{fig:Antenna_18} (central antenna) and~\ref{fig:Antenna_7} (antenna n. 7)

        Our results show that the misalignment does not introduce significant deviations in the co-polar patterns and there is still a good agreement with the measurement. The cross polarization pattern, instead, is sensitive to the alignment (as expected) and we can appreciate a difference between the two simulated patterns. This effect, however, is not enough to reproduce the measured cross polarization level, and this result is consistent with the one found in section~\ref{sec_esimate_misalignment}, where we found that we need at least 0.03\,mm misalignment to reproduce the data.

        In conclusion, the deviation of the pin diameters from their nominal value has probably caused a deterioration in the cross polarization of the array but this does not account for all the observed effect. We still need to assess the impact of the cross polarization induced by the transition from circular to rectangular waveguide, and this will be done by future measurement in which we will replace the transition with a custom-designed orthomode transducer.

\end{document}